\documentclass[11pt,a4paper]{article}
\usepackage{jheppub}
\usepackage[english]{babel}
\usepackage{graphicx}
\usepackage{framed}

\usepackage{dsfont}
\usepackage{listings}
\usepackage{systeme}
\usepackage{ytableau}
\usepackage{tensor}
\usepackage{tikz}

\usepackage{xcolor}

\definecolor{codegreen}{rgb}{0,0.6,0}
\definecolor{codegray}{rgb}{0.5,0.5,0.5}
\definecolor{codepurple}{rgb}{0.58,0,0.82}
\definecolor{backcolour}{rgb}{0.95,0.95,0.92}
\definecolor{light-gray}{gray}{0.95} 

\lstdefinestyle{mystyle}{
    backgroundcolor=\color{light-gray},   
    commentstyle=\color{codegreen},
    keywordstyle=\color{magenta},
    numberstyle=\tiny\color{codegray},
    stringstyle=\color{codepurple},
    basicstyle=\ttfamily\footnotesize,
    breakatwhitespace=false,         
    breaklines=false,   
    captionpos=b,   
    keepspaces=true,
    numbersep=5pt,                  
    showspaces=false,                
    showstringspaces=false,
    showtabs=false,                  
    tabsize=2
}

\usepackage{latexsym}
\usepackage[normalem]{ulem}
\usepackage{amsmath}
\usepackage{amsthm}
\usepackage{amssymb}
\usepackage{amsfonts}
\usepackage{enumerate}
\usepackage[utf8]{inputenc}
\usepackage{color}
\theoremstyle{plain}

\newtheorem*{remark2}{Remark}
\theoremstyle{remark}

\newcommand{\Z}{\mathbb{Z}}

\title{Unitarity bounds for all symmetry-constrained 3HDMs}
\author[a]{Miguel P. Bento,}
\author[a]{Jorge C. Romão,}
\author[a]{Jo\~{a}o P. Silva,}
\affiliation[a]{CFTP, Departamento de F\'{i}sica, Instituto Superior T\'{e}cnico,
Universidade de Lisboa,\\
Avenida Rovisco Pais 1, 1049 Lisboa, Portugal}
\emailAdd{miguel.pedra.bento@tecnico.ulisboa.pt}
\emailAdd{jorge.romao@tecnico.ulisboa.pt}
\emailAdd{jpsilva@cftp.ist.utl.pt}

\abstract{Models with three Higgs doublets (3HDM) are the source of much recent
activity, for they are key components of many solutions to the problems of
the Standard Model; from extra sources of CP violation to Dark Matter
candidates. We compute explicitly the theoretical bounds for all
symmetry-constrained 3HDM arising from the perturbative unitarity of
two-to-two scattering amplitudes. In addition, we propose a method based
on principal minors that foregoes diagonalization and which is preferable
in all models (not only 3HDM) dealing with large scattering matrices.}

\begin{document}
\maketitle

\section{Introduction}

The discovery at LHC \cite{ATLAS:2012yve,CMS:2012qbp} of a scalar particle with 125 GeV
has inaugurated the era of experimental exploration of the spontaneous
symmetry breaking (SSB) mechanism.
Questions which are being addressed include the following.
Is there only one scalar particle?
Since there are multiple fermion families,
perhaps there are also more scalar families,
naturally urging one to study N Higgs doublet models (NHDM) ---
for reviews see, for example, \cite{Gunion:1989we,Branco:2011iw,Ivanov:2017dad}.
Such models, besides more scalars, usually involve also
couplings of the 125 GeV scalar to gauge bosons and to fermions
at odds with the Standard Model (SM).
How close are the measured couplings from those SM values?
Can NHDM fix problems currently unsolved by the SM?
Indeed,
new sources of CP violation in the scalar sector can
explain the observed baryon asymmetry in the universe,
which cannot be accommodated in the SM.
Moreover,
many NHDM can accommodate one (or more)
dark matter particles.

NHDM usually involve a very large parameter space.
It is customary to reduce the number of parameters through the use of 
symmetries acting on the space of scalar fields.
This is done for several reasons.
First, such symmetries reduce the number of independent parameters, making
it easier to explore the range of possibilities in a given model.
Second, 
when extended to the fermion sector, NHDM usually have flavour changing neutral
scalar interactions, which are severely constrained by experiments in flavour physics.
Some family symmetries set these flavour changing neutral
scalar coupling to zero in a natural way.
The most well known case is the preclusion of such couplings
via a $\mathbb{Z}_2$ symmetry in the 2HDM \cite{Glashow:1976nt,Paschos:1976ay}.
Finally,
when both the Lagrangian and the vacuum respect a given symmetry,
the particle spectrum has the same symmetry;
by setting all SM particles in a sector with no ``charge'' under the
discrete symmetry,
a neutral lightest particle in a sector ``charged'' under the
discrete symmetry is a candidate for dark matter.
The classification of all symmetry-constrained 2HDM can be found in
\cite{Ivanov:2007de} and for the 3HDM in
\cite{Ivanov:2012fp,Ivanov:2014doa,deMedeirosVarzielas:2019rrp,Darvishi:2019dbh}.
This is summarized in section~\ref{sec:list_3HDM}.
The full classification has not yet been achieved for NHDM with $N \ge 4$.

The large parameter space of NHDM is further reduced by
constraints of a theoretical nature, including conditions for
bounded from below potential
\cite{Deshpande:1977rw,Ivanov:2006yq,Maniatis:2006fs,Kannike:2012pe,Maniatis:2015gma},
for the chosen vacuum to constitute indeed the
absolute minimum of the theory
\cite{Ivanov:2006yq,Maniatis:2015gma,Ivanov:2015nea},
and for the scattering matrices to exhibit perturbative unitarity.
These constrains are mandatory in order for the theory and
any phenomenology consequences derived therefrom to make any sense.
This article is dedicated to the study of perturbative unitarity
for all symmetry-constrained 3HDM.
In section~\ref{sec:unitbound},
we write explicitly all scattering sub-matrices,
except for the $\mathbb{Z}_2^{(\mathrm{CP})}$ symmetric 3HDM,
which involves a $9\times9$ scattering matrix.
We also present,
in section~\ref{sec:procedure},
several techniques which are applicable to matrices
of arbitrary dimension, involving the study of principal minors,
and which enable faster numerical studies,
when compared with the numerical determination of the eigenvalues.
The important results of section~\ref{sec:procedure},
are illustrated in section~\ref{sec:applications}
with applications based on some of the matrices obtained
in section~\ref{sec:unitbound}.
In conjunction, we cover all symmetry-constrained 3HDM.

Perturbative unitarity has been thoroughly studied in the context
of the Standard Model in a method championed by Lee, Quigg and Thacker
\cite{Lee:1977eg,Lee:1977yc}.
In the 2HDM, it was computed for
a model with $\mathbb{Z}_2$ symmetry \cite{Ginzburg:2003fe}
and, later, for the general case \cite{Ginzburg:2005dt,Kanemura:2015ska}.
In the 3HDM, it has been studied with an $S_3 \rtimes \mathbb{Z}_2^{(\mathrm{CP})}$ symmetry
\cite{Das:2014fea}, $CP4$ and $\mathbb{Z}_3$ symmetries
\cite{Bento:2017eti} and in the case of $\mathbb{Z}_2 \times \mathbb{Z}_2
\times \mathbb{Z}_2^{(\mathrm{CP})}$
\cite{Moretti:2015cwa}. In the former and latter cases, the authors
started from a Higgs family and then imposed that all complex coefficients
are real,
effectively enlarging the symmetry group.

Concentrating on special cases,
refs.~\cite{Ginzburg:2003fe,Moretti:2015cwa} explored
the use of both the electric charge and the Abelian charges
of the discrete symmetries to classify the scattering matrices.
Here, we use both the hypercharge $\mathcal{Y}$
and electric charge $Q$, following \cite{Bento:2017eti}.
We combine this with a simple algorithm to block diagonalize
the matrices with permutations, presented in appendix~\ref{app:block}.
With this algorithm, we automatically separate
the Abelian charges of the global symmetries that are imposed. Thus,
we often obtain the minimal form for the scattering
matrices, for every possible symmetry.
We show in appendix~\ref{app:M20=M2++} that 
some scattering matrices always coincide, thus simplifying the analysis.

We include the simplest explicit formulae for any particular symmetry-constrained 3HDM,
despite the fact that some models can be obtained as limits of models with a smaller symmetry.
We do this for three reasons. First, the reader can simply concentrate on the particular
model of interest and its notation, without having to set, sometimes error-prone 
limits (see reason three).
Second, higher symmetries usually turn a large matrix into its smaller blocks,
where exact formulae for the eigenvalues then become possible.
Third,
consider a subgroup $G'$ of a larger symmetry $G$.
It is often the case that the potential invariant under $G'$
is simpler to see (or more commonly studied in the literature) in a basis
where the extension to $G$ becomes quite complicated.
Said otherwise,
the natural basis to study the $G$-invariant potential
and the natural basis to study the $G'$-invariant potential
are often at odds with each other.
This problem is discussed in detail in appendix~\ref{app:arrows}.

Throughout the paper, we will use the notation of \cite{Ferreira:2008zy},
which denotes the real (complex) coefficients by $r_i$ ($c_i$).
We summarize the notations in appendix~\ref{app:notations},
by stating some common alternatives.

\section{\label{sec:list_3HDM}Symmetry-constrained 3HDMs}

The scalar potential of the most general 3HDM is given by
\begin{equation}
V_H =  \mu_{ij} (\Phi^\dagger_i \Phi_j) + z_{ij,kl} 
(\Phi^\dagger_i \Phi_j)(\Phi^\dagger_k \Phi_l) 
= - \mathcal{L}_\mathrm{Higgs} \, ,
\label{eq:VH}
\end{equation}
with $i,j,k,l$ running from $1$ to $3$.
Before using the freedom to perform unitary transformations in the space of
scalar fields, one has the following independent parameters
\cite{Ferreira:2008zy} in the potential \eqref{eq:VH}:
$\mu_{ij}$ has 3 real and 3 complex (6 magnitudes and 3 phases);
$z_{ij,kl}$ has 9 real and 18 complex (27 magnitudes and 18 phases).
The first counting is trivial, since $\mu_{ij}$ is a $3\times3$ Hermitian matrix,
while that for $z_{ij,kl}$ is easily seen from the parametrization
in \eqref{eq:FS_param}.
Thus, the most general 3HDM has
12 real and 21 complex (33 magnitudes and 21 phases) parameters.
However, one can choose a different parametrization for the scalar
fields, using a $3\times3$ unitary transformation, which keeps the kinetic
terms invariant.
Such a transformation can be used to take out 3 magnitudes and 5 phases from
the parameters of $V_H$ (one further overall phase in the unitary
transformation has no impact on $V_H$).
This leaves 30 independent magnitudes and 16 independent phases in
$V_H$.

In the 3HDM,
many symmetries may be imposed on the potential
as to prevent flavour changing neutral currents (FCNC),
model dark matter or impose CP properties in the theory.

\begin{figure}[!htb]
   \centering
\begin{tikzpicture}
  \node (E) {$\{e\}$};
  \node (Z2) [above of=E] {$\Z_2$};
  \node (Z2Z2) [above of=Z2] {$\Z_2\times\Z_2$};
  \node (Z3) [node distance=2cm, right of=Z2Z2] {$\Z_3$};
  \node (Z4) [node distance=2cm, left of=Z2Z2] {$\Z_4$};
  \node (D4) [above of=Z4] {$D_4$};
  \node (A4) [above of=Z2Z2] {$A_4$};
  \node (S4) [node distance=1.5cm, above of=A4] {$S_4$};
  \node (S3)[node distance=2cm, right of=A4]{$S_3$};
  \node (D54) [above of=S3] {$\Delta(54)$};
  \node (S36) [above of=D54] {$\Sigma(36)$};
  \draw[->] (E) to node {} (Z2);
  \draw[->] (E) to node {} (Z3);
  \draw[->] (Z2) to node {} (Z4);
  \draw[->] (Z2) to node {} (Z2Z2);
  \draw[->] (Z2) to node {} (S3);
  \draw[->] (Z3) to node {} (S3);
  \draw[->] (Z3) to node {} (A4);
  \draw[->] (Z4) to node {} (D4);
  \draw[->] (Z4) to node {} (S36);
  \draw[->] (Z2Z2) to node {} (D4);
  \draw[->] (Z2Z2) to node {} (A4);
  \draw[->] (A4) to node {} (S4);
  \draw[->] (D4) to node {} (S4);
  \draw[->] (S3) to node {} (S4);
  \draw[->] (S3) to node {} (D54);
  \draw[->] (D54) to node {} (S36);
\end{tikzpicture}
\caption{Tree of finite realizable groups of Higgs-family transformations in 3HDM.}
   \label{fig:tree}
\end{figure}
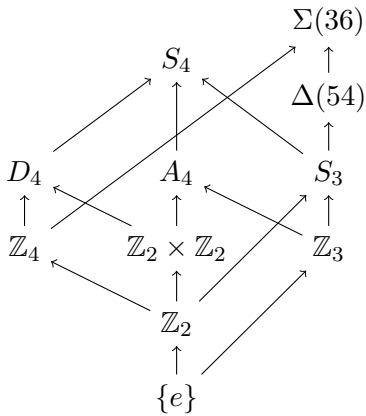

The study of symmetries in the 3HDM has been thoroughly performed
in \cite{Ivanov:2012fp,Ivanov:2014doa,deMedeirosVarzielas:2019rrp,Darvishi:2019dbh}.
In fig.~\ref{fig:tree},
we illustrate the map of realizable discrete Higgs-family symmetries obtained in
\cite{Ivanov:2012fp}.\footnote{We note that throughout this paper we distinguish
the semidirect and direct products by commutativity of the involved symmetries. Thus,
$A \rtimes B$ and $A \times B$ are the same if the generators of $A$ and $B$ commute.}
By ``realizable'' symmetry we mean a symmetry which, when imposed on the potential,
does not yield a potential with a larger symmetry.
To be specific, consider the 2HDM.
Imposing $\mathbb{Z}_3$ on the 2HDM scalar potential,
it becomes immediately invariant under the full Peccei-Quinn $U(1)$ symmetry.
Thus, there is \textit{no} realizable $\mathbb{Z}_3$ 2HDM.
In contrast,
imposing $\mathbb{Z}_3$ on the 3HDM scalar potential does not
lead to a potential invariant under a larger symmetry.
Thus, \textit{there exists} a realizable $\mathbb{Z}_3$ 3HDM.
The full list of realizable discrete symmetries in the 3HDM was composed in
\cite{Ivanov:2012fp} which we summarize in table~\ref{table:1}.
In table~\ref{table:2} we summarize the continuous groups in the 3HDM.

\begin{table}[h!]
\centering
\begin{tabular}{ |c||c| }
\hline
\multicolumn{2}{|c|}{Discrete symmetries in the 3HDM} \\
\hline
\hline
Unitary & $\mathbb{Z}_2 , \mathbb{Z}_3, \mathbb{Z}_2 \times \mathbb{Z}_2,
\mathbb{Z}_4, S_3 , D_4, A_4, S_4, \Delta(54) , \Sigma(36)$ \\
\hline
\hline
Anti-unitary (GCP) & $\mathbb{Z}_2^{(\mathrm{CP})} , \mathbb{Z}_2 \times \mathbb{Z}_2^{(\mathrm{CP})}
, \mathbb{Z}_2 \times \mathbb{Z}_2 \times \mathbb{Z}_2^{(\mathrm{CP})},
CP4,\mathbb{Z}_3 \rtimes \mathbb{Z}_2^{(\mathrm{CP})}$, 
\\
&
$S_3 
\times \mathbb{Z}_2^{(\mathrm{CP})} 
, \Delta(54) \rtimes \mathbb{Z}_2^{(\mathrm{CP})}$ \\ 
\hline
\end{tabular}
\caption{Full list of discrete symmetries in the 3HDM,
where $\mathbb{Z}_2^{(\mathrm{CP})}$ stands for the usual CP.}
\label{table:1}
\end{table}

\begin{table}[h!]
\centering
\begin{tabular}{ |c||c| }
\hline
\multicolumn{2}{|c|}{Continuous symmetries in the 3HDM} \\
\hline
\hline
Abelian & $\mathrm{U}(1)_1, \mathrm{U}(1)_2 , \mathrm{U}(1)_2 \times \mathbb{Z}_2 ,
\mathrm{U}(1) \times \mathrm{U}(1)$ \\
\hline
\hline
Non-abelian & $\mathrm{U}(2), \mathrm{O}(2) , \mathrm{SU}(3) , \mathrm{SO}(3)$ \\ 
\hline
\end{tabular}
\caption{List of continuous symmetries in the 3HDM.}
\label{table:2}
\end{table}

\section{\label{sec:procedure}Optimized unitarity bounds}

In this section we will provide for both necessary
conditions for unitarity in any theory and a procedure
that greatly improves the usual method.

In the literature, the standard route for unitarity
bounds (and the one we will pursue later, in section~\ref{sec:unitbound})
is to build the scattering matrices and
diagonalize them. Then, one proceeds to impose a
bound on the eigenvalues such that for a scattering
matrix $A$, its eigenvalues $\lambda_i$ are bounded
by $|\lambda_i| < 8 \pi$. This method was spearheaded in
\cite{Lee:1977eg, Lee:1977yc}.

\subsection{Unitarity bounds without diagonalization}

As stated before, the standard method relies heavily 
on diagonalization,
which (barring an explicit formula,
impossible for matrices larger than $4\times4$)
has to be performed numerically for each point in the parameter space of the model.
But, this is not the most efficient method.
In this section we propose an approach which is
based on lemma 10.4.1 of \cite{algebra}.
If $A$ is an Hermitian matrix with eigenvalues
$\lambda_i$, then $A + c I$ has eigenvalues
$\lambda_i + c$. Then, we can use this simple statement
and Sylvester's criterion involving principal
minors~\footnote{Principal minors
have also been used by \cite{Kannike:2012pe} in the different context of searching
for bounded from below conditions in scalar potentials.}
to state the following remark.

\begin{remark2}
Let $A$ be an $n\times n$ Hermitian matrix and $\lambda_i$ its eigenvalues.
Then the following statements are equivalent:
\begin{enumerate}
        \item The eigenvalues are bounded as $|\lambda_i| < c$;
        \item The determinants of all the upper left k-by-k submatrices of $A + c I$
        and $c I - A$ are positive;
        \item The leading principal minors $D_k(A + cI)$ and
        $D_k(cI - A)$ are positive.
\end{enumerate}
\end{remark2}

Thus, if $A$ is a scattering matrix with bounds on its eigenvalues
$|\lambda_i| < 8 \pi$, then
\begin{align}
    D_k(A + 8 \pi I) > 0 \quad \text{and} \quad D_k(8 \pi I - A) > 0 \, ,
\end{align}
such that
\begin{equation}
    D_k(A + 8 \pi I)
    =
    \begin{vmatrix}
     A_{11} + 8 \pi & \cdots & A_{1k} \\
     \vdots & \ddots & \vdots \\
     A_{k1} & \cdots & A_{kk} + 8 \pi
    \end{vmatrix} \, .
\end{equation}
In particular, $D_1(A + 8 \pi I) = A_{11}+8\pi$ and $D_n(A + 8 \pi I) = \det (A+8\pi I)$.

Although not needed, one may also add further conditions.
Specifically, if the leading principal minors $D_k(A +  c I)$ and $D_k(c I - A)$ are positive,
then all of its principal minors, not
just the leading ones, are positive. A direct consequence
of this assertion is that $|\lambda_i| < c$ also implies
the following remark:
\begin{remark2}
Let $A$ be an $n\times n$ Hermitian matrix and $\lambda_i$ its eigenvalues.
Then if the eigenvalues are bounded as $|\lambda_i| < c$,
it is a necessary condition that
\begin{equation}\label{eq:diagonal_bound}
    |A_{i i}| < c \, , \quad i = 1,2,\cdots, n \, .
\end{equation}
\end{remark2}
This has already been pointed out, through a different argument,
in \cite{Ginzburg:2005dt},
where unitarity bounds for larger matrices were being considered.

\subsection{Necessary conditions for unitarity in a NHDM}

When dealing with some NHDM model with many parameters,
some general bounds may be extracted by
looking at scattering matrices and using the conditions
of eq.~\eqref{eq:diagonal_bound}.
In a general NHDM, we have that\footnote{We note the important
fact that in some cases, unitarity supersedes perturbativity,
as evidenced by $|\lambda_{ii,ii}| < \frac{4 \pi}{3}$.}
\begin{align}\label{eq:reals_bounds_nhdm}
    &|\lambda_{ii,ii}| < \frac{4 \pi}{3} \, , \nonumber \\[2mm]
    &|\lambda_{ii,jj}| < 4 \pi \, , \nonumber \\[2mm]
    &|\lambda_{ii,jj}+2 \lambda_{ij,ji}| < 4 \pi \, .
\end{align}
In particular, for any 3HDM we have the necessary
(but not sufficient) unitarity constraints
\begin{align}\label{eq:reals_bounds}
    &|r_1|, \, |r_2|, \, |r_3| < \frac{4 \pi}{3} \, , \nonumber \\[2mm]
    &|r_4|, \, |r_5|, \, |r_6| < 4 \pi \, , \nonumber \\[2mm]
    &|r_4+2 r_7| , \, |r_5+2 r_8| , \, |r_6+2 r_9| < 4 \pi \, ,
\end{align}
confirming the particular result of eq.~(3.4) of \cite{Hernandez-Sanchez:2020aop},
obtained for the case of the $\mathbb{Z}_2 \times \mathbb{Z}_2 \times
\mathbb{Z}_2^{(CP)}$.

\subsection{An improved procedure}

Although it may not seem at first hand an improvement
to the standard method, the technique with
the leading principal minors yields four main advantages:
\begin{itemize}
\item Determinants are polynomial in nature and therefore
more numerically stable, as root problems may occur in diagonalizations;
\item Both determinants and diagonalization are tipically
$\mathcal{O}(n^3)$, although the former is much faster;
\item The use of eq.~\eqref{eq:diagonal_bound} enables
a timely choice of the random matrices. We only consider
random matrices which check $|A_{ii}| < c$;
\item As it is much faster, it enables a much more thorough and (thus) reliable scan
of the parameter space;
\item Analytical inequalities are trivial to compute,
regardless of the size of the scattering matrix.
\end{itemize}
Thus, we present an example of the use of this technique
with the following procedure:
\begin{enumerate}
\item Sample a very large number of random Hermitian matrices
by making them check eq.~\eqref{eq:diagonal_bound};
\item Loop through the random Hermitian matrices calculating the determinants
$D_2(A + cI)$ and $D_2(cI - A)$;
\item Check positivity of the determinants;
\item Trim the remaining Hermitian matrices;
\item Go to step 2, but now compute $D_3(A + cI)$ and $D_3(cI - A)$
until we reach the full n-by-n determinants.
\end{enumerate}
When finished, the remaining matrices are valid
scattering matrices through unitarity.

We tested the comparison between the methods with minors and
eigenvalues with a python code, which we include in the paper
as an ancillary file. In this test, we ran unitarity through
400000 symmetric matrices with size $5\times 5$. We concluded
that our method runs about four times faster in
this example.\footnote{The generalization from symmetric to
Hermitian matrices is trivial to perform.}

The procedure proposed here is interesting even for simple $3\times3$ matrices.
We will illustrate this point in section~\ref{sec:applications},
using some $3\times3$ matrices which show up in our discussion
of the scattering matrices for all symmetry-constrained 3HDM models,
to be performed later, in section~\ref{sec:unitbound}.

\section{\label{sec:applications}Conditions for larger matrices}

In the symmetry-constrained 3HDM cases to be presented in section~\ref{sec:unitbound} below,
we will find many matrices of large dimension.
Even in the case of $3\times3$ matrices,
we can use the formula for solutions of cubic equations or, else,
we can utilize the new procedure described in section~\ref{sec:procedure}.
In this section, we provide a few examples of the latter.
Though simple,
they illustrate well how powerful the procedure in section~\ref{sec:procedure} is.

\subsection{The $\mathbb{Z}_2 \times \mathbb{Z}_2$ symmetry}

In this case we will find,
\begin{equation}
    M^{++}_2 \supset A = 2
    \begin{pmatrix}
     r_1 & c_3 & c_5 \\
     c_3^* & r_2 & c_{17} \\
     c_5^* & c_{17}^* & r_3
    \end{pmatrix} \, ,
\end{equation}
and the conditions are
\begin{align}
    D_1(A + 8 \pi I) > 0 \Rightarrow&\, r_1 > -4 \pi \, , \nonumber \\[2mm]
    D_2(A + 8 \pi I) > 0 \Rightarrow&\, \left(r_1+4 \pi \right) 
    \left(r_2+4 \pi \right) - \left| c_3\right|^2 > 0 \, , \nonumber \\[2mm]
    D_3(A + 8 \pi I) > 0 \Rightarrow&\, 
    2 \Re \left[ c_3 c_{17} c_5^* \right]+\left(r_1+4 \pi \right) 
    \left(r_2+4 \pi \right) \left(r_3+4 \pi \right) \nonumber \\
    &- \left(r_3+4 \pi \right) \left| c_3\right|^2
    -\left(r_1+4 \pi \right) \left| c_{17}\right|^2
    -\left(r_2+4 \pi \right) \left| c_5\right|^2 > 0 \nonumber \\
    &\Leftrightarrow \det \left( A + 8 \pi I \right) > 0 \, , \nonumber \\[2mm]
\end{align}
and
\begin{align}
    D_1(8 \pi I - A) > 0 \Rightarrow&\, r_1 < 4 \pi \, , \nonumber \\[2mm]
    D_2(8 \pi I - A) > 0 \Rightarrow&\, \left(r_1 - 4 \pi \right) 
    \left(r_2 - 4 \pi \right) - \left| c_3\right|^2 > 0 \, , \nonumber \\[2mm]
    D_3(8 \pi I - A) > 0 \Rightarrow&\, 
    \det \left( 8 \pi I - A \right) > 0 \, . \nonumber \\[2mm]
\end{align}
With these six conditions we have necessary and sufficient conditions
for unitarity. We may also add, in consequence of eq.~\eqref{eq:diagonal_bound},
that $|r_2| < 4 \pi$ and $|r_3| < 4 \pi$, althought it does not yield
any new information.

The next matrix is
\begin{equation}
    M^{+}_0 \supset A = 2
    \begin{pmatrix}
     r_1 & r_7 & r_8 \\
     r_7 & r_2 & r_9 \\
     r_8 & r_9 & r_3
    \end{pmatrix} \, ,
\end{equation}
and the conditions are
\begin{align}
    D_1(A + 8 \pi I) > 0 \Rightarrow&\, r_1 > -4 \pi \, , \nonumber \\[2mm]
    D_2(A + 8 \pi I) > 0 \Rightarrow&\, \left(r_1+4 \pi \right) 
    \left(r_2+4 \pi \right) - r_7^2 > 0 \, , \nonumber \\[2mm]
    D_3(A + 8 \pi I) > 0 \Rightarrow&\, 
    2 r_7 r_8 r_9 +\left(r_1+4 \pi \right) 
    \left(r_2+4 \pi \right) \left(r_3+4 \pi \right) \nonumber \\
    &- \left(r_3+4 \pi \right) r_7^2
    -\left(r_1+4 \pi \right) r_9^2
    -\left(r_2+4 \pi \right) r_8^2 > 0 \nonumber \\
    &\Leftrightarrow \det \left( A + 8 \pi I \right) > 0 \, , \nonumber \\[2mm]
\end{align}
and
\begin{align}
    D_1(8 \pi I - A) > 0 \Rightarrow&\, r_1 < 4 \pi \, , \nonumber \\[2mm]
    D_2(8 \pi I - A) > 0 \Rightarrow&\, \left(r_1 - 4 \pi \right) 
    \left(r_2 - 4 \pi \right) - r_7^2 > 0 \, , \nonumber \\[2mm]
    D_3(8 \pi I - A) > 0 \Rightarrow&\, 
    \det \left( 8 \pi I - A \right) > 0 \, . \nonumber \\[2mm]
\end{align}
The next matrix is
\begin{equation}
    M^{0}_0 \supset A = 2
    \begin{pmatrix}
     3 r_1 & 2 r_4 + r_7 & 2 r_5 + r_8 \\
     2 r_4 + r_7 & 3 r_2 & 2 r_6 + r_9 \\
     2 r_5 + r_8 & 2 r_6 + r_9 & 3 r_3
    \end{pmatrix} \, ,
\end{equation}
and the conditions are
\begin{align}
    D_1(A + 8 \pi I) > 0 \Rightarrow&\, 3 r_1 > -4 \pi \, , \nonumber \\[2mm]
    D_2(A + 8 \pi I) > 0 \Rightarrow&\, \left(3 r_1+4 \pi \right) 
    \left(3 r_2+4 \pi \right) - \left(2 r_4+r_7\right)^2 > 0 \, , \nonumber \\[2mm]
    D_3(A + 8 \pi I) > 0 \Rightarrow&\, \det \left( A + 8 \pi I \right) > 0 
    \, , \nonumber \\[2mm]
\end{align}
and
\begin{align}
    D_1(8 \pi I - A) > 0 \Rightarrow&\, 3 r_1 < 4 \pi \, , \nonumber \\[2mm]
    D_2(8 \pi I - A) > 0 \Rightarrow&\, \left(3 r_1-4 \pi \right) 
    \left(3 r_2-4 \pi \right) - \left(2 r_4+r_7\right)^2 > 0  \, , \nonumber \\[2mm]
    D_3(8 \pi I - A) > 0 \Rightarrow&\, 
    \det \left( 8 \pi I - A \right) > 0 \, . \nonumber \\[2mm]
\end{align}
We note that this matrix yields a stronger bound on $r_1 , r_2 , r_3$
than the previous ones. We have $|r_i| < 4 \pi / 3$ for $i=1,2,3$.

\subsection{The $S_3$ symmetry}

In the case of this symmetry,
we will find
\begin{equation}
    M^{+}_0 \supset A = 2
    \begin{pmatrix}
     r_4 & c_{12} & c_{12}^* \\
     c_{12}^* & r_5 & c_{11} \\
     c_{12} & c_{11}^* & r_5
    \end{pmatrix} \, ,
\end{equation}
and the conditions are
\begin{align}
    D_1(A + 8 \pi I) > 0 \Rightarrow&\, r_4 > -4 \pi \, , \nonumber \\[2mm]
    D_2(A + 8 \pi I) > 0 \Rightarrow&\, \left(r_4+4 \pi \right) 
    \left(r_5+4 \pi \right) - \left| c_{12}\right|^2 > 0 \, , \nonumber \\[2mm]
    D_3(A + 8 \pi I) > 0 \Rightarrow&\, 
    2 \Re \left[ c_{11} c_{12}^2 \right]+\left(r_1+4 \pi \right) 
    \left(r_5+4 \pi \right)^2 \nonumber \\
    &- \left(r_4+4 \pi \right) \left| c_{11}\right|^2
    -2 \left(r_5+4 \pi \right) \left| c_{12}\right|^2 > 0 \nonumber \\
    &\Leftrightarrow \det \left( A + 8 \pi I \right) > 0 \, , \nonumber \\[2mm]
\end{align}
and
\begin{align}
    D_1(8 \pi I - A) > 0 \Rightarrow&\, r_4 < 4 \pi \, , \nonumber \\[2mm]
    D_2(8 \pi I - A) > 0 \Rightarrow&\, \left(r_4 - 4 \pi \right) 
    \left(r_5 - 4 \pi \right) - \left| c_{11}\right|^2 > 0 \, , \nonumber \\[2mm]
    D_3(8 \pi I - A) > 0 \Rightarrow&\, 
    \det \left( 8 \pi I - A \right) > 0 \, . \nonumber \\[2mm]
\end{align}
The next matrix is
\begin{equation}
    M^{0}_0 \supset A = 2
    \begin{pmatrix}
     r_4+2 r_7 & 3 c_{12} & 3 c_{12}^* \\
     3 c_{12}^* & r_5+2 r_8 & 3 c_{11} \\
     3 c_{12} & 3 c_{11}^* & r_5+2 r_8
    \end{pmatrix} \, ,
\end{equation}
and the conditions are
\begin{align}
    D_1(A + 8 \pi I) > 0 \Rightarrow&\, (r_4+2 r_7)+4 \pi >0 \, , \nonumber \\[2mm]
    D_2(A + 8 \pi I) > 0 \Rightarrow&\, \left(r_4+2 r_7+4 \pi \right)
    \left(r_5+2 r_8+4 \pi \right) - 9 \left| c_{12}\right|^2 > 0 \, , \nonumber \\[2mm]
    D_3(A + 8 \pi I) > 0 \Rightarrow&\, \det \left( A + 8 \pi I \right) > 0 
    \, , \nonumber \\[2mm]
\end{align}
and
\begin{align}
    D_1(8 \pi I - A) > 0 \Rightarrow&\, 4 \pi - (r_4+2 r_7) > 0
    \, , \nonumber \\[2mm]
    D_2(8 \pi I - A) > 0 \Rightarrow&\, \left(r_4+2 r_7 - 4 \pi \right)
    \left(r_5+2 r_8 - 4 \pi \right) - 9 \left| c_{12}\right|^2 > 0  
    \, , \nonumber \\[2mm]
    D_3(8 \pi I - A) > 0 \Rightarrow&\, 
    \det \left( 8 \pi I - A \right) > 0 \, . \nonumber \\[2mm]
\end{align}

\section{\label{sec:unitbound}Unitarity bounds for all symmetry-constrained 3HDM}

In NHDM, there are neutral scalars and charge $\pm$ scalars (in units
of the positron charge).
Thus,
in $2 \rightarrow 2$ scattering,
the initial (and final) charges can be $0$, $+$ (same scattering matrices as $-$),
or $++$ (same matrices as $- -$).
Following the method provided in \cite{Bento:2017eti} for tree-level
unitarity bounds, we present the eigenvalues for the matrices
$M_2^{++},M_2^{+},M_0^{+},M_2^{0},M_0^{0}$, where
$M_{2 \mathcal{Y}}^{Q}$ are scattering matrices
with hypercharge $2\mathcal{Y}$
and~\footnote{In our notation, $Q=T_3+\mathcal{Y}$,
where $T_3$ is the third component of weak isospin.}
 electric charge $Q$.
We will state that
a matrix $M$ is ``equal" ($=$) to its block-diagonal form
by only making use of permutations.
To identify the relevant permutations, we use the algorithm presented in
appendix~\ref{app:block}.
For $M_0^{0}$ we will also use orthogonal matrices of the type
\begin{equation}
    \mathcal{O} = \frac{1}{\sqrt{2}}
    \begin{pmatrix}
     -\mathds{1} & \mathds{1} \\
     \mathds{1} & \mathds{1}
    \end{pmatrix} \, ,
\end{equation}
to further reduce the size of the matrices. For this operation
we will use the symbol of ``similar" ($\sim \,$). These operations will
simplify the presentation, while at the same time preserving
the final results. It is then without loss of generality that we use them.

We will denote the eigenvalues as $\Lambda^{Q,2 \mathcal{Y}}_i$
for the $i$th eigenvalue of charge $Q$ and hypercharge $\mathcal{Y}$.\footnote{Since
when $Q=++$ only $2\mathcal{Y}=2$ exists,
we suppress the explicit reference to the hypercharge in the corresponding
eigenvalues: $\Lambda^{++,2}_i \rightarrow \Lambda^{++}_i$.}
These correspond to the eigenvalues of the corresponding matrices
$M_{2 \mathcal{Y}}^{Q}$.
The unitarity bounds provided by using $|\Lambda| < 8 \pi$
for symmetry-constrained 3HDMs are given in the following subsections.

\subsection{The $\mathbb{Z}_2^{(\mathrm{CP})}$ symmetry}

By imposing $G = \mathbb{Z}_2^{(\mathrm{CP})}$ we get the most general
3HDM but now with real coefficients. This is the smallest symmetry
possible. In general, we must contend with $9\times9$ irreducible scattering matrices
and, thus, its unitarity bounds should be obtained numerically.
As mentioned in section~\ref{sec:procedure}, for these cases we
advocate a faster procedure based on principle minors.

\subsection{The $\mathbb{Z}_2$ symmetry}

By imposing $G = \mathbb{Z}_2$ with representation $\mathrm{diag}(1,1,-1)$
we get the quartic potential
\begin{align}
    V_{\mathbb{Z}_2} =&
    \sum_{i=1}^3 r_i |\phi_i|^4 + 2 r_4 (\phi_1^\dagger \phi_1)(\phi_2^\dagger \phi_2)
    + 2 r_5 (\phi_1^\dagger \phi_1)(\phi_3^\dagger \phi_3)
    + 2 r_6 (\phi_2^\dagger \phi_2)(\phi_3^\dagger \phi_3)
    \nonumber \\[2mm]
    & + 2 r_7 |\phi_1^\dagger \phi_2|^2
    + 2 r_8 |\phi_1^\dagger \phi_3|^2 + 2 r_9 |\phi_2^\dagger \phi_3|^2
    + \Big[
	2 c_1 (\phi_1^\dagger \phi_1)(\phi_1^\dagger \phi_2) + c_3 (\phi_1^\dagger \phi_2)^2
    \nonumber \\[2mm]
    &+ c_5 (\phi_1^\dagger \phi_3)^2 + 2 c_7 (\phi_1^\dagger \phi_2)(\phi_2^\dagger \phi_2)
    + 2 c_{11} (\phi_1^\dagger \phi_3)(\phi_2^\dagger \phi_3)
    \nonumber \\[2mm]
    &+ 2 c_{13} (\phi_1^\dagger \phi_2)(\phi_3^\dagger \phi_3)
    + 2 c_{14} (\phi_1^\dagger \phi_3)(\phi_3^\dagger \phi_2)
    + c_{17} (\phi_2^\dagger \phi_3)^2 + h.c. \Big]\, ,
\label{V_Z2}
\end{align}
with the following scattering matrices.

\subsubsection*{The matrix $M_2^{++}$}

From $M_2^{++}$ we get
\begin{equation}\label{eq:example_alg}
    \frac{1}{2} \, M_2^{++} = \mathrm{diag} \left\{
    \begin{pmatrix}
     r_1 & \sqrt{2} c_1 & c_3 & c_5 \\
     \sqrt{2} c_1^* & r_4+r_7 & \sqrt{2} c_7 & \sqrt{2} c_{11} \\
     c_3^* & \sqrt{2} c_7^* & r_2 & c_{17} \\
     c_5^* & \sqrt{2} c_{11}^* & c_{17}^* & r_3
    \end{pmatrix}
    ,
    \begin{pmatrix}
     r_5+r_8 & c_{13}+c_{14} \\
     c_{13}^*+c_{14}^* & r_6+r_9
    \end{pmatrix} \right\}
\end{equation}
and thus we get the eigenvalues of $M_2^{++}$:
\begin{align}
    &\Lambda^{++}_{1-4} = \text{Eigenvalues of first matrix} \, , \nonumber \\[2mm]
    &\Lambda^{++}_{5,6} = \pm \sqrt{4 \left|c_{13}+c_{14}\right|^2
    +\left(r_5-r_6+r_8-r_9\right){}^2}+r_5+r_6+r_8+r_9 \, .
\end{align}

\subsubsection*{The matrix $M_2^+$}

From $M_2^{+}$ we get
\begin{equation}
    \frac{1}{2} \, M_2^{+} = \mathrm{diag} \left\{
    \begin{pmatrix}
     r_1 & c_1 & c_1 & c_3 & c_5 \\
     c_1^* & r_4 & r_7 & c_7 & c_{11} \\
     c_1^* & r_7 & r_4 & c_7 & c_{11} \\
     c_3^* & c_7^* & c_7^* & r_2 & c_{17} \\
     c_5^* & c_{11}^* & c_{11}^* & c_{17}^* & r_3
    \end{pmatrix}
    ,
    \begin{pmatrix}
     r_5 & c_{13} & r_8 & c_{14} \\
     c_{13}^* & r_6 & c_{14}^* & r_9 \\
     r_8 & c_{14} & r_5 & c_{13} \\
     c_{14}^* & r_9 & c_{13}^* & r_6
    \end{pmatrix} \right\} \, ,
\end{equation}
with eigenvalues of $M_2^{+}$:
\begin{align}
    &\Lambda^{+,2}_{1-5} = \text{Eigenvalues of first matrix} \, , \nonumber \\[2mm]
    &\Lambda^{+,2}_{6,7} = \Lambda^{++}_{5,6} \, , \nonumber \\[2mm]
    &\Lambda^{+,2}_{8,9} = \pm \sqrt{4 \left|c_{13}-c_{14}\right|^2
    +\left(r_5-r_6-r_8+r_9\right){}^2}+r_5+r_6-r_8-r_9 \, .
\end{align}

\subsubsection*{The matrix $M_0^+$}

From $M_0^{+}$ we get
\begin{equation}
    \frac{1}{2} \, M_0^{+} = \mathrm{diag} \left\{
    \begin{pmatrix}
     r_1 & c_1^* & c_1 & r_7 & r_8 \\
     c_1 & r_4 & c_3 & c_7 & c_{14} \\
     c_1^* & c_3^* & r_4 & c_7^* & c_{14}^* \\
     r_7 & c_7^* & c_7 & r_2 & r_9 \\
     r_8 & c_{14}^* & c_{14} & r_9 & r_3
    \end{pmatrix}
    ,
    \begin{pmatrix}
     r_5 & c_{13} & c_5 & c_{11} \\
     c_{13}^* & r_6 & c_{11} & c_{17} \\
     c_5^* & c_{11}^* & r_5 & c_{13}^* \\
     c_{11}^* & c_{17}^* & c_{13} & r_6
    \end{pmatrix} \right\} \, ,
\end{equation}
with eigenvalues of $M_0^{+}$:
\begin{align}
    &\Lambda^{+,0}_{1-5} = \text{Eigenvalues of first matrix} \, , \nonumber \\[2mm]
    &\Lambda^{+,0}_{6-9} = \text{Eigenvalues of second matrix} \, .
\end{align}

\subsubsection*{The matrix $M_2^0$}

From $M_2^{0}$ we get
\begin{align}
    \frac{1}{2} \, M_2^{0} = \frac{1}{2} \, M_2^{++} \ ,
\end{align}
with eigenvalues of $M_2^{0}$:
\begin{align}
    &\Lambda^{0,2}_{1-6} = \Lambda^{++}_{1-6} \, .
\end{align}

\subsubsection*{The matrix $M_0^0$}

From $M_0^{0}$ we get
\begin{equation}
    \frac{1}{2} \, M_0^{0} \sim \,
    \mathrm{diag} \left\{\frac{1}{2} \, M_0^+ , A , B \right\} \, ,
\end{equation}
with
\begin{equation}
A=
\begin{pmatrix}
 3 r_1 & 3 c_1^* & 3 c_1 & 2 r_4+r_7 & 2 r_5+r_8 \\
 3 c_1 & r_4+2 r_7 & 3 c_3 & 3 c_7 & 2 c_{13}+c_{14} \\
 3 c_1^* & 3 c_3^* & r_4+2 r_7 & 3 c_7^* & 2 c_{13}^*+c_{14}^* \\
 2 r_4+r_7 & 3 c_7^* & 3 c_7 & 3 r_2 & 2 r_6+r_9 \\
 2 r_5+r_8 & 2 c_{13}^*+c_{14}^* & 2 c_{13}+c_{14} & 2 r_6+r_9 & 3 r_3
\end{pmatrix} \, ,
\end{equation}

\begin{equation}
B=
\begin{pmatrix}
 r_5+2 r_8 & c_{13}+2 c_{14} & 3 c_5 & 3 c_{11} \\
 c_{13}^*+2 c_{14}^* & r_6+2 r_9 & 3 c_{11} & 3 c_{17} \\
 3 c_5^* & 3 c_{11}^* & r_5+2 r_8 & c_{13}^*+2 c_{14}^* \\
 3 c_{11}^* & 3 c_{17}^* & c_{13}+2 c_{14} & r_6+2 r_9 \\
\end{pmatrix} \, ,
\end{equation}
with eigenvalues of $M_0^{0}$:
\begin{align}
    &\Lambda^{0,0}_{1-5} = \text{Eigenvalues of second matrix} \, , \nonumber \\[2mm]
    &\Lambda^{0,0}_{6-9} = \text{Eigenvalues of third matrix} \, .
\end{align}

\subsection{The $\mathbb{Z}_2 \times \mathbb{Z}_2^{(\mathrm{CP})}$ symmetry}

By imposing $G = \mathbb{Z}_2 \times \mathbb{Z}_2^{(\mathrm{CP})}$
we get the quartic potential
\begin{align}
    V_{\mathbb{Z}_2 \times \mathbb{Z}_2^{(\mathrm{CP})}} =&
    \sum_{i=1}^3 r_i |\phi_i|^4 + 2 r_4 (\phi_1^\dagger \phi_1)(\phi_2^\dagger \phi_2)
    + 2 r_5 (\phi_1^\dagger \phi_1)(\phi_3^\dagger \phi_3)
    + 2 r_6 (\phi_2^\dagger \phi_2)(\phi_3^\dagger \phi_3)
    \nonumber \\[2mm]
    & + 2 r_7 |\phi_1^\dagger \phi_2|^2
    + 2 r_8 |\phi_1^\dagger \phi_3|^2 + 2 r_9 |\phi_2^\dagger \phi_3|^2
    + \Big[
	2 r_{10} (\phi_1^\dagger \phi_1)(\phi_1^\dagger \phi_2) + r_{11} (\phi_1^\dagger \phi_2)^2
    \nonumber \\[2mm]
    &+ r_{12} (\phi_1^\dagger \phi_3)^2 + 2 r_{13} (\phi_1^\dagger \phi_2)(\phi_2^\dagger \phi_2)
    + 2 r_{14} (\phi_1^\dagger \phi_3)(\phi_2^\dagger \phi_3)
    \nonumber \\[2mm]
    &+ 2 r_{15} (\phi_1^\dagger \phi_2)(\phi_3^\dagger \phi_3)
    + 2 r_{16} (\phi_1^\dagger \phi_3)(\phi_3^\dagger \phi_2)
    + r_{17} (\phi_2^\dagger \phi_3)^2 + h.c. \Big]\, ,
\end{align}
with the following scattering matrices.
This case is obtained from eq.~\eqref{V_Z2} by making all coefficients real.
Here, and in similar cases below,
we stress the fact that all parameters are real by
changing the $c_k$ in the notation of \cite{Ferreira:2008zy},
into $r_j$ with $j\ge 10$.
Specifically, in this case,
we do
$(c_{1}, c_{3}, c_{5}, c_{7}, c_{11}, c_{13}, c_{14}, c_{17})
\rightarrow (r_{10}, r_{11}, r_{12}, r_{13}, r_{14}, r_{15}, r_{16}, r_{17})$.

\subsubsection*{The matrix $M_2^{++}$}

From $M_2^{++}$ we get
\begin{equation}
    \frac{1}{2} \, M_2^{++} = \mathrm{diag} \left\{
    \begin{pmatrix}
     r_1 & \sqrt{2} r_{10} & r_{11} & r_{12} \\
     \sqrt{2} r_{10} & r_4+r_7 & \sqrt{2} r_{13} & \sqrt{2} r_{14} \\
     r_{11} & \sqrt{2} r_{13} & r_2 & r_{17} \\
     r_{12} & \sqrt{2} r_{14} & r_{17} & r_3
    \end{pmatrix}
    ,
    \begin{pmatrix}
     r_5+r_8 & r_{15}+r_{16} \\
     r_{15}+r_{16} & r_6+r_9
    \end{pmatrix} \right\} \, ,
\end{equation}
and thus we get the eigenvalues of $M_2^{++}$:
\begin{align}
    &\Lambda^{++}_{1-4} = \text{Eigenvalues of first matrix} \, , \nonumber \\[2mm]
    &\Lambda^{++}_{5,6} = \pm \sqrt{4 \left(r_{15}+r_{16}\right)^2
    +\left(r_5-r_6+r_8-r_9\right){}^2}+r_5+r_6+r_8+r_9 \, .
\end{align}

\subsubsection*{The matrix $M_2^+$}

From $M_2^{+}$ we get
\begin{equation}
    \frac{1}{2} \, M_2^{+} = \mathrm{diag} \left\{
    \begin{pmatrix}
     r_1 & r_{10} & r_{10} & r_{11} & r_{12} \\
     r_{10} & r_4 & r_7 & r_{13} & r_{14} \\
     r_{10} & r_7 & r_4 & r_{13} & r_{14} \\
     r_{11} & r_{13} & r_{13} & r_2 & r_{17} \\
     r_{12} & r_{14} & r_{14} & r_{17} & r_3
    \end{pmatrix}
    ,
    \begin{pmatrix}
     r_5 & r_{15} & r_8 & r_{16} \\
     r_{15} & r_6 & r_{16} & r_9 \\
     r_8 & r_{16} & r_5 & r_{15} \\
     r_{16} & r_9 & r_{15} & r_6
    \end{pmatrix} \right\} \, ,
\end{equation}
with eigenvalues of $M_2^{+}$:
\begin{align}
    &\Lambda^{+,2}_{1-5} = \text{Eigenvalues of first matrix} \, , \nonumber \\[2mm]
    &\Lambda^{+,2}_{6,7} = \Lambda^{++}_{5,6} \, , \nonumber \\[2mm]
    &\Lambda^{+,2}_{8,9} = \pm \sqrt{4 \left(r_{15}-r_{16}\right)^2
    +\left(r_5-r_6-r_8+r_9\right){}^2}+r_5+r_6-r_8-r_9 \, .
\end{align}

\subsubsection*{The matrix $M_0^+$}

From $M_0^{+}$ we get
\begin{equation}
    \frac{1}{2} \, M_0^{+} = \mathrm{diag} \left\{
    \begin{pmatrix}
     r_1 & r_{10} & r_{10} & r_7 & r_8 \\
     r_{10} & r_4 & r_{11} & r_{13} & r_{16} \\
     r_{10} & r_{11} & r_4 & r_{13} & r_{16} \\
     r_7 & r_{13} & r_{13} & r_2 & r_9 \\
     r_8 & r_{16} & r_{16} & r_9 & r_3
    \end{pmatrix}
    ,
    \begin{pmatrix}
     r_5 & r_{15} & r_{12} & r_{14} \\
     r_{15} & r_6 & r_{14} & r_{17} \\
     r_{12} & r_{14} & r_5 & r_{15} \\
     r_{14} & r_{17} & r_{15} & r_6
    \end{pmatrix} \right\} \, ,
\end{equation}
with eigenvalues of $M_0^{+}$:
\begin{align}
    &\Lambda^{+,0}_{1-5} = \text{Eigenvalues of first matrix} \, , \nonumber \\[2mm]
    &\Lambda^{+,0}_{6,7} = \pm \sqrt{4 \left(r_{14}+r_{15}\right){}^2
    +\left(r_5-r_6+r_{12}-r_{17}\right){}^2}+r_5+r_6+r_{12}+r_{17} \, , \nonumber \\[2mm]
    &\Lambda^{+,0}_{8,9} = \pm \sqrt{4 \left(r_{14}-r_{15}\right){}^2
    +\left(r_5-r_6-r_{12}+r_{17}\right){}^2}+r_5+r_6-r_{12}-r_{17} \, .
\end{align}

\subsubsection*{The matrix $M_2^0$}

From $M_2^{0}$ we get
\begin{align}
    \frac{1}{2} \, M_2^{0} = \frac{1}{2} \, M_2^{++} \ ,
\end{align}
with eigenvalues of $M_2^{0}$:
\begin{align}
    &\Lambda^{0,2}_{1-6} = \Lambda^{++}_{1-6} \, .
\end{align}

\subsubsection*{The matrix $M_0^0$}

From $M_0^{0}$ we get
\begin{equation}
    \frac{1}{2} \, M_0^{0} \sim \,
    \mathrm{diag} \left\{\frac{1}{2} \, M_0^+ , A , B \right\} \, ,
\end{equation}
with
\begin{equation}
A=
\begin{pmatrix}
 3 r_1 & 3 r_{10} & 3 r_{10} & 2 r_4+r_7 & 2 r_5+r_8 \\
 3 r_{10} & r_4+2 r_7 & 3 r_{11} & 3 r_{13} & 2 r_{15}+r_{16} \\
 3 r_{10} & 3 r_{11} & r_4+2 r_7 & 3 r_{13} & 2 r_{15}+r_{16} \\
 2 r_4+r_7 & 3 r_{13} & 3 r_{13} & 3 r_2 & 2 r_6+r_9 \\
 2 r_5+r_8 & 2 r_{15}+r_{16} & 2 r_{15}+r_{16} & 2 r_6+r_9 & 3 r_3
\end{pmatrix} \, ,
\end{equation}

\begin{equation}
B=
\begin{pmatrix}
 r_5+2 r_8 & r_{15}+2 r_{16} & 3 r_{12} & 3 r_{14} \\
 r_{15}+2 r_{16} & r_6+2 r_9 & 3 r_{14} & 3 r_{17} \\
 3 r_{12} & 3 r_{14} & r_5+2 r_8 & r_{15}+2 r_{16} \\
 3 r_{14} & 3 r_{17} & r_{15}+2 r_{16} & r_6+2 r_9 \\
\end{pmatrix} \, ,
\end{equation}
with eigenvalues of $M_0^{0}$:
\begin{align}
    &\Lambda^{0,0}_{1-9} = \Lambda^{+,0}_{1-9} \, , \nonumber \\[2mm]
    &\Lambda^{0,0}_{10-14} = \text{Eigenvalues of first matrix} \, , \nonumber \\[2mm]
    &\Lambda^{0,0}_{15,16} = \pm\sqrt{4 \left(-3 r_{14}+r_{15}+2 r_{16}\right){}^2
    +\left(r_5-r_6+2 r_8-2 r_9-3 r_{12}+3 r_{17}\right){}^2} \nonumber \\
    &\quad \quad \quad \quad +r_5+r_6+2 r_8+2 r_9-3 \left(r_{12}+r_{17}\right) \, , \nonumber \\[2mm]
    &\Lambda^{0,0}_{17,18} = \pm \sqrt{4 \left(3 r_{14}+r_{15}+2 r_{16}\right){}^2
    +\left(r_5-r_6+2 r_8-2 r_9+3 r_{12}-3 r_{17}\right){}^2} \nonumber \\
    &\quad \quad \quad \quad +r_5+r_6+2 r_8+2 r_9+3 \left(r_{12}+r_{17}\right) \, .
\end{align}

\subsection{The $\mathbb{Z}_4$ symmetry}

By imposing $G = \mathbb{Z}_4$ with representation $\mathrm{diag}(i,-i,1)$
we get the quartic potential
\begin{align}
    V_{\mathbb{Z}_4} =&
    \sum_{i=1}^3 r_i |\phi_i|^4 + 2 r_4 (\phi_1^\dagger \phi_1)(\phi_2^\dagger \phi_2)
    + 2 r_5 (\phi_1^\dagger \phi_1)(\phi_3^\dagger \phi_3)
    + 2 r_6 (\phi_2^\dagger \phi_2)(\phi_3^\dagger \phi_3)
    \nonumber \\[2mm]
    & + 2 r_7 |\phi_1^\dagger \phi_2|^2
    + 2 r_8 |\phi_1^\dagger \phi_3|^2 + 2 r_9 |\phi_2^\dagger \phi_3|^2
    + r_{10} \left[ (\phi_1^\dagger \phi_2)^2 + h.c. \right]
    \nonumber \\[2mm]
    &+ 2 r_{11} \left[ (\phi_1^\dagger \phi_3)(\phi_2^\dagger \phi_3) + h.c. \right] \, ,
\end{align}
which can be easily achieved by setting from $\mathbb{Z}_2$
the constraints $\{ c_1, c_5, c_7, c_{13}, c_{14}, c_{17} \} \rightarrow 0$.

Thus, we get the following scattering matrices.

\subsubsection*{The matrix $M_2^{++}$}

From $M_2^{++}$ we get
\begin{equation}
    \frac{1}{2} \, M_2^{++} = \mathrm{diag} \left\{
    \begin{pmatrix}
     r_1 & r_{10} \\
     r_{10} & r_2
    \end{pmatrix}
    ,
    \begin{pmatrix}
     r_4+r_7 & \sqrt{2} r_{11} \\
     \sqrt{2} r_{11} & r_3
    \end{pmatrix}
    ,
    (r_5 + r_8)
    ,
    (r_6 + r_9) \right\} \, ,
\end{equation}
and thus we get the eigenvalues of $M_2^{++}$:
\begin{align}
    &\Lambda^{++}_{1,2} = \pm \sqrt{4 r_{10}^2 +\left(r_1-r_2\right){}^2}+r_1+r_2 
    \, , \nonumber \\[2mm]
    &\Lambda^{++}_{3,4} = \pm \sqrt{8 r_{11}^2 +\left(-r_3+r_4+r_7\right){}^2}
    +r_3+r_4+r_7 \, , \nonumber \\[2mm]
    &\Lambda^{++}_{5} = 2 (r_5 + r_8) \, , \nonumber \\[2mm]
    &\Lambda^{++}_{6} = 2 (r_6 + r_9) \, .
\end{align}

\subsubsection*{The matrix $M_2^+$}

From $M_2^{+}$ we get
\begin{equation}
    \frac{1}{2} \, M_2^{+} = \mathrm{diag} \left\{
    \begin{pmatrix}
     r_4 & r_7 & r_{11} \\
     r_7 & r_4 & r_{11} \\
     r_{11} & r_{11} & r_3
    \end{pmatrix}
    ,
    \begin{pmatrix}
     r_1 & r_{10} \\
     r_{10} & r_2
    \end{pmatrix}
    ,
    \begin{pmatrix}
     r_5 & r_8 \\
     r_8 & r_5
    \end{pmatrix}
    ,
    \begin{pmatrix}
     r_6 & r_9 \\
     r_9 & r_6
    \end{pmatrix} \right\} \, ,
\end{equation}
with eigenvalues of $M_2^{+}$:
\begin{align}
    &\Lambda^{+,2}_{1,2} = \Lambda^{++}_{3,4} \, , \nonumber \\[2mm]
    &\Lambda^{+,2}_{3} = 2 (r_4 - r_7) \, , \nonumber \\[2mm]
    &\Lambda^{+,2}_{4,5} = \Lambda^{++}_{1,2} \, , \nonumber \\[2mm]
    &\Lambda^{+,2}_{6,7} = 2(r_5 \pm r_8) \, \nonumber \\[2mm]
    &\Lambda^{+,2}_{8,9} = 2(r_6 \pm r_9) \, .
\end{align}

\subsubsection*{The matrix $M_0^+$}

From $M_0^{+}$ we get
\begin{equation}
    \frac{1}{2} \, M_0^{+} = \mathrm{diag} \left\{
    \begin{pmatrix}
     r_1 & r_7 & r_8 \\
     r_7 & r_2 & r_9 \\
     r_8 & r_9 & r_3
    \end{pmatrix}
    ,
    \begin{pmatrix}
     r_4 & r_{10} \\
     r_{10} & r_4
    \end{pmatrix}
    ,
    \begin{pmatrix}
     r_5 & r_{11} \\
     r_{11} & r_6
    \end{pmatrix}
    ,
    \begin{pmatrix}
     r_6 & r_{11} \\
     r_{11} & r_5
    \end{pmatrix} \right\} \, ,
\end{equation}
with eigenvalues of $M_0^{+}$:
\begin{align}
    &\Lambda^{+,0}_{1-3} = \text{Roots of:} \nonumber \\
    &\quad \quad \quad \quad x^3 - 2(r_1 + r_2 + r_3)x^2
    +4(-r_7^2-r_8^2-r_9^2+r_1 r_2+r_1 r_3+r_2 r_3)x \nonumber \\
    &\quad \quad \quad \quad +8 ( r_3 r_7^2-2 r_8 r_9 r_7+r_2 r_8^2+r_1 r_9^2-r_1 r_2 r_3) 
    = 0 \, , \nonumber \\[2mm]
    &\Lambda^{+,0}_{4,5} = 2 \left(r_4 \pm r_{10} \right) \, , \nonumber \\[2mm]
    &\Lambda^{+,0}_{6-9} = 
    \pm \sqrt{4 r_{11}^2 +\left(r_5-r_6\right){}^2}+r_5+r_6 \, .
\end{align}

\subsubsection*{The matrix $M_2^0$}

As shown in complete generality in appendix~\ref{app:M20=M2++},
\begin{align}
    \frac{1}{2} \, M_2^{0} = \frac{1}{2} \, M_2^{++} \ ,
\end{align}
and, thus, the eigenvalues of $M_2^{0}$ and $M_2^{++}$ coincide:
\begin{align}
    &\Lambda^{0,2}_{1-6} = \Lambda^{++}_{1-6} \, .
\end{align}

\subsubsection*{The matrix $M_0^0$}

From $M_0^{0}$ we get
\begin{align}
    \frac{1}{2} \, M_0^{0} \sim \, &
    \mathrm{diag} \left\{
    \frac{1}{2} M_0^{+}
    ,
    \begin{pmatrix}
     3 r_1 & 2 r_4+r_7 & 2 r_5+r_8 \\
     2 r_4+r_7 & 3 r_2 & 2 r_6+r_9 \\
     2 r_5+r_8 & 2 r_6+r_9 & 3 r_3
    \end{pmatrix}
    ,
    \begin{pmatrix}
     r_4+2 r_7 & 3 r_{10} \\
     r_{10} & r_4+2 r_7
    \end{pmatrix} ,
    \right.
    \nonumber \\[10mm]
    &
    \left.
    \begin{pmatrix}
     r_5+2 r_8 & 3 r_{11} \\
     3 r_{11} & r_6+2 r_9
    \end{pmatrix}
    ,
    \begin{pmatrix}
     r_6+2 r_9 & 3 r_{11} \\
     3 r_{11} & r_5+2 r_8
    \end{pmatrix} \right\} \, ,
\end{align}
with eigenvalues of $M_0^{0}$:
\begin{align}
    &\Lambda^{0,0}_{1-9} = \Lambda^{+,0}_{1-9} \, , \nonumber \\[2mm]
    &\Lambda^{0,0}_{10-12} = \text{Roots of:} \nonumber \\
    &\quad \quad \quad \quad x^3 + 2 \left(-3 r_1-3 r_2-3 r_3\right) x^2 + 4 \left(-4 r_4^2-4 r_7 r_4-4 r_5^2-4 r_6^2-r_7^2-r_8^2-r_9^2 \right.
    \nonumber \\
    &\left. \quad \quad \quad \quad +9 r_1 r_2+9 r_1 r_3+9 r_2 r_3-4 r_5 r_8-4 r_6 r_9\right)x
    +8 \left( 12 r_3 r_4^2+12 r_2 r_5^2+12 r_1 r_6^2 \right.\nonumber \\
    &\quad \quad \quad \quad +3 r_3 r_7^2+3 r_2 r_8^2+3 r_1 r_9^2-27 r_1 r_2 r_3-16 r_4 r_5 r_6+12 r_3 r_4 r_7-8 r_5 r_6 r_7 \nonumber \\
    &\quad \quad \quad \quad +12 r_2 r_5 r_8-8 r_4 r_6 r_8-4 r_6 r_7 r_8-8 r_4 r_5 r_9+12 r_1 r_6 r_9-4 r_5 r_7 r_9-4 r_4 r_8 r_9 \nonumber \\
    &\quad \quad \quad \quad \left. -2 r_7 r_8 r_9 \right) = 0 \, , \nonumber \\[2mm]
    &\Lambda^{0,0}_{13,14} =  2 \left(r_4+2 r_7 \pm 3 r_{10}\right) \, , \nonumber \\[2mm]
    &\Lambda^{0,0}_{15-18} = \pm \sqrt{36 r_{11}^2 +\left(r_5-r_6+2 r_8-2 r_9\right){}^2}
    +r_5+r_6+2 r_8+2 r_9 \, .
\end{align}

\subsection{The $\mathbb{Z}_3$ symmetry}

By imposing $G = \mathbb{Z}_3$ with representation 
$\mathrm{diag}(e^{\frac{2 \pi i}{3}},e^{\frac{-2 \pi i}{3}},1)$
we get the quartic potential
\begin{align}
    V_{\mathbb{Z}_3} =&
    \sum_{i=1}^3 r_i |\phi_i|^4 + 2 r_4 (\phi_1^\dagger \phi_1)(\phi_2^\dagger \phi_2)
    + 2 r_5 (\phi_1^\dagger \phi_1)(\phi_3^\dagger \phi_3)
    + 2 r_6 (\phi_2^\dagger \phi_2)(\phi_3^\dagger \phi_3)
    \nonumber \\[2mm]
    & + 2 r_7 |\phi_1^\dagger \phi_2|^2
    + 2 r_8 |\phi_1^\dagger \phi_3|^2 + 2 r_9 |\phi_2^\dagger \phi_3|^2
    + \Big[
	2 c_4 (\phi_1^\dagger \phi_2)(\phi_1^\dagger \phi_3)
    \nonumber \\[2mm]
    &+ 2 c_{11} (\phi_1^\dagger \phi_3)(\phi_2^\dagger \phi_3)
    + 2 c_{12} (\phi_1^\dagger \phi_2)(\phi_3^\dagger \phi_2) + h.c. \Big]\, ,
\label{V_Z3_one}
\end{align}
with the following scattering matrices.

\subsubsection*{The matrix $M_2^{++}$}

From $M_2^{++}$ we get
\begin{equation}
    \frac{1}{2} \, M_2^{++} = \mathrm{diag} \left\{
    \begin{pmatrix}
     r_1 & \sqrt{2} c_4 \\
     \sqrt{2} c_4^* & r_6 + r_9
    \end{pmatrix}
    ,
    \begin{pmatrix}
     r_4 + r_7 & \sqrt{2} c_{11} \\
     \sqrt{2} c_{11}^* & r_3
    \end{pmatrix}
    ,
    \begin{pmatrix}
     r_5 + r_8 & \sqrt{2} c_{12} \\
     \sqrt{2} c_{12}^* & r_2
    \end{pmatrix} \right\} \, ,
\end{equation}
and thus we get the eigenvalues of $M_2^{++}$:
\begin{align}
    &\Lambda^{++}_{1,2} = \pm \sqrt{8 |c_4|^2 +\left(-r_1+r_6+r_9\right){}^2}+r_1+r_6+r_9 \, , \nonumber \\[2mm]
    &\Lambda^{++}_{3,4} = \pm \sqrt{8 |c_{11}|^2 +\left(-r_3+r_4+r_7\right){}^2}+r_3+r_4+r_7 \, , \nonumber \\[2mm]
    &\Lambda^{++}_{5,6} = \pm \sqrt{8 |c_{12}|^2 +\left(-r_2+r_5+r_8\right){}^2}+r_2+r_5+r_8 \, .
\end{align}

\subsubsection*{The matrix $M_2^+$}

From $M_2^{+}$ we get
\begin{equation}
    \frac{1}{2} \, M_2^{+} = \mathrm{diag} \left\{
    \begin{pmatrix}
     r_1 & c_4 & c_4 \\
     c_4^* & r_6 & r_9 \\
     c_4^* & r_9 & r_6
    \end{pmatrix}
    ,
    \begin{pmatrix}
     r_4 & r_7 & c_{11} \\
     r_7 & r_4 & c_{11} \\
     c_{11}^* & c_{11}^* & r_3
    \end{pmatrix}
    ,
    \begin{pmatrix}
     r_5 & c_{12} & r_8 \\
     c_{12}^* & r_2 & c_{12}^* \\
     r_8 & c_{12} & r_5
    \end{pmatrix} \right\} \, ,
\end{equation}
with eigenvalues of $M_2^{+}$:
\begin{align}
    &\Lambda^{+,2}_{1} = 2(r_6 - r_9) \, , \nonumber \\[2mm]
    &\Lambda^{+,2}_{2,3} = \Lambda^{++}_{1,2} \, , \nonumber \\[2mm]
    &\Lambda^{+,2}_{4} = 2(r_4 - r_7) \, , \nonumber \\[2mm]
    &\Lambda^{+,2}_{5,6} = \Lambda^{++}_{3,4} \, \nonumber \\[2mm]
    &\Lambda^{+,2}_{7} = 2(r_5 - r_8) \, \nonumber \\[2mm]
    &\Lambda^{+,2}_{8,9} = \Lambda^{++}_{5,6} \, .
\end{align}

\subsubsection*{The matrix $M_0^+$}

From $M_0^{+}$ we get
\begin{equation}
    \frac{1}{2} \, M_0^{+} = \mathrm{diag} \left\{
    \begin{pmatrix}
     r_1 & r_7 & r_8 \\
     r_7 & r_2 & r_9 \\
     r_8 & r_9 & r_3
    \end{pmatrix}
    ,
    \begin{pmatrix}
     r_4 & c_{12} & c_4 \\
     c_{12}^* & r_6 & c_{11} \\
     c_4^* & c_{11}^* & r_5
    \end{pmatrix}
    ,
    \begin{pmatrix}
     r_5 & c_4 & c_{11} \\
     c_4^* & r_4 & c_{12}^* \\
     c_{11}^* & c_{12} & r_6
    \end{pmatrix} \right\} \, ,
\end{equation}
with eigenvalues of $M_0^{+}$:
\begin{align}
    &\Lambda^{+,0}_{1-3} = \text{Roots of:} \nonumber \\
    &\quad \quad \quad \quad x^3 - 2(r_1 + r_2 + r_3)x^2
    +4(-r_7^2-r_8^2-r_9^2+r_1 r_2+r_1 r_3+r_2 r_3)x \nonumber \\
    &\quad \quad \quad \quad +8 ( r_3 r_7^2-2 r_8 r_9 r_7+r_2 r_8^2+r_1 r_9^2-r_1 r_2 r_3) 
    = 0 \, , \nonumber \\[2mm]
    &\Lambda^{+,0}_{4-6} =  \text{Roots of:} \nonumber \\
    &\quad \quad \quad \quad x^3 - 2(r_4 + r_5 + r_6)x^2
    + 4(- |c_{4}|^2 - |c_{11}|^2 - |c_{12}|^2 +r_4 r_5 + r_4 r_6 + r_5 r_6 )x \nonumber \\
    &\quad \quad \quad \quad + 8 \left( r_6 |c_4|^2 + r_4 |c_{11}|^2 + r_5 |c_{12}|^2 - 2 \Re (c_4 c_{11}^* c_{12}^*)
    - r_4 r_5 r_6 \right) = 0 \, , \nonumber \\[2mm]
    &\Lambda^{+,0}_{7-9} = \Lambda^{+,0}_{4-6} \, .
\end{align}

\subsubsection*{The matrix $M_2^0$}

From $M_2^{0}$ we get
\begin{align}
    \frac{1}{2} \, M_2^{0} = \frac{1}{2} \, M_2^{++} \ ,
\end{align}
with eigenvalues of $M_2^{0}$:
\begin{align}
    &\Lambda^{0,2}_{1-6} = \Lambda^{++}_{1-6} \, .
\end{align}

\subsubsection*{The matrix $M_0^0$}

From $M_0^{0}$ we get
\begin{align}
    \frac{1}{2} \, M_0^{0} \sim \, &
    \mathrm{diag} \left\{
    \frac{1}{2} M_0^{+}
    ,
    \begin{pmatrix}
     3 r_1 & 2 r_4+r_7 & 2 r_5+r_8 \\
     2 r_4+r_7 & 3 r_2 & 2 r_6+r_9 \\
     2 r_5+r_8 & 2 r_6+r_9 & 3 r_3
    \end{pmatrix}
    ,
    \begin{pmatrix}
     r_4+2 r_7 & 3 c_{12} & 3 c_4 \\
     3 c_{12}^* & r_6+2 r_9 & 3 c_{11} \\
     3 c_4^* & 3 c_{11}^* & r_5+2 r_8
    \end{pmatrix} ,
    \right.
    \nonumber \\[3mm]
    &
\hspace{12mm}
    \left.
    \begin{pmatrix}
     r_5+2 r_8 & 3 c_4 & 3 c_{11} \\
     3 c_4^* & r_4+2 r_7 & 3 c_{12}^* \\
     3 c_{11}^* & 3 c_{12} & r_6+2 r_9
    \end{pmatrix} \right\} \, ,
\end{align}
with eigenvalues of $M_0^{0}$:
\begin{align}
    &\Lambda^{0,0}_{1-9} = \Lambda^{+,0}_{1-9} \, , \nonumber \\[2mm]
    &\Lambda^{0,0}_{10-12} = \text{Roots of:} \nonumber \\
    &\quad \quad \quad \quad x^3 + 2 \left(-3 r_1-3 r_2-3 r_3\right) x^2 
    + 4 \left(-4 r_4^2-4 r_7 r_4-4 r_5^2-4 r_6^2-r_7^2-r_8^2-r_9^2 \right.
    \nonumber \\
    &\left. \quad \quad \quad \quad +9 r_1 r_2+9 r_1 r_3+9 r_2 r_3-4 r_5 r_8-4 r_6 r_9\right)x
    + 8\left(12 r_3 r_4^2+12 r_2 r_5^2+12 r_1 r_6^2 \right. \nonumber \\
    &\quad \quad \quad \quad +3 r_3 r_7^2+3 r_2 r_8^2+3 r_1 r_9^2-27 r_1 r_2 r_3-16 r_4 r_5 r_6+12 r_3 r_4 r_7-8 r_5 r_6 r_7 \nonumber \\
    &\quad \quad \quad \quad +12 r_2 r_5 r_8-8 r_4 r_6 r_8-4 r_6 r_7 r_8-8 r_4 r_5 r_9+12 r_1 r_6 r_9-4 r_5 r_7 r_9-4 r_4 r_8 r_9 \nonumber \\
    &\quad \quad \quad \quad \left. -2 r_7 r_8 r_9 \right) = 0 \, , \nonumber \\[2mm]
    &\Lambda^{0,0}_{13-15} = \text{Roots of:} \nonumber \\
    &\quad \quad \quad \quad x^3 + 2 \left(-r_4-r_5-r_6-2 r_7-2 r_8-2 r_9\right) x^2 
    + 4 \left(-9 |c_4|^2 -9 |c_{11}|^2-9 |c_{12}|^2
    \right. \nonumber \\
    &\quad \quad \quad \quad +r_4 r_5+r_4 r_6+r_5 r_6+2 r_5 r_7+2 r_6 r_7+2 r_4 r_8+2 r_6 r_8+4 r_7 r_8+2 r_4 r_9 \nonumber \\
    &\quad \quad \quad \quad \left. +2 r_5 r_9+4 r_7 r_9+4 r_8 r_9 \right) x +
    8 \left( 9 r_6 |c_4|^2 +18 r_9 |c_4|^2 +9 r_4 |c_{11}|^2 +9 r_5 |c_{12}|^2 \right. \nonumber \\
    &\quad \quad \quad \quad +18 r_7 |c_{11}|^2 +18 r_8 |c_{12}|^2 -54 \Re( c_4 c_{11}^* c_{12}^*)
    -r_4 r_5 r_6-2 r_5 r_6 r_7-2 r_4 r_6 r_8 \nonumber \\
    &\quad \quad \quad \quad \left. -4 r_6 r_7 r_8-2 r_4 r_5 r_9-4 r_5 r_7 r_9-4 r_4 r_8 r_9-8 r_7 r_8 r_9
    \right) = 0 \, , \nonumber \\[2mm]
    &\Lambda^{0,0}_{16-18} = \Lambda^{0,0}_{13-15} \, .
\end{align}

\subsection{The $\mathbb{Z}_3 \rtimes \mathbb{Z}_2^{(\mathrm{CP})}$ symmetry}

By imposing $G = \mathbb{Z}_3 \rtimes \mathbb{Z}_2^{(\mathrm{CP})}$
we get the quartic potential
\begin{align}
    V_{\mathbb{Z}_3 \rtimes \mathbb{Z}_2^{(\mathrm{CP})}} =&
    \sum_{i=1}^3 r_i |\phi_i|^4 + 2 r_4 (\phi_1^\dagger \phi_1)(\phi_2^\dagger \phi_2)
    + 2 r_5 (\phi_1^\dagger \phi_1)(\phi_3^\dagger \phi_3)
    + 2 r_6 (\phi_2^\dagger \phi_2)(\phi_3^\dagger \phi_3)
    \nonumber \\[2mm]
    & + 2 r_7 |\phi_1^\dagger \phi_2|^2
    + 2 r_8 |\phi_1^\dagger \phi_3|^2 + 2 r_9 |\phi_2^\dagger \phi_3|^2
    + 2 r_{10} \left[ (\phi_1^\dagger \phi_2)(\phi_1^\dagger \phi_3) + h.c. \right]
    \nonumber \\[2mm]
    &+ 2 r_{11} \left[ (\phi_1^\dagger \phi_3)(\phi_2^\dagger \phi_3) + h.c. \right]
    + 2 r_{12} \left[ (\phi_1^\dagger \phi_2)(\phi_3^\dagger \phi_2) + h.c. \right] \, ,
\end{align}
which can be easily achieved by setting from $\mathbb{Z}_3$ the constraints
$\{ c_4 , c_{11} , c_{12} \} \in \mathbb{R}$.
Thus, we get the following eigenvalues.

\subsubsection*{The matrix $M_2^{++}$}

From $M_2^{++}$ we get
\begin{equation}
    \frac{1}{2} \, M_2^{++} = \mathrm{diag} \left\{
    \begin{pmatrix}
     r_1 & \sqrt{2} r_{10} \\
     \sqrt{2} r_{10} & r_6 + r_9
    \end{pmatrix}
    ,
    \begin{pmatrix}
     r_4 + r_7 & \sqrt{2} r_{11} \\
     \sqrt{2} r_{11} & r_3
    \end{pmatrix}
    ,
    \begin{pmatrix}
     r_5 + r_8 & \sqrt{2} r_{12} \\
     \sqrt{2} r_{12} & r_2
    \end{pmatrix} \right\} \, ,
\end{equation}
and thus we get the eigenvalues of $M_2^{++}$:
\begin{align}
    &\Lambda^{++}_{1,2} = \pm \sqrt{8 r_{10}^2 +\left(-r_1+r_6+r_9\right){}^2}+r_1+r_6+r_9 \, , \nonumber \\[2mm]
    &\Lambda^{++}_{3,4} = \pm \sqrt{8 r_{11}^2 +\left(-r_3+r_4+r_7\right){}^2}+r_3+r_4+r_7 \, , \nonumber \\[2mm]
    &\Lambda^{++}_{5,6} = \pm \sqrt{8 r_{12}^2 +\left(-r_2+r_5+r_8\right){}^2}+r_2+r_5+r_8 \, .
\end{align}

\subsubsection*{The matrix $M_2^+$}

From $M_2^{+}$ we get
\begin{equation}
    \frac{1}{2} \, M_2^{+} = \mathrm{diag} \left\{
    \begin{pmatrix}
     r_1 & r_{10} & r_{10} \\
     r_{10} & r_6 & r_9 \\
     r_{10} & r_9 & r_6
    \end{pmatrix}
    ,
    \begin{pmatrix}
     r_4 & r_7 & r_{11} \\
     r_7 & r_4 & r_{11} \\
     r_{11} & r_{11} & r_3
    \end{pmatrix}
    ,
    \begin{pmatrix}
     r_5 & r_{12} & r_8 \\
     r_{12} & r_2 & r_{12} \\
     r_8 & r_{12} & r_5
    \end{pmatrix} \right\} \, ,
\end{equation}
with eigenvalues of $M_2^{+}$:
\begin{align}
    &\Lambda^{+,2}_{1} = 2(r_6 - r_9) \, , \nonumber \\[2mm]
    &\Lambda^{+,2}_{2,3} = \Lambda^{++}_{1,2} \, , \nonumber \\[2mm]
    &\Lambda^{+,2}_{4} = 2(r_4 - r_7) \, , \nonumber \\[2mm]
    &\Lambda^{+,2}_{5,6} = \Lambda^{++}_{3,4} \, \nonumber \\[2mm]
    &\Lambda^{+,2}_{7} = 2(r_5 - r_8) \, \nonumber \\[2mm]
    &\Lambda^{+,2}_{8,9} = \Lambda^{++}_{5,6} \, .
\end{align}

\subsubsection*{The matrix $M_0^+$}

From $M_0^{+}$ we get
\begin{equation}
    \frac{1}{2} \, M_0^{+} = \mathrm{diag} \left\{
    \begin{pmatrix}
     r_1 & r_7 & r_8 \\
     r_7 & r_2 & r_9 \\
     r_8 & r_9 & r_3
    \end{pmatrix}
    ,
    \begin{pmatrix}
     r_4 & r_{12} & r_{10} \\
     r_{12} & r_6 & r_{11} \\
     r_{10} & r_{11} & r_5
    \end{pmatrix}
    ,
    \begin{pmatrix}
     r_5 & r_{10} & r_{11} \\
     r_{10} & r_4 & r_{12} \\
     r_{11} & r_{12} & r_6
    \end{pmatrix} \right\} \, ,
\end{equation}
with eigenvalues of $M_0^{+}$:
\begin{align}
    &\Lambda^{+,0}_{1-3} = \text{Roots of:} \nonumber \\
    &\quad \quad \quad \quad x^3 - 2(r_1 + r_2 + r_3)x^2
    +4(-r_7^2-r_8^2-r_9^2+r_1 r_2+r_1 r_3+r_2 r_3)x \nonumber \\
    &\quad \quad \quad \quad +8 ( r_3 r_7^2-2 r_8 r_9 r_7+r_2 r_8^2+r_1 r_9^2-r_1 r_2 r_3) 
    = 0 \, , \nonumber \\[2mm]
    &\Lambda^{+,0}_{4-6} =  \text{Roots of:} \nonumber \\
    &\quad \quad \quad \quad x^3 - 2(r_4 + r_5 + r_6)x^2
    + 4(- r_{10}^2 - r_{11}^2 - r_{12}^2 +r_4 r_5 + r_4 r_6 + r_5 r_6 )x \nonumber \\
    &\quad \quad \quad \quad + 8 \left( r_6 r_{10}^2 + r_4 r_{11}^2 + r_5 r_{12}^2 
    - 2 r_{10} r_{11} r_{12}
    - r_4 r_5 r_6 \right) = 0 \, , \nonumber \\[2mm]
    &\Lambda^{+,0}_{7-9} = \Lambda^{+,0}_{4-6} \, .
\end{align}

\subsubsection*{The matrix $M_2^0$}

From $M_2^{0}$ we get
\begin{align}
    \frac{1}{2} \, M_2^{0} = \frac{1}{2} \, M_2^{++} \ ,
\end{align}
with eigenvalues of $M_2^{0}$:
\begin{align}
    &\Lambda^{0,2}_{1-6} = \Lambda^{++}_{1-6} \, .
\end{align}

\subsubsection*{The matrix $M_0^0$}

From $M_0^{0}$ we get
\begin{align}
    \frac{1}{2} \, M_0^{0} \sim \, &
    \mathrm{diag} \left\{
    \frac{1}{2} M_0^{+}
    ,
    \begin{pmatrix}
     3 r_1 & 2 r_4+r_7 & 2 r_5+r_8 \\
     2 r_4+r_7 & 3 r_2 & 2 r_6+r_9 \\
     2 r_5+r_8 & 2 r_6+r_9 & 3 r_3
    \end{pmatrix}
    ,
    \begin{pmatrix}
     r_4+2 r_7 & 3 r_{12} & 3 r_{10} \\
     3 r_{12} & r_6+2 r_9 & 3 r_{11} \\
     3 r_{10} & 3 r_{11} & r_5+2 r_8
    \end{pmatrix} ,
    \right.
    \nonumber \\[10mm]
    &
    \left.
    \begin{pmatrix}
     r_5+2 r_8 & 3 r_{10} & 3 r_{11} \\
     3 r_{10} & r_4+2 r_7 & 3 r_{12} \\
     3 r_{11} & 3 r_{12} & r_6+2 r_9
    \end{pmatrix} \right\} \, ,
\end{align}
with eigenvalues of $M_0^{0}$:
\begin{align}
    &\Lambda^{0,0}_{1-9} = \Lambda^{+,0}_{1-9} \, , \nonumber \\[2mm]
    &\Lambda^{0,0}_{10-12} = \text{Roots of:} \nonumber \\
    &\quad \quad \quad \quad x^3 + 2 \left(-3 r_1-3 r_2-3 r_3\right) x^2 
    + 4 \left(-4 r_4^2-4 r_7 r_4-4 r_5^2-4 r_6^2-r_7^2-r_8^2-r_9^2 \right.
    \nonumber \\
    &\left. \quad \quad \quad \quad +9 r_1 r_2+9 r_1 r_3+9 r_2 r_3-4 r_5 r_8-4 r_6 r_9\right)x
    + 8\left(12 r_3 r_4^2+12 r_2 r_5^2+12 r_1 r_6^2 \right. \nonumber \\
    &\quad \quad \quad \quad +3 r_3 r_7^2+3 r_2 r_8^2+3 r_1 r_9^2-27 r_1 r_2 r_3-16 r_4 r_5 r_6+12 r_3 r_4 r_7-8 r_5 r_6 r_7 \nonumber \\
    &\quad \quad \quad \quad +12 r_2 r_5 r_8-8 r_4 r_6 r_8-4 r_6 r_7 r_8-8 r_4 r_5 r_9+12 r_1 r_6 r_9-4 r_5 r_7 r_9-4 r_4 r_8 r_9 \nonumber \\
    &\quad \quad \quad \quad \left. -2 r_7 r_8 r_9 \right) = 0 \, , \nonumber \\[2mm]
    &\Lambda^{0,0}_{13-15} = \text{Roots of:} \nonumber \\
    &\quad \quad \quad \quad x^3 + 2 \left(-r_4-r_5-r_6-2 r_7-2 r_8-2 r_9\right) x^2 
    + 4 \left(-9 r_{10}^2 -9 r_{11}^2-9 r_{12}^2
    \right. \nonumber \\
    &\quad \quad \quad \quad +r_4 r_5+r_4 r_6+r_5 r_6+2 r_5 r_7+2 r_6 r_7+2 r_4 r_8+2 r_6 r_8+4 r_7 r_8+2 r_4 r_9 \nonumber \\
    &\quad \quad \quad \quad \left. +2 r_5 r_9+4 r_7 r_9+4 r_8 r_9 \right) x +
    8 \left( 9 r_6 r_{10}^2 +18 r_9 r_{10}^2 +9 r_4 r_{11}^2 +9 r_5 r_{12}^2 \right. \nonumber \\
    &\quad \quad \quad \quad +18 r_7 r_{11}^2 +18 r_8 r_{12}^2 -54 r_{10} r_{11} r_{12}
    -r_4 r_5 r_6-2 r_5 r_6 r_7-2 r_4 r_6 r_8 \nonumber \\
    &\quad \quad \quad \quad \left. -4 r_6 r_7 r_8-2 r_4 r_5 r_9-4 r_5 r_7 r_9-4 r_4 r_8 r_9-8 r_7 r_8 r_9
    \right) = 0 \, , \nonumber \\[2mm]
    &\Lambda^{0,0}_{16-18} = \Lambda^{0,0}_{13-15} \, .
\end{align}

\subsection{The $\mathrm{U}(1)_2$ symmetry}

By imposing $G = \mathrm{U}(1)_2$ with representation $\mathrm{diag}(1,1,e^{i \alpha})$,
with $\alpha \neq \{ 0, \pi \}$,
we get the quartic potential
\begin{align}
    V_{\mathrm{U}(1)_2} =&
    \sum_{i=1}^3 r_i |\phi_i|^4 + 2 r_4 (\phi_1^\dagger \phi_1)(\phi_2^\dagger \phi_2)
    + 2 r_5 (\phi_1^\dagger \phi_1)(\phi_3^\dagger \phi_3)
    + 2 r_6 (\phi_2^\dagger \phi_2)(\phi_3^\dagger \phi_3)
    \nonumber \\[2mm]
    & + 2 r_7 |\phi_1^\dagger \phi_2|^2
    + 2 r_8 |\phi_1^\dagger \phi_3|^2 + 2 r_9 |\phi_2^\dagger \phi_3|^2
    + \Big[
	2 c_1 (\phi_1^\dagger \phi_1)(\phi_1^\dagger \phi_2) + c_3 (\phi_1^\dagger \phi_2)^2
    \nonumber \\[2mm]
    &+ 2 c_7 (\phi_1^\dagger \phi_2)(\phi_2^\dagger \phi_2)
    + 2 c_{13} (\phi_1^\dagger \phi_2)(\phi_3^\dagger \phi_3)
    + 2 c_{14} (\phi_1^\dagger \phi_3)(\phi_3^\dagger \phi_2)
    + h.c.  \Big]\, ,
\end{align}
which can be easily achieved by setting from $\mathbb{Z}_2$
the constraints $\{c_5, c_{17} \} \rightarrow 0$.

Thus, we get the following scattering matrices.

\subsubsection*{The matrix $M_2^{++}$}

From $M_2^{++}$ we get
\begin{equation}
    \frac{1}{2} \, M_2^{++} = \mathrm{diag} \left\{
    \begin{pmatrix}
     r_1 & \sqrt{2} c_1 & c_3 \\
     \sqrt{2} c_1^* & r_4+r_7 & \sqrt{2} c_7 \\
     c_3^* & \sqrt{2} c_7^* & r_2
    \end{pmatrix}
    ,
    \begin{pmatrix}
     r_5+r_8 & c_{13}+c_{14} \\
     c_{13}^*+c_{14}^* & r_6+r_9
    \end{pmatrix}
    ,
    r_3 \right\} \, ,
\end{equation}
and thus we get the eigenvalues of $M_2^{++}$:
\begin{align}
    &\Lambda^{++}_{1-3} = \text{Roots of:} \nonumber \\
    &\quad \quad \quad \quad x^3 + 2 x^2 \left(-r_1-r_2-r_4-r_7\right)+4 x \left(-2 |c_1|^2-|c_3|^2-2 |c_7|^2+r_1 r_2+r_1 r_4 \right. 
    \nonumber \\[2mm]
    &\quad \quad \quad \quad \left. +r_2 r_4+r_1 r_7+r_2 r_7\right)
    + 8 \left( 2 r_1 |c_7|^2+2 r_2 |c_1|^2+(r_4 + r_7)|c_3|^2 -
    4 \Re( c_1 c_3^* c_7) \right. \nonumber \\[2mm]
    &\quad \quad \quad \quad \left. -r_1 r_2 r_4-r_1 r_2 r_7 \right) = 0 \, , \nonumber \\[2mm]
    &\Lambda^{++}_{4,5} = \pm \sqrt{4 | c_{13}+c_{14} |^2 +\left(r_5-r_6+r_8-r_9\right){}^2}+r_5+r_6+r_8+r_9 \, , \nonumber \\[2mm]
    &\Lambda^{++}_{6} =  2 r_3 \, .
\end{align}

\subsubsection*{The matrix $M_2^+$}

From $M_2^{+}$ we get
\begin{equation}
    \frac{1}{2} \, M_2^{+} = \mathrm{diag} \left\{
    \begin{pmatrix}
     r_1 & c_1 & c_1 & c_3 \\
     c_1^* & r_4 & r_7 & c_7 \\
     c_1^* & r_7 & r_4 & c_7 \\
     c_3^* & c_7^* & c_7^* & r_2
    \end{pmatrix}
    ,
    \begin{pmatrix}
     r_5 & c_{13} & r_8 & c_{14} \\
     c_{13}^* & r_6 & c_{14}^* & r_9 \\
     r_8 & c_{14} & r_5 & c_{13} \\
     c_{14}^* & r_9 & c_{13}^* & r_6
    \end{pmatrix}
    ,
    r_3 \right\} \, ,
\end{equation}
with eigenvalues of $M_2^{+}$:
\begin{align}
    &\Lambda^{+,2}_{1-3} = \Lambda^{++}_{1-3} \, , \nonumber \\[2mm]
    &\Lambda^{+,2}_{4} = 2 (r_4-r_7) \, , \nonumber \\[2mm]
    &\Lambda^{+,2}_{5,6} = \Lambda^{++}_{4,5} \, , \nonumber \\[2mm]
    &\Lambda^{+,2}_{7,8} = \pm \sqrt{4 |c_{13}-c_{14}|^2
    +\left(r_5-r_6-r_8+r_9\right){}^2}+r_5+r_6-r_8-r_9 \, \nonumber \\[2mm]
    &\Lambda^{+,2}_{9} = \Lambda^{++}_{6} \, .
\end{align}

\subsubsection*{The matrix $M_0^+$}

From $M_0^{+}$ we get
\begin{equation}
    \frac{1}{2} \, M_0^{+} = \mathrm{diag} \left\{
    \begin{pmatrix}
     r_1 & c_1^* & c_1 & r_7 & r_8 \\
     c_1 & r_4 & c_3 & c_7 & c_{14} \\
     c_1^* & c_3^* & r_4 & c_7^* & c_{14}^* \\
     r_7 & c_7^* & c_7 & r_2 & r_9 \\
     r_8 & c_{14}^* & c_{14} & r_9 & r_3
    \end{pmatrix}
    ,
    \begin{pmatrix}
     r_5 & c_{13} \\
     c_{13}^* & r_6
    \end{pmatrix}
    ,
    \begin{pmatrix}
     r_5 & c_{13}^* \\
     c_{13} & r_6
    \end{pmatrix} \right\} \, ,
\end{equation}
with eigenvalues of $M_0^{+}$:
\begin{align}
    &\Lambda^{+,0}_{1-5} = \text{Eigenvalues of first matrix} \, , \nonumber \\[2mm]
    &\Lambda^{+,0}_{6,7} = \pm \sqrt{4 |c_{13}|^2+\left(r_5-r_6\right){}^2}
    +r_5+r_6 \, , \nonumber \\[2mm]
    &\Lambda^{+,0}_{8,9} = \Lambda^{+,0}_{6,7} \, .
\end{align}

\subsubsection*{The matrix $M_2^0$}

From $M_2^{0}$ we get
\begin{align}
    \frac{1}{2} \, M_2^{0} = \frac{1}{2} \, M_2^{++} \ ,
\end{align}
with eigenvalues of $M_2^{0}$:
\begin{align}
    &\Lambda^{0,2}_{1-6} = \Lambda^{++}_{1-6} \, .
\end{align}

\subsubsection*{The matrix $M_0^0$}

From $M_0^{0}$ we get
\begin{align}
    \frac{1}{2} \, M_0^{0} \sim \, &
    \mathrm{diag} \left\{
    \frac{1}{2} M_0^{+}
    ,
    \begin{pmatrix}
     3 r_1 & 3 c_1^* & 3 c_1 & 2 r_4+r_7 & 2 r_5+r_8 \\
     3 c_1 & r_4+2 r_7 & 3 c_3 & 3 c_7 & 2 c_{13}+c_{14} \\
     3 c_1^* & 3 c_3^* & r_4+2 r_7 & 3 c_7^* & 2 c_{13}^*+c_{14}^* \\
     2 r_4+r_7 & 3 c_7^* & 3 c_7 & 3 r_2 & 2 r_6+r_9 \\
     2 r_5+r_8 & 2 c_{13}^*+c_{14}^* & 2 c_{13}+c_{14} & 2 r_6+r_9 & 3 r_3
    \end{pmatrix} , \right.
    \nonumber \\[3mm]
    &
\hspace{15mm}
\left.
    \begin{pmatrix}
     r_5+2 r_8 & c_{13}+2 c_{14} \\
     c_{13}^*+2 c_{14}^* & r_6+2 r_9
    \end{pmatrix}
    ,
    \begin{pmatrix}
     r_5+2 r_8 & c_{13}^*+2 c_{14}^* \\
     c_{13}+2 c_{14} & r_6+2 r_9
    \end{pmatrix} \right\} \, .
\end{align}
with eigenvalues of $M_0^{0}$:
\begin{align}
    &\Lambda^{0,0}_{1-9} = \Lambda^{+,0}_{1-9} \, , \nonumber \\[2mm]
    &\Lambda^{0,0}_{10-14} = \text{Eigenvalues of second matrix} 
    \, , \nonumber \\[2mm]
    &\Lambda^{0,0}_{15,16} = \pm \sqrt{4 |c_{13} + 2 c_{14}|^2
    +\left(r_5-r_6+2 r_8-2 r_9\right){}^2}+r_5+r_6+2 r_8+2 r_9
    \, , \nonumber \\[2mm]
    &\Lambda^{0,0}_{17-18} = \Lambda^{0,0}_{15-16} \, .
\end{align}

\subsection{The $\mathrm{U}(1)_1$ symmetry}

By imposing $G = \mathrm{U}(1)_1$ with representation 
$\mathrm{diag}(e^{i \alpha},e^{- i \alpha},1)$,
with $\alpha \neq \{ 0, \pi/2, 2\pi/3, \pi \}$,
we get the quartic potential
\begin{align}
    V_{\mathrm{U}(1)_1} =&
    \sum_{i=1}^3 r_i |\phi_i|^4 + 2 r_4 (\phi_1^\dagger \phi_1)(\phi_2^\dagger \phi_2)
    + 2 r_5 (\phi_1^\dagger \phi_1)(\phi_3^\dagger \phi_3)
    + 2 r_6 (\phi_2^\dagger \phi_2)(\phi_3^\dagger \phi_3)
    \nonumber \\[2mm]
    & + 2 r_7 |\phi_1^\dagger \phi_2|^2
    + 2 r_8 |\phi_1^\dagger \phi_3|^2 + 2 r_9 |\phi_2^\dagger \phi_3|^2
	+ 2 r_{11} \left[ (\phi_1^\dagger \phi_3)(\phi_2^\dagger \phi_3) + h.c. \right] \, ,
\end{align}
which can be easily achieved by setting from $\mathbb{Z}_4$
the constraint $r_{10} \rightarrow 0$.

Thus, we get the following scattering matrices.

\subsubsection*{The matrix $M_2^{++}$}

From $M_2^{++}$ we get
\begin{equation}
    \frac{1}{2} \, M_2^{++} = \mathrm{diag} \left\{
    \begin{pmatrix}
     r_4+r_7 & \sqrt{2} r_{11} \\
     \sqrt{2} r_{11} & r_3
    \end{pmatrix}
    ,
    r_1
    ,
    r_2
    ,
    (r_5 + r_8)
    ,
    (r_6 + r_9) \right\} \, ,
\end{equation}
and thus we get the eigenvalues of $M_2^{++}$:
\begin{align}
    &\Lambda^{++}_{1,2} = \pm \sqrt{8 r_{11}^2
    +\left(-r_3+r_4+r_7\right){}^2}+r_3+r_4+r_7 \, ,
    \nonumber \\[2mm]
    &\Lambda^{++}_{3} = 2 r_1\, , \nonumber \\[2mm]
    &\Lambda^{++}_{4} =  2 r_2 \, , \nonumber \\[2mm]
    &\Lambda^{++}_{5} =  2 (r_5 + r_8) \, , \nonumber \\[2mm]
    &\Lambda^{++}_{6} =  2 (r_6 + r_9) \, .
\end{align}

\subsubsection*{The matrix $M_2^+$}

From $M_2^{+}$ we get
\begin{equation}
    \frac{1}{2} \, M_2^{+} = \mathrm{diag} \left\{
    \begin{pmatrix}
     r_4 & r_7 & r_{11} \\
     r_7 & r_4 & r_{11} \\
     r_{11} & r_{11} & r_3
    \end{pmatrix}
    ,
    \begin{pmatrix}
     r_5 & r_8 \\
     r_8 & r_5
    \end{pmatrix}
    ,
    \begin{pmatrix}
     r_6 & r_9 \\
     r_9 & r_6
    \end{pmatrix}
    ,
    r_1
    ,
    r_2 \right\} \, ,
\end{equation}
with eigenvalues of $M_2^{+}$:
\begin{align}
    &\Lambda^{+,2}_{1,2} = \Lambda^{++}_{1,2} \, , \nonumber \\[2mm]
    &\Lambda^{+,2}_{3} = 2 \left(r_4-r_7\right) \, , \nonumber \\[2mm]
    &\Lambda^{+,2}_{4,5} = 2 (r_5 \pm r_8) \, , \nonumber \\[2mm]
    &\Lambda^{+,2}_{6,7} = 2 (r_6 \pm r_9) \, , \nonumber \\[2mm]
     &\Lambda^{+,2}_{8,9} = \Lambda^{++}_{3,4} \, .
\end{align}

\subsubsection*{The matrix $M_0^+$}

From $M_0^{+}$ we get
\begin{equation}
    \frac{1}{2} \, M_0^{+} = \mathrm{diag} \left\{
    \begin{pmatrix}
     r_1 & r_7 & r_8 \\
     r_7 & r_2 & r_9 \\
     r_8 & r_9 & r_3
    \end{pmatrix}
    ,
    \begin{pmatrix}
     r_5 & r_{11} \\
     r_{11} & r_6
    \end{pmatrix}
    ,
    \begin{pmatrix}
     r_6 & r_{11} \\
     r_{11} & r_5
    \end{pmatrix}
    ,
    r_4
    ,
    r_4 \right\} \, ,
\end{equation}
with eigenvalues of $M_0^{+}$:
\begin{align}
    &\Lambda^{+,0}_{1-3} = \text{Roots of:} \nonumber \\
    &\quad \quad \quad \quad x^3 - 2(r_1 + r_2 + r_3)x^2
    +4(-r_7^2-r_8^2-r_9^2+r_1 r_2+r_1 r_3+r_2 r_3)x \nonumber \\
    &\quad \quad \quad \quad +8 ( r_3 r_7^2-2 r_8 r_9 r_7+r_2 r_8^2+r_1 r_9^2-r_1 r_2 r_3) 
    = 0 \, , \nonumber \\[2mm]
    &\Lambda^{+,0}_{4,5} = \pm \sqrt{4 r_{11}^2 +\left(r_5-r_6\right){}^2}
    +r_5+r_6 \, , \nonumber \\[2mm]
     &\Lambda^{+,0}_{6,7} = \Lambda^{+,0}_{4,5} \, , \nonumber \\[2mm]
    &\Lambda^{+,0}_{8,9} = 2 r_4 \, .
\end{align}

\subsubsection*{The matrix $M_2^0$}

From $M_2^{0}$ we get
\begin{align}
    \frac{1}{2} \, M_2^{0} = \frac{1}{2} \, M_2^{++} \ ,
\end{align}
with eigenvalues of $M_2^{0}$:
\begin{align}
    &\Lambda^{0,2}_{1-6} = \Lambda^{++}_{1-6} \, .
\end{align}

\subsubsection*{The matrix $M_0^0$}

From $M_0^{0}$ we get
\begin{align}
    \frac{1}{2} \, M_0^{0} \sim \, &
    \mathrm{diag} \left\{
    \frac{1}{2} M_0^{+}
    ,
    \begin{pmatrix}
     3 r_1 & 2 r_4+r_7 & 2 r_5+r_8 \\
     2 r_4+r_7 & 3 r_2 & 2 r_6+r_9 \\
     2 r_5+r_8 & 2 r_6+r_9 & 3 r_3
    \end{pmatrix}
    ,
    \begin{pmatrix}
     r_5+2 r_8 & 3 r_{11} \\
     3 r_{11} & r_6+2 r_9
    \end{pmatrix} , \right.
    \nonumber \\[3mm]
    &
\hspace{15mm}
    \left.
    \begin{pmatrix}
     r_6+2 r_9 & 3 r_{11} \\
     3 r_{11} & r_5+2 r_8
    \end{pmatrix}
    ,
    (r_4+2 r_7) , (r_4+2 r_7) \right\} \, .
\end{align}
with eigenvalues of $M_0^{0}$:
\begin{align}
    &\Lambda^{0,0}_{1-9} = \Lambda^{+,0}_{1-9} \, , \nonumber \\[2mm]
    &\Lambda^{0,0}_{10-12} = \text{Roots of:} \nonumber \\
    &\quad \quad \quad \quad x^3 + 2 \left(-3 r_1-3 r_2-3 r_3\right) x^2 
    + 4 \left(-4 r_4^2-4 r_7 r_4-4 r_5^2-4 r_6^2-r_7^2-r_8^2-r_9^2 \right.
    \nonumber \\
    &\left. \quad \quad \quad \quad +9 r_1 r_2+9 r_1 r_3+9 r_2 r_3-4 r_5 r_8-4 r_6 r_9\right)x
    + 8 \left( 12 r_3 r_4^2+12 r_2 r_5^2+12 r_1 r_6^2 \right. \nonumber \\
    &\quad \quad \quad \quad +3 r_3 r_7^2+3 r_2 r_8^2+3 r_1 r_9^2-27 r_1 r_2 r_3-16 r_4 r_5 r_6+12 r_3 r_4 r_7-8 r_5 r_6 r_7 \nonumber \\
    &\quad \quad \quad \quad +12 r_2 r_5 r_8-8 r_4 r_6 r_8-4 r_6 r_7 r_8-8 r_4 r_5 r_9+12 r_1 r_6 r_9-4 r_5 r_7 r_9-4 r_4 r_8 r_9 \nonumber \\
    &\quad \quad \quad \quad \left. -2 r_7 r_8 r_9 \right) = 0 \, , \nonumber \\[2mm]
    &\Lambda^{0,0}_{13,14} = \pm \sqrt{36 r_{11}^2+
    \left(r_5-r_6+2 r_8-2 r_9\right){}^2}+r_5+r_6+2 \left(r_8+r_9\right) \, , \nonumber \\[2mm]
    &\Lambda^{0,0}_{15,16} = \Lambda^{0,0}_{13,14} \, , \nonumber \\[2mm]
    &\Lambda^{0,0}_{17,18} = 2 (r_4+2 r_7) \, .
\end{align}

\subsection{The $\mathrm{U}(1) \times \mathbb{Z}_2$ symmetry}

By imposing $G = \mathrm{U}(1) \times \mathbb{Z}_2$ 
with representation $\mathrm{diag}(1,-1,e^{i \alpha})$,
with $\alpha \neq k \pi/2, k \in \mathbb{Z}$,
we get the quartic potential\footnote{In \cite{Ferreira:2008zy},
the authors state that $\alpha \neq \{ 0, \pi \}$ but if 
$\alpha \neq k \pi/2, k \in \mathbb{Z}$ we also get a generator for
$\mathbb{Z}_4$.}
\begin{align}
    V_{\mathrm{U}(1) \times \mathbb{Z}_2} =&
    \sum_{i=1}^3 r_i |\phi_i|^4 + 2 r_4 (\phi_1^\dagger \phi_1)(\phi_2^\dagger \phi_2)
    + 2 r_5 (\phi_1^\dagger \phi_1)(\phi_3^\dagger \phi_3)
    + 2 r_6 (\phi_2^\dagger \phi_2)(\phi_3^\dagger \phi_3)
    \nonumber \\[2mm]
    & + 2 r_7 |\phi_1^\dagger \phi_2|^2
    + 2 r_8 |\phi_1^\dagger \phi_3|^2 + 2 r_9 |\phi_2^\dagger \phi_3|^2
    + r_{10} \left[ (\phi_1^\dagger \phi_2)^2 + h.c. \right] \, ,
\end{align}
which can be easily achieved by setting from $\mathbb{Z}_4$
the constraint $r_{11} \rightarrow 0$.

Thus, we get the following scattering matrices.

\subsubsection*{The matrix $M_2^{++}$}

From $M_2^{++}$ we get
\begin{equation}
    \frac{1}{2} \, M_2^{++} = \mathrm{diag} \left\{
    \begin{pmatrix}
     r_1 & r_{10} \\
     r_{10} & r_2
    \end{pmatrix}
    ,
    (r_4 + r_7)
    ,
    (r_5 + r_8)
    ,
    (r_6+r_9)
    ,
    r_3 \right\} \, ,
\end{equation}
and thus we get the eigenvalues of $M_2^{++}$:
\begin{align}
    &\Lambda^{++}_{1,2} = \pm \sqrt{4 r_{10}^2 +\left(r_1-r_2\right){}^2}+r_1+r_2 \, ,
    \nonumber \\[2mm]
    &\Lambda^{++}_{3} = 2 (r_4 + r_7) \, , \nonumber \\[2mm]
    &\Lambda^{++}_{4} =  2 (r_5 + r_8) \, , \nonumber \\[2mm]
    &\Lambda^{++}_{5} =  2 (r_6+r_9) \, , \nonumber \\[2mm]
    &\Lambda^{++}_{6} =  2 r_3 \, .
\end{align}

\subsubsection*{The matrix $M_2^+$}

From $M_2^{+}$ we get
\begin{equation}
    \frac{1}{2} \, M_2^{+} = \mathrm{diag} \left\{
    \begin{pmatrix}
     r_1 & r_{10} \\
     r_{10} & r_2
    \end{pmatrix}
    ,
    \begin{pmatrix}
     r_4 & r_7 \\
     r_7 & r_4
    \end{pmatrix}
    ,
    \begin{pmatrix}
     r_5 & r_8 \\
     r_8 & r_5
    \end{pmatrix}
    ,
    \begin{pmatrix}
     r_6 & r_9 \\
     r_9 & r_6
    \end{pmatrix}
    ,
    r_3 \right\} \, ,
\end{equation}
with eigenvalues of $M_2^{+}$:
\begin{align}
    &\Lambda^{+,2}_{1,2} = \Lambda^{++}_{1,2} \, , \nonumber \\[2mm]
    &\Lambda^{+,2}_{3,4} = 2 (r_4 \pm r_7) \, , \nonumber \\[2mm]
    &\Lambda^{+,2}_{5,6} = 2 (r_5 \pm r_8) \, , \nonumber \\[2mm]
    &\Lambda^{+,2}_{7,8} = 2 (r_6 \pm r_9) \, , \nonumber \\[2mm]
    &\Lambda^{+,2}_{9} = \Lambda^{++}_{6} \, .
\end{align}

\subsubsection*{The matrix $M_0^+$}

From $M_0^{+}$ we get
\begin{equation}
    \frac{1}{2} \, M_0^{+} = \mathrm{diag} \left\{
    \begin{pmatrix}
     r_1 & r_7 & r_8 \\
     r_7 & r_2 & r_9 \\
     r_8 & r_9 & r_3
    \end{pmatrix}
    ,
    \begin{pmatrix}
     r_4 & r_{10} \\
     r_{10} & r_4
    \end{pmatrix}
    ,
    r_5
    ,
    r_5
    ,
    r_6
    ,
    r_6 \right\} \, ,
\end{equation}
with eigenvalues of $M_0^{+}$:
\begin{align}
    &\Lambda^{+,0}_{1-3} = \text{Roots of:} \nonumber \\
    &\quad \quad \quad \quad x^3 - 2(r_1 + r_2 + r_3)x^2
    +4(-r_7^2-r_8^2-r_9^2+r_1 r_2+r_1 r_3+r_2 r_3)x \nonumber \\
    &\quad \quad \quad \quad +8 ( r_3 r_7^2-2 r_8 r_9 r_7+r_2 r_8^2+r_1 r_9^2-r_1 r_2 r_3) 
    = 0 \, , \nonumber \\[2mm]
    &\Lambda^{+,0}_{4,5} = 2 \left(r_4 \pm r_{10}\right)
    \, , \nonumber \\[2mm]
     &\Lambda^{+,0}_{6,7} = 2 r_5 \, , \nonumber \\[2mm]
    &\Lambda^{+,0}_{8,9} = 2 r_6 \, .
\end{align}

\subsubsection*{The matrix $M_2^0$}

From $M_2^{0}$ we get
\begin{align}
    \frac{1}{2} \, M_2^{0} = \frac{1}{2} \, M_2^{++} \ ,
\end{align}
with eigenvalues of $M_2^{0}$:
\begin{align}
    &\Lambda^{0,2}_{1-6} = \Lambda^{++}_{1-6} \, .
\end{align}

\subsubsection*{The matrix $M_0^0$}

From $M_0^{0}$ we get
\begin{align}
    \frac{1}{2} \, M_0^{0} \sim \, &
    \mathrm{diag} \left\{
    \frac{1}{2} M_0^{+}
    ,
    \begin{pmatrix}
     3 r_1 & 2 r_4+r_7 & 2 r_5+r_8 \\
     2 r_4+r_7 & 3 r_2 & 2 r_6+r_9 \\
     2 r_5+r_8 & 2 r_6+r_9 & 3 r_3
    \end{pmatrix}
    ,
    \begin{pmatrix}
     r_4+2 r_7 & 3 r_{10} \\
     3 r_{10} & r_4+2 r_7
    \end{pmatrix} , \right.
    \nonumber \\[3mm]
    &
\hspace{15mm}
    (r_5 + 2r_8)
    ,
    (r_5 + 2r_8)
    , 
    (r_6 + 2 r_9) , (r_6 + 2 r_9) \Big\} \, .
\end{align}
with eigenvalues of $M_0^{0}$:
\begin{align}
    &\Lambda^{0,0}_{1-9} = \Lambda^{+,0}_{1-9} \, , \nonumber \\[2mm]
    &\Lambda^{0,0}_{10-12} = \text{Roots of:} \nonumber \\
    &\quad \quad \quad \quad x^3 + 2 \left(-3 r_1-3 r_2-3 r_3\right) x^2 
    + 4 \left(-4 r_4^2-4 r_7 r_4-4 r_5^2-4 r_6^2-r_7^2-r_8^2-r_9^2 \right.
    \nonumber \\
    &\left. \quad \quad \quad \quad +9 r_1 r_2+9 r_1 r_3+9 r_2 r_3-4 r_5 r_8-4 r_6 r_9\right)x
    + 8 \left( 12 r_3 r_4^2+12 r_2 r_5^2+12 r_1 r_6^2 \right. \nonumber \\
    &\quad \quad \quad \quad +3 r_3 r_7^2+3 r_2 r_8^2+3 r_1 r_9^2-27 r_1 r_2 r_3-16 r_4 r_5 r_6+12 r_3 r_4 r_7-8 r_5 r_6 r_7 \nonumber \\
    &\quad \quad \quad \quad +12 r_2 r_5 r_8-8 r_4 r_6 r_8-4 r_6 r_7 r_8-8 r_4 r_5 r_9+12 r_1 r_6 r_9-4 r_5 r_7 r_9-4 r_4 r_8 r_9 \nonumber \\
    &\quad \quad \quad \quad \left. -2 r_7 r_8 r_9 \right) = 0 \, , \nonumber \\[2mm]
    &\Lambda^{0,0}_{13,14} = 2 \left(\pm 3 r_{10} +r_4+2 r_7\right)
    \, , \nonumber \\[2mm]
    &\Lambda^{0,0}_{15,16} = 2 (r_5 + 2 r_8) \, , \nonumber \\[2mm]
    &\Lambda^{0,0}_{17,18} = 2 (r_6 + 2 r_9) \, .
\end{align}

\subsection{The $\mathbb{Z}_2 \times \mathbb{Z}_2$ symmetry}

By imposing $G = \mathbb{Z}_2 \times \mathbb{Z}_2$
we get the quartic potential of the model originally proposed by
Weinberg \cite{Weinberg:1976hu},
\begin{align}
    V_{\mathbb{Z}_2 \times \mathbb{Z}_2} =&
    \sum_{i=1}^3 r_i |\phi_i|^4 + 2 r_4 (\phi_1^\dagger \phi_1)(\phi_2^\dagger \phi_2)
    + 2 r_5 (\phi_1^\dagger \phi_1)(\phi_3^\dagger \phi_3)
    + 2 r_6 (\phi_2^\dagger \phi_2)(\phi_3^\dagger \phi_3)
    \nonumber \\[2mm]
    & + 2 r_7 |\phi_1^\dagger \phi_2|^2
    + 2 r_8 |\phi_1^\dagger \phi_3|^2 + 2 r_9 |\phi_2^\dagger \phi_3|^2
    + \Big[
	c_3 (\phi_1^\dagger \phi_2)^2 + c_{5} (\phi_1^\dagger \phi_3)^2  \nonumber \\[2mm]
    &+ c_{17} (\phi_2^\dagger \phi_3)^2 + h.c. \Big]\, ,
\end{align}
which can be easily achieved by setting from the $\mathbb{Z}_2$ symmetric 3HDM potential
the constraints $\{c_1, c_7, c_{11}, c_{13}, c_{14} \} \rightarrow 0$.

Thus, we get the following scattering matrices which
reproduce in the limit of real coefficients the conditions
(91)--(100) of ref.~\cite{Moretti:2015cwa}.

\subsubsection*{The matrix $M_2^{++}$}

From $M_2^{++}$ we get
\begin{equation}
    \frac{1}{2} \, M_2^{++} = \mathrm{diag} \left\{
    \begin{pmatrix}
     r_1 & c_3 & c_5 \\
     c_3^* & r_2 & c_{17} \\
     c_5^* & c_{17}^* & r_3
    \end{pmatrix}
    ,
    (r_4 + r_7)
    ,
    (r_5 + r_8)
    ,
    (r_6+r_9) \right\} \, ,
\end{equation}
and thus we get the eigenvalues of $M_2^{++}$:
\begin{align}
    &\Lambda^{++}_{1-3} = \text{Roots of:} \nonumber \\
    &\quad \quad \quad \quad x^3 + 2 \left(-r_1-r_2-r_3\right) x^2 
    + 4 \left( -|c_3|^2 -|c_5|^2- |c_{17}|^2 +r_1 r_2+r_1 r_3+r_2 r_3 \right) x
    \nonumber \\
    &\quad \quad \quad \quad + 8 \left( r_3 |c_3|^2 + r_2 |c_5|^2 + r_1 |c_{17}|^2
    - 2 \mathrm{Re}(c_3 c_5^* c_{17}) -r_1 r_2 r_3
    \right)
    = 0 \, , \nonumber \\[2mm]
    &\Lambda^{++}_{4} = 2 (r_4 + r_7) \, , \nonumber \\[2mm]
    &\Lambda^{++}_{5} =  2 (r_5 + r_8) \, , \nonumber \\[2mm]
    &\Lambda^{++}_{6} =  2 (r_6+r_9) \, .
\end{align}

\subsubsection*{The matrix $M_2^+$}

From $M_2^{+}$ we get
\begin{equation}
    \frac{1}{2} \, M_2^{+} = \mathrm{diag} \left\{
    \begin{pmatrix}
     r_1 & c_3 & c_5 \\
     c_3^* & r_2 & c_{17} \\
     c_5^* & c_{17}^* & r_3
    \end{pmatrix}
    ,
    \begin{pmatrix}
     r_4 & r_7 \\
     r_7 & r_4
    \end{pmatrix}
    ,
    \begin{pmatrix}
     r_5 & r_8 \\
     r_8 & r_5
    \end{pmatrix}
    ,
    \begin{pmatrix}
     r_6 & r_9 \\
     r_9 & r_6
    \end{pmatrix} \right\} \, ,
\end{equation}
with eigenvalues of $M_2^{+}$:
\begin{align}
    &\Lambda^{+,2}_{1-3} = \Lambda^{++}_{1-3} \, , \nonumber \\[2mm]
    &\Lambda^{+,2}_{4,5} = 2 (r_4 \pm r_7) \, , \nonumber \\[2mm]
    &\Lambda^{+,2}_{6,7} = 2 (r_5 \pm r_8) \, , \nonumber \\[2mm]
    &\Lambda^{+,2}_{8,9} = 2 (r_6 \pm r_9) \, .
\end{align}

\subsubsection*{The matrix $M_0^+$}

From $M_0^{+}$ we get
\begin{equation}
    \frac{1}{2} \, M_0^{+} = \mathrm{diag} \left\{
    \begin{pmatrix}
     r_1 & r_7 & r_8 \\
     r_7 & r_2 & r_9 \\
     r_8 & r_9 & r_3
    \end{pmatrix}
    ,
    \begin{pmatrix}
     r_4 & c_3 \\
     c_3^* & r_4
    \end{pmatrix}
    ,
    \begin{pmatrix}
     r_5 & c_5 \\
     c_5^* & r_5
    \end{pmatrix}
    ,
    \begin{pmatrix}
     r_6 & c_{17} \\
     c_{17}^* & r_6
    \end{pmatrix} \right\} \, ,
\end{equation}
with eigenvalues of $M_0^{+}$:
\begin{align}
    &\Lambda^{+,0}_{1-3} = \text{Roots of:} \nonumber \\
    &\quad \quad \quad \quad x^3 - 2(r_1 + r_2 + r_3)x^2
    +4(-r_7^2-r_8^2-r_9^2+r_1 r_2+r_1 r_3+r_2 r_3)x \nonumber \\
    &\quad \quad \quad \quad +8 ( r_3 r_7^2-2 r_8 r_9 r_7+r_2 r_8^2+r_1 r_9^2-r_1 r_2 r_3) 
    = 0 \, , \nonumber \\[2mm]
    &\Lambda^{+,0}_{4,5} = 2 \left(r_4 \pm \sqrt{c_3} \sqrt{c_3^*}\right)
    \, , \nonumber \\[2mm]
     &\Lambda^{+,0}_{6,7} = 2 \left(r_5 \pm \sqrt{c_5} \sqrt{c_5^*}\right)
     \, , \nonumber \\[2mm]
    &\Lambda^{+,0}_{8,9} = 2 \left(r_6 \pm \sqrt{c_{17}} \sqrt{c_{17}^*}\right)
    \, .
\end{align}

\subsubsection*{The matrix $M_2^0$}

From $M_2^{0}$ we get
\begin{align}
    \frac{1}{2} \, M_2^{0} = \frac{1}{2} \, M_2^{++} \ ,
\end{align}
with eigenvalues of $M_2^{0}$:
\begin{align}
    &\Lambda^{0,2}_{1-6} = \Lambda^{++}_{1-6} \, .
\end{align}

\subsubsection*{The matrix $M_0^0$}

From $M_0^{0}$ we get
\begin{align}
    \frac{1}{2} \, M_0^{0} \sim \, &
    \mathrm{diag} \left\{
    \frac{1}{2} M_0^{+}
    ,
    \begin{pmatrix}
     3 r_1 & 2 r_4+r_7 & 2 r_5+r_8 \\
     2 r_4+r_7 & 3 r_2 & 2 r_6+r_9 \\
     2 r_5+r_8 & 2 r_6+r_9 & 3 r_3
    \end{pmatrix}
    ,
    \begin{pmatrix}
     r_4+2 r_7 & 3 c_3 \\
     3 c_3^* & r_4+2 r_7
    \end{pmatrix} , \right.
    \nonumber \\[3mm]
    &
\hspace{15mm}
    \left.
    \begin{pmatrix}
     r_5+2 r_8 & 3 c_5 \\
     3 c_5^* & r_5+2 r_8
    \end{pmatrix}
    ,
    \begin{pmatrix}
     r_6+2 r_9 & 3 c_{17} \\
     3 c_{17}^* & r_6+2 r9
    \end{pmatrix} \right\} \, .
\end{align}
with eigenvalues of $M_0^{0}$:
\begin{align}
    &\Lambda^{0,0}_{1-9} = \Lambda^{+,0}_{1-9} \, , \nonumber \\[2mm]
    &\Lambda^{0,0}_{10-12} = \text{Roots of:} \nonumber \\
    &\quad \quad \quad \quad x^3 + 2 \left(-3 r_1-3 r_2-3 r_3\right) x^2 
    + 4 \left(-4 r_4^2-4 r_7 r_4-4 r_5^2-4 r_6^2-r_7^2-r_8^2-r_9^2 \right.
    \nonumber \\
    &\left. \quad \quad \quad \quad +9 r_1 r_2+9 r_1 r_3+9 r_2 r_3-4 r_5 r_8-4 r_6 r_9\right)x
    + 8 \left( 12 r_3 r_4^2+12 r_2 r_5^2+12 r_1 r_6^2 \right. \nonumber \\
    &\quad \quad \quad \quad +3 r_3 r_7^2+3 r_2 r_8^2+3 r_1 r_9^2-27 r_1 r_2 r_3-16 r_4 r_5 r_6+12 r_3 r_4 r_7-8 r_5 r_6 r_7 \nonumber \\
    &\quad \quad \quad \quad +12 r_2 r_5 r_8-8 r_4 r_6 r_8-4 r_6 r_7 r_8-8 r_4 r_5 r_9+12 r_1 r_6 r_9-4 r_5 r_7 r_9-4 r_4 r_8 r_9 \nonumber \\
    &\quad \quad \quad \quad \left. -2 r_7 r_8 r_9 \right) = 0 \, , \nonumber \\[2mm]
    &\Lambda^{0,0}_{13,14} = 2 \left(\pm 3 \sqrt{c_3} \sqrt{c_3^*}+r_4+2 r_7\right)
    \, , \nonumber \\[2mm]
    &\Lambda^{0,0}_{15,16} = 2 \left(\pm 3 \sqrt{c_5} \sqrt{c_5^*}+r_5+2 r_8\right) 
    \, , \nonumber \\[2mm]
    &\Lambda^{0,0}_{17,18} = 2 \left(-3 \sqrt{c_{17}} \sqrt{c_{17}^*}+r_6+2 r_9\right)
    \, .
\end{align}

\subsection{The $\mathbb{Z}_2 \times \mathbb{Z}_2 \times \mathbb{Z}_2^{(\mathrm{CP})}$ symmetry}

By imposing $G = \mathbb{Z}_2 \times \mathbb{Z}_2 \times \mathbb{Z}_2^{(\mathrm{CP})}$,
the so called Branco model \cite{Branco:1980sz},
we get the quartic potential
\begin{align}
    V_{\mathbb{Z}_2 \times \mathbb{Z}_2 \times \mathbb{Z}_2^{(\mathrm{CP})}} =&
    \sum_{i=1}^3 r_i |\phi_i|^4 + 2 r_4 (\phi_1^\dagger \phi_1)(\phi_2^\dagger \phi_2)
    + 2 r_5 (\phi_1^\dagger \phi_1)(\phi_3^\dagger \phi_3)
    + 2 r_6 (\phi_2^\dagger \phi_2)(\phi_3^\dagger \phi_3)
    \nonumber \\[2mm]
    & + 2 r_7 |\phi_1^\dagger \phi_2|^2
    + 2 r_8 |\phi_1^\dagger \phi_3|^2 + 2 r_9 |\phi_2^\dagger \phi_3|^2
    + r_{10} \left[ (\phi_1^\dagger \phi_2)^2 + h.c. \right]   \nonumber \\[2mm]
    &+ r_{11} \left[ (\phi_1^\dagger \phi_3)^2 + h.c. \right] 
    + r_{12} \left[ (\phi_2^\dagger \phi_3)^2 + h.c. \right] \, ,
\end{align}
which can be easily achieved by setting from $\mathbb{Z}_2 \times \mathbb{Z}_2$
the constraints $\{c_3, c_5, c_{17} \} \in \mathbb{R}$.

Thus, we get the following scattering matrices which
reproduce the conditions
(91)--(100) of ref.~\cite{Moretti:2015cwa, Hernandez-Sanchez:2020aop}.

\subsubsection*{The matrix $M_2^{++}$}

From $M_2^{++}$ we get
\begin{equation}
    \frac{1}{2} \, M_2^{++} = \mathrm{diag} \left\{
    \begin{pmatrix}
     r_1 & r_{10} & r_{11} \\
     r_{10} & r_2 & r_{12} \\
     r_{11} & r_{12} & r_3
    \end{pmatrix}
    ,
    (r_4 + r_7)
    ,
    (r_5 + r_8)
    ,
    (r_6+r_9) \right\} \, ,
\end{equation}
and thus we get the eigenvalues of $M_2^{++}$:
\begin{align}
    &\Lambda^{++}_{1-3} = \text{Roots of:} \nonumber \\
    &\quad \quad \quad \quad x^3 + 2 \left(-r_1-r_2-r_3\right) x^2 
    + 4 \left( -r_{10}^2 -r_{11}^2- r_{12}^2 +r_1 r_2+r_1 r_3+r_2 r_3 \right) x
    \nonumber \\
    &\quad \quad \quad \quad + 8 \left( r_3 r_{10}^2 + r_2 r_{11}^2 + r_1 r_{12}^2
    - 2 r_{10} r_{11} r_{12} -r_1 r_2 r_3
    \right)
    = 0 \, , \nonumber \\[2mm]
    &\Lambda^{++}_{4} = 2 (r_4 + r_7) \, , \nonumber \\[2mm]
    &\Lambda^{++}_{5} =  2 (r_5 + r_8) \, , \nonumber \\[2mm]
    &\Lambda^{++}_{6} =  2 (r_6+r_9) \, .
\end{align}

\subsubsection*{The matrix $M_2^+$}

From $M_2^{+}$ we get
\begin{equation}
    \frac{1}{2} \, M_2^{+} = \mathrm{diag} \left\{
    \begin{pmatrix}
     r_1 & r_{10} & r_{11} \\
     r_{10} & r_2 & r_{12} \\
     r_{11} & r_{12} & r_3
    \end{pmatrix}
    ,
    \begin{pmatrix}
     r_4 & r_7 \\
     r_7 & r_4
    \end{pmatrix}
    ,
    \begin{pmatrix}
     r_5 & r_8 \\
     r_8 & r_5
    \end{pmatrix}
    ,
    \begin{pmatrix}
     r_6 & r_9 \\
     r_9 & r_6
    \end{pmatrix} \right\} \, ,
\end{equation}
with eigenvalues of $M_2^{+}$:
\begin{align}
    &\Lambda^{+,2}_{1-3} = \Lambda^{++}_{1-3} \, , \nonumber \\[2mm]
    &\Lambda^{+,2}_{4,5} = 2 (r_4 \pm r_7) \, , \nonumber \\[2mm]
    &\Lambda^{+,2}_{6,7} = 2 (r_5 \pm r_8) \, , \nonumber \\[2mm]
    &\Lambda^{+,2}_{8,9} = 2 (r_6 \pm r_9) \, .
\end{align}

\subsubsection*{The matrix $M_0^+$}

From $M_0^{+}$ we get
\begin{equation}
    \frac{1}{2} \, M_0^{+} = \mathrm{diag} \left\{
    \begin{pmatrix}
     r_1 & r_7 & r_8 \\
     r_7 & r_2 & r_9 \\
     r_8 & r_9 & r_3
    \end{pmatrix}
    ,
    \begin{pmatrix}
     r_4 & r_{10} \\
     r_{10} & r_4
    \end{pmatrix}
    ,
    \begin{pmatrix}
     r_5 & r_{11} \\
     r_{11} & r_5
    \end{pmatrix}
    ,
    \begin{pmatrix}
     r_6 & r_{12} \\
     r_{12} & r_6
    \end{pmatrix} \right\} \, ,
\end{equation}
with eigenvalues of $M_0^{+}$:
\begin{align}
    &\Lambda^{+,0}_{1-3} = \text{Roots of:} \nonumber \\
    &\quad \quad \quad \quad x^3 - 2(r_1 + r_2 + r_3)x^2
    +4(-r_7^2-r_8^2-r_9^2+r_1 r_2+r_1 r_3+r_2 r_3)x \nonumber \\
    &\quad \quad \quad \quad +8 ( r_3 r_7^2-2 r_8 r_9 r_7+r_2 r_8^2+r_1 r_9^2-r_1 r_2 r_3) 
    = 0 \, , \nonumber \\[2mm]
    &\Lambda^{+,0}_{4,5} = 2 \left(r_4 \pm r_{10} \right)
    \, , \nonumber \\[2mm]
     &\Lambda^{+,0}_{6,7} = 2 \left(r_5 \pm r_{11} \right)
     \, , \nonumber \\[2mm]
    &\Lambda^{+,0}_{8,9} = 2 \left(r_6 \pm r_{12} \right)
    \, .
\end{align}

\subsubsection*{The matrix $M_2^0$}

From $M_2^{0}$ we get
\begin{align}
    \frac{1}{2} \, M_2^{0} = \frac{1}{2} \, M_2^{++} \ ,
\end{align}
with eigenvalues of $M_2^{0}$:
\begin{align}
    &\Lambda^{0,2}_{1-6} = \Lambda^{++}_{1-6} \, .
\end{align}

\subsubsection*{The matrix $M_0^0$}

From $M_0^{0}$ we get
\begin{align}
    \frac{1}{2} \, M_0^{0} \sim \, &
    \mathrm{diag} \left\{
    \frac{1}{2} M_0^{+}
    ,
    \begin{pmatrix}
     3 r_1 & 2 r_4+r_7 & 2 r_5+r_8 \\
     2 r_4+r_7 & 3 r_2 & 2 r_6+r_9 \\
     2 r_5+r_8 & 2 r_6+r_9 & 3 r_3
    \end{pmatrix}
    ,
    \begin{pmatrix}
     r_4+2 r_7 & 3 r_{10} \\
     3 r_{10} & r_4+2 r_7
    \end{pmatrix} , \right.
    \nonumber \\[3mm]
    &
\hspace{15mm}
    \left.
    \begin{pmatrix}
     r_5+2 r_8 & 3 r_{11} \\
     3 r_{11} & r_5+2 r_8
    \end{pmatrix}
    ,
    \begin{pmatrix}
     r_6+2 r_9 & 3 r_{12} \\
     3 r_{12} & r_6+2 r9
    \end{pmatrix} \right\} \, .
\end{align}
with eigenvalues of $M_0^{0}$:
\begin{align}
    &\Lambda^{0,0}_{1-9} = \Lambda^{+,0}_{1-9} \, , \nonumber \\[2mm]
    &\Lambda^{0,0}_{10-12} = \text{Roots of:} \nonumber \\
    &\quad \quad \quad \quad x^3 + 2 \left(-3 r_1-3 r_2-3 r_3\right) x^2 
    + 4 \left(-4 r_4^2-4 r_7 r_4-4 r_5^2-4 r_6^2-r_7^2-r_8^2-r_9^2 \right.
    \nonumber \\
    &\left. \quad \quad \quad \quad +9 r_1 r_2+9 r_1 r_3+9 r_2 r_3-4 r_5 r_8-4 r_6 r_9\right)x
    + 8 \left( 12 r_3 r_4^2+12 r_2 r_5^2+12 r_1 r_6^2 \right. \nonumber \\
    &\quad \quad \quad \quad +3 r_3 r_7^2+3 r_2 r_8^2+3 r_1 r_9^2-27 r_1 r_2 r_3-16 r_4 r_5 r_6+12 r_3 r_4 r_7-8 r_5 r_6 r_7 \nonumber \\
    &\quad \quad \quad \quad +12 r_2 r_5 r_8-8 r_4 r_6 r_8-4 r_6 r_7 r_8-8 r_4 r_5 r_9+12 r_1 r_6 r_9-4 r_5 r_7 r_9-4 r_4 r_8 r_9 \nonumber \\
    &\quad \quad \quad \quad \left. -2 r_7 r_8 r_9 \right) = 0 \, , \nonumber \\[2mm]
    &\Lambda^{0,0}_{13,14} = 2 \left(r_4+2 r_7 \pm 3 r_{10}\right)
    \, , \nonumber \\[2mm]
    &\Lambda^{0,0}_{15,16} = 2 \left(r_5+2 r_8 \pm 3 r_{11}\right)
    \, , \nonumber \\[2mm]
    &\Lambda^{0,0}_{17,18} = 2 \left(r_6+2 r_9 \pm 3 r_{12}\right)
    \, .
\end{align}

\subsection{The $\mathrm{U}(1) \times \mathrm{U}(1)$ symmetry}

By imposing $G = \mathrm{U}(1) \times \mathrm{U}(1)$ 
with representation $\mathrm{diag}(1,e^{i \alpha},e^{i \beta})$,
with $\alpha \neq \{ 0, \pi \}$ and $\beta \neq \{ 0, \pi, \pm \alpha \}$,
we get the quartic potential
\begin{align}
    V_{\mathrm{U}(1) \times \mathrm{U}(1)} =&
    \sum_{i=1}^3 r_i |\phi_i|^4 + 2 r_4 (\phi_1^\dagger \phi_1)(\phi_2^\dagger \phi_2)
    + 2 r_5 (\phi_1^\dagger \phi_1)(\phi_3^\dagger \phi_3)
    + 2 r_6 (\phi_2^\dagger \phi_2)(\phi_3^\dagger \phi_3)
    \nonumber \\[2mm]
    & + 2 r_7 |\phi_1^\dagger \phi_2|^2
    + 2 r_8 |\phi_1^\dagger \phi_3|^2 + 2 r_9 |\phi_2^\dagger \phi_3|^2 \, ,
\end{align}
which can be easily achieved by setting from $\mathrm{U}(1)_1$
the constraint $c_{11} \rightarrow 0$.

\subsubsection*{The matrix $M_2^{++}$}

From $M_2^{++}$ we get
\begin{equation}
    \frac{1}{2} \, M_2^{++} = \mathrm{diag} \left\{
    r_1
    ,
    r_2
    ,
    r_3
    ,
    (r_4 + r_7)
    ,
    (r_5 + r_8)
    ,
    (r_6+r_9) \right\} \, ,
\end{equation}
and thus we get the eigenvalues of $M_2^{++}$:
\begin{align}
    &\Lambda^{++}_{1} = 2 r_1 \, ,
    \nonumber \\[2mm]
    &\Lambda^{++}_{2} =  2 r_2 \, , \nonumber \\[2mm]
    &\Lambda^{++}_{3} =  2 r_3 \, , \nonumber \\[2mm]
    &\Lambda^{++}_{4} = 2 (r_4 + r_7) \, , \nonumber \\[2mm]
    &\Lambda^{++}_{5} =  2 (r_5 + r_8) \, , \nonumber \\[2mm]
    &\Lambda^{++}_{6} =  2 (r_6+r_9) \, .
\end{align}

\subsubsection*{The matrix $M_2^+$}

From $M_2^{+}$ we get
\begin{equation}
    \frac{1}{2} \, M_2^{+} = \mathrm{diag} \left\{
    \begin{pmatrix}
     r_4 & r_7 \\
     r_7 & r_4
    \end{pmatrix}
    ,
    \begin{pmatrix}
     r_5 & r_8 \\
     r_8 & r_5
    \end{pmatrix}
    ,
    \begin{pmatrix}
     r_6 & r_9 \\
     r_9 & r_6
    \end{pmatrix}
    ,
    r_1
    ,
    r_2
    ,
    r_3 \right\} \, ,
\end{equation}
with eigenvalues of $M_2^{+}$:
\begin{align}
    &\Lambda^{+,2}_{1,2} = 2 (r_4 \pm r_7) \, , \nonumber \\[2mm]
    &\Lambda^{+,2}_{3,4} = 2 (r_5 \pm r_8) \, , \nonumber \\[2mm]
    &\Lambda^{+,2}_{5,6} = 2 (r_6 \pm r_9) \, , \nonumber \\[2mm]
    &\Lambda^{+,2}_{7} = \Lambda^{++}_{1} \, , \nonumber \\[2mm]
    &\Lambda^{+,2}_{8} = \Lambda^{++}_{2} \, , \nonumber \\[2mm]
    &\Lambda^{+,2}_{9} = \Lambda^{++}_{3} \, .
\end{align}

\subsubsection*{The matrix $M_0^+$}

From $M_0^{+}$ we get
\begin{equation}
    \frac{1}{2} \, M_0^{+} = \mathrm{diag} \left\{
    \begin{pmatrix}
     r_1 & r_7 & r_8 \\
     r_7 & r_2 & r_9 \\
     r_8 & r_9 & r_3
    \end{pmatrix}
    ,
    r_4
    ,
    r_4
    ,
    r_5
    ,
    r_5
    ,
    r_6
    ,
    r_6 \right\} \, ,
\end{equation}
with eigenvalues of $M_0^{+}$:
\begin{align}
    &\Lambda^{+,0}_{1-3} = \text{Roots of:} \nonumber \\
    &\quad \quad \quad \quad x^3 - 2(r_1 + r_2 + r_3)x^2
    +4(-r_7^2-r_8^2-r_9^2+r_1 r_2+r_1 r_3+r_2 r_3)x \nonumber \\
    &\quad \quad \quad \quad +8 ( r_3 r_7^2-2 r_8 r_9 r_7+r_2 r_8^2+r_1 r_9^2-r_1 r_2 r_3) 
    = 0 \, , \nonumber \\[2mm]
    &\Lambda^{+,0}_{4,5} = 2 r_4
    \, , \nonumber \\[2mm]
     &\Lambda^{+,0}_{6,7} = 2 r_5 \, , \nonumber \\[2mm]
    &\Lambda^{+,0}_{8,9} = 2 r_6 \, .
\end{align}

\subsubsection*{The matrix $M_2^0$}

From $M_2^{0}$ we get
\begin{align}
    \frac{1}{2} \, M_2^{0} = \frac{1}{2} \, M_2^{++} \ ,
\end{align}
with eigenvalues of $M_2^{0}$:
\begin{align}
    &\Lambda^{0,2}_{1-6} = \Lambda^{++}_{1-6} \, .
\end{align}

\subsubsection*{The matrix $M_0^0$}

From $M_0^{0}$ we get
\begin{align}
    \frac{1}{2} \, M_0^{0} \sim \, &
    \mathrm{diag} \left\{
    \frac{1}{2} M_0^{+}
    ,
    \begin{pmatrix}
     3 r_1 & 2 r_4+r_7 & 2 r_5+r_8 \\
     2 r_4+r_7 & 3 r_2 & 2 r_6+r_9 \\
     2 r_5+r_8 & 2 r_6+r_9 & 3 r_3
    \end{pmatrix}
    ,
    (r_4 + 2r_7)
    ,
    (r_4 + 2r_7) , \right.
    \nonumber \\[3mm]
    &
\hspace{15mm}
    (r_5 + 2r_8)
    ,
    (r_5 + 2r_8)
    , 
    (r_6 + 2 r_9) , (r_6 + 2 r_9) \Big\} \, .
\end{align}
with eigenvalues of $M_0^{0}$:
\begin{align}
    &\Lambda^{0,0}_{1-9} = \Lambda^{+,0}_{1-9} \, , \nonumber \\[2mm]
    &\Lambda^{0,0}_{10-12} = \text{Roots of:} \nonumber \\
    &\quad \quad \quad \quad x^3 + 2 \left(-3 r_1-3 r_2-3 r_3\right) x^2 
    + 4 \left(-4 r_4^2-4 r_7 r_4-4 r_5^2-4 r_6^2-r_7^2-r_8^2-r_9^2 \right.
    \nonumber \\
    &\left. \quad \quad \quad \quad +9 r_1 r_2+9 r_1 r_3+9 r_2 r_3-4 r_5 r_8-4 r_6 r_9\right)x
    + 8 \left( 12 r_3 r_4^2+12 r_2 r_5^2+12 r_1 r_6^2 \right. \nonumber \\
    &\quad \quad \quad \quad +3 r_3 r_7^2+3 r_2 r_8^2+3 r_1 r_9^2-27 r_1 r_2 r_3-16 r_4 r_5 r_6+12 r_3 r_4 r_7-8 r_5 r_6 r_7 \nonumber \\
    &\quad \quad \quad \quad +12 r_2 r_5 r_8-8 r_4 r_6 r_8-4 r_6 r_7 r_8-8 r_4 r_5 r_9+12 r_1 r_6 r_9-4 r_5 r_7 r_9-4 r_4 r_8 r_9 \nonumber \\
    &\quad \quad \quad \quad \left. -2 r_7 r_8 r_9 \right) = 0 \, , \nonumber \\[2mm]
    &\Lambda^{0,0}_{13,14} = 2 (r_4 + 2r_7)
    \, , \nonumber \\[2mm]
    &\Lambda^{0,0}_{15,16} = 2 (r_5 + 2 r_8) \, , \nonumber \\[2mm]
    &\Lambda^{0,0}_{17,18} = 2 (r_6 + 2 r_9) \, .
\end{align}

\subsection{The $\mathrm{U}(2)$ symmetry}

By imposing $G = \mathrm{U}(2)$
we get the quartic potential
\begin{align}
    V_{\mathrm{U}(2)} =&
    r_1 \left[ (\phi_1^\dagger \phi_1) + (\phi_2^\dagger \phi_2) \right]^2
    + r_3 |\phi_3|^4 + 2 r_5 (\phi_1^\dagger \phi_1
    + \phi_2^\dagger \phi_2)(\phi_3^\dagger \phi_3)
    \nonumber \\[2mm]
    & + 2 r_7 \left[ |\phi_1^\dagger \phi_2|^2 - 
    (\phi_1^\dagger \phi_1)(\phi_2^\dagger \phi_2) \right]
    + 2 r_8 \left[ |\phi_1^\dagger \phi_3|^2 + |\phi_2^\dagger \phi_3|^2 \right]\, ,
\end{align}
with the following scattering matrices.

\subsubsection*{The matrix $M_2^{++}$}

From $M_2^{++}$ we get
\begin{equation}
    \frac{1}{2} \, M_2^{++} = \mathrm{diag} \left\{
    r_1
    ,
    r_1
    ,
    r_1
    ,
    r_3
    ,
    (r_5 + r_8)
    ,
    (r_5 + r_8) \right\} \, ,
\end{equation}
and thus we get the eigenvalues of $M_2^{++}$:
\begin{align}
    &\Lambda^{++}_{1,2,3} = 2 r_1 \, ,
    \nonumber \\[2mm]
    &\Lambda^{++}_{4} =  2 r_3 \, , \nonumber \\[2mm]
    &\Lambda^{++}_{5,6} =  2 (r_5+r_8) \, .
\end{align}

\subsubsection*{The matrix $M_2^+$}

From $M_2^{+}$ we get
\begin{equation}
    \frac{1}{2} \, M_2^{+} = \mathrm{diag} \left\{
    \begin{pmatrix}
     r_1 - r_7 & r_7 \\
     r_7 & r_1 - r_7
    \end{pmatrix}
    ,
    \begin{pmatrix}
     r_5 & r_8 \\
     r_8 & r_5
    \end{pmatrix}
    ,
    \begin{pmatrix}
     r_5 & r_8 \\
     r_8 & r_5
    \end{pmatrix}
    ,
    r_1
    ,
    r_1
    ,
    r_3 \right\} \, ,
\end{equation}
with eigenvalues of $M_2^{+}$:
\begin{align}
    &\Lambda^{+,2}_{1} = 2 (r_1 - 2 r_7) \, , \nonumber \\[2mm]
    &\Lambda^{+,2}_{2-5} = 2 (r_5 \pm r_8) \, , \nonumber \\[2mm]
    &\Lambda^{+,2}_{6-8} = 2 r_1 \, , \nonumber \\[2mm]
    &\Lambda^{+,2}_{9} = 2 r_3 \, .
\end{align}

\subsubsection*{The matrix $M_0^+$}

From $M_0^{+}$ we get
\begin{equation}
    \frac{1}{2} \, M_0^{+} = \mathrm{diag} \left\{
    \begin{pmatrix}
     r_1 & r_7 & r_8 \\
     r_7 & r_1 & r_8 \\
     r_8 & r_8 & r_3
    \end{pmatrix}
    ,
    (r_1 - r_7)
    ,
    (r_1 - r_7)
    ,
    r_5
    ,
    r_5
    ,
    r_5
    ,
    r_5 \right\} \, ,
\end{equation}
with eigenvalues of $M_0^{+}$:
\begin{align}
    &\Lambda^{+,0}_{1,2} = \pm \sqrt{r_1^2-2 \left(r_3-r_7\right) 
    r_1+r_3^2+r_7^2+8 r_8^2-2 r_3 r_7}+r_1+r_3+r_7 \, , \nonumber \\[2mm]
    &\Lambda^{+,0}_{3-5} = 2 (r_1 - r_7)
    \, , \nonumber \\[2mm]
    &\Lambda^{+,0}_{6-9} = 2 r_5 \, .
\end{align}

\subsubsection*{The matrix $M_2^0$}

From $M_2^{0}$ we get
\begin{align}
    \frac{1}{2} \, M_2^{0} = \frac{1}{2} \, M_2^{++} \ ,
\end{align}
with eigenvalues of $M_2^{0}$:
\begin{align}
    &\Lambda^{0,2}_{1-6} = \Lambda^{++}_{1-6} \, .
\end{align}

\subsubsection*{The matrix $M_0^0$}

From $M_0^{0}$ we get
\begin{align}
    \frac{1}{2} \, M_0^{0} \sim \, &
    \mathrm{diag} \left\{
    \frac{1}{2} M_0^{+}
    ,
    \begin{pmatrix}
     3 r_1 & 2 r_1 - r_7 & 2 r_5+r_8 \\
     2 r_1 - r_7 & 3 r_1 & 2 r_5+r_8 \\
     2 r_5+r_8 & 2 r_5+r_8 & 3 r_3
    \end{pmatrix}
    ,
    (r_1 + r_7)
    ,
    (r_1 + r_7) , \right.
    \nonumber \\[3mm]
    &
\hspace{15mm}
    (r_5 + 2r_8)
    ,
    (r_5 + 2r_8)
    , 
    (r_5 + 2r_8) , (r_5 + 2r_8) \Big\} \, .
\end{align}
with eigenvalues of $M_0^{0}$:
\begin{align}
    &\Lambda^{0,0}_{1-9} = \Lambda^{+,0}_{1-9} \, , \nonumber \\[2mm]
    &\Lambda^{0,0}_{10,11} = \pm \sqrt{\left(-5 r_1-3 r_3+r_7\right){}^2+4 
    \left(8 r_5^2+8 r_8 r_5+2 r_8^2-15 r_1 r_3+3 r_3 r_7\right)}+5 r_1+3 r_3-r_7 \, , \nonumber \\[2mm]
    &\Lambda^{0,0}_{12-14} = 2 (r_1 + r_7) \, , \nonumber \\[2mm]
    &\Lambda^{0,0}_{15-18} = 2 (r_5 + 2 r_8) \, .
\end{align}

\subsection{The $\mathrm{O}(2)$ symmetry}

By imposing $G = \mathrm{O}(2)$
we get the quartic potential
\begin{align}
    V_{\mathrm{O}(2)} =&
    r_1 \left[ (\phi_1^\dagger \phi_1)^2 + (\phi_2^\dagger \phi_2)^2 \right]
    + r_3 |\phi_3|^4 + 2 r_4 (\phi_1^\dagger \phi_1)(\phi_2^\dagger \phi_2)
    \nonumber \\[2mm]
    &+ 2 r_5 (\phi_1^\dagger \phi_1
    + \phi_2^\dagger \phi_2)(\phi_3^\dagger \phi_3) + 2 r_7 |\phi_1^\dagger \phi_2|^2
    \nonumber \\[2mm]
    &+ 2 r_8 \left[ |\phi_1^\dagger \phi_3|^2 + |\phi_2^\dagger \phi_3|^2 \right] 
    + 2 r_{10} \left[ (\phi_1^\dagger \phi_3)(\phi_2^\dagger \phi_3) + h.c. \right] \, ,
\end{align}
with the following scattering matrices.

\subsubsection*{The matrix $M_2^{++}$}

From $M_2^{++}$ we get
\begin{equation}
    \frac{1}{2} \, M_2^{++} = \mathrm{diag} \left\{
    \begin{pmatrix}
     r_4+r_7 & \sqrt{2} r_{10} \\
     \sqrt{2} r_{10} & r_3 
    \end{pmatrix}
    ,
    (r_5 + r_8)
    ,
    (r_5 + r_8)
    ,
    r_1
    ,
    r_1 \right\} \, ,
\end{equation}
and thus we get the eigenvalues of $M_2^{++}$:
\begin{align}
    &\Lambda^{++}_{1,2} = \pm \sqrt{8 r_{10}^2 +\left(-r_3+r_4+r_7\right){}^2}+r_3+r_4+r_7 \, ,
    \nonumber \\[2mm]
    &\Lambda^{++}_{3,4} =  2 (r_5 + r_8) \, , \nonumber \\[2mm]
    &\Lambda^{++}_{5,6} =  2 r_1 \, .
\end{align}

\subsubsection*{The matrix $M_2^+$}

From $M_2^{+}$ we get
\begin{equation}
    \frac{1}{2} \, M_2^{+} = \mathrm{diag} \left\{
    \begin{pmatrix}
     r_4 & r_7 & r_{10} \\
     r_7 & r_4 & r_{10} \\
     r_{10} & r_{10} & r_3
    \end{pmatrix}
    ,
    \begin{pmatrix}
     r_5 & r_8 \\
     r_8 & r_5
    \end{pmatrix}
    ,
    \begin{pmatrix}
     r_5 & r_8 \\
     r_8 & r_5
    \end{pmatrix}
    ,
    r_1
    ,
    r_1 \right\} \, ,
\end{equation}
with eigenvalues of $M_2^{+}$:
\begin{align}
    &\Lambda^{+,2}_{1, 2} = \pm\sqrt{8 r_{10}^2
    +\left(-r_3+r_4+r_7\right){}^2}+r_3+r_4+r_7 \, , \nonumber \\[2mm]
    &\Lambda^{+,2}_{3} = 2 (r_4 - r_7) \, , \nonumber \\[2mm]
    &\Lambda^{+,2}_{4-7} = 2 (r_5 \pm r_8) \, , \nonumber \\[2mm]
    &\Lambda^{+,2}_{8,9} = 2 r_1 \, .
\end{align}

\subsubsection*{The matrix $M_0^+$}

From $M_0^{+}$ we get
\begin{equation}
    \frac{1}{2} \, M_0^{+} = \mathrm{diag} \left\{
    \begin{pmatrix}
     r_1 & r_7 & r_8 \\
     r_7 & r_1 & r_8 \\
     r_8 & r_8 & r_3
    \end{pmatrix}
    ,
    \begin{pmatrix}
     r_5 & r_{10} \\
     r_{10} & r_5
    \end{pmatrix}
    ,
    \begin{pmatrix}
     r_5 & r_{10} \\
     r_{10} & r_5
    \end{pmatrix}
    ,
    r_4
    ,
    r_4 \right\} \, ,
\end{equation}
with eigenvalues of $M_0^{+}$:
\begin{align}
    &\Lambda^{+,0}_{1,2} = \pm \sqrt{\left(r_1-r_3+r_7\right){}^2+8 r_8^2}
    +r_1+r_3+r_7 \, , \nonumber \\[2mm]
    &\Lambda^{+,0}_{3} = 2 (r_1 - r_7)
    \, , \nonumber \\[2mm]
    &\Lambda^{+,0}_{4-7} = 2 \left(r_5 \pm r_{10} \right)
    \, , \nonumber \\[2mm]
    &\Lambda^{+,0}_{8,9} = 2 r_4 \, .
\end{align}

\subsubsection*{The matrix $M_2^0$}

From $M_2^{0}$ we get
\begin{align}
    \frac{1}{2} \, M_2^{0} = \frac{1}{2} \, M_2^{++} \ ,
\end{align}
with eigenvalues of $M_2^{0}$:
\begin{align}
    &\Lambda^{0,2}_{1-6} = \Lambda^{++}_{1-6} \, .
\end{align}

\subsubsection*{The matrix $M_0^0$}

From $M_0^{0}$ we get
\begin{align}
    \frac{1}{2} \, M_0^{0} \sim \, &
    \mathrm{diag} \left\{
    \frac{1}{2} M_0^{+}
    ,
    \begin{pmatrix}
     3 r_1 & 2 r_4 + r_7 & 2 r_5+r_8 \\
     r_4 + r_7 & 3 r_1 & 2 r_5+r_8 \\
     2 r_5+r_8 & 2 r_5+r_8 & 3 r_3
    \end{pmatrix}
    ,
    \begin{pmatrix}
     r_5+2 r_8 & 3 r_{10} \\
     3 r_{10} & r_5+2 r_8
    \end{pmatrix} , \right.
    \nonumber \\[3mm]
    & 
\hspace{15mm}
\left.
    \begin{pmatrix}
     r_5+2 r_8 & 3 r_{10} \\
     3 r_{10} & r_5+2 r_8
    \end{pmatrix}
    ,
    (r_4 + 2r_7)
    ,
    (r_4 + 2r_7) \right\} \, .
\end{align}
with eigenvalues of $M_0^{0}$:
\begin{align}
    &\Lambda^{0,0}_{1-9} = \Lambda^{+,0}_{1-9} \, , \nonumber \\[2mm]
    &\Lambda^{0,0}_{10,11} = \pm \sqrt{8 \left(2 r_5+r_8\right){}^2
    + \left(3 r_1-3 r_3+2 r_4+r_7\right){}^2} \nonumber \\
    &\quad \quad \quad \quad +3 r_1+3 r_3+2 r_4+r_7 \, , \nonumber \\[2mm]
    &\Lambda^{0,0}_{12} = 6 r_1-2 \left(2 r_4+r_7\right) \, , \nonumber \\[2mm]
    &\Lambda^{0,0}_{13-16} = 2 \left( r_5+2 r_8
    \pm 3 r_{10} \right) \, , \nonumber \\[2mm]
    &\Lambda^{0,0}_{17,18} = 2 (r_4 + 2 r_7) \, .
\end{align}

\subsection{The $D_4$ symmetry}

By imposing $G =D_4$
we get the quartic potential
\begin{align}
    V_{D_4} =&
    r_1 \left[ (\phi_1^\dagger \phi_1)^2 + (\phi_2^\dagger \phi_2)^2 \right]
    + r_3 |\phi_3|^4 + 2 r_4 (\phi_1^\dagger \phi_1)(\phi_2^\dagger \phi_2)
    \nonumber \\[2mm]
    &+ 2 r_5 (\phi_1^\dagger \phi_1
    + \phi_2^\dagger \phi_2)(\phi_3^\dagger \phi_3) + 2 r_7 |\phi_1^\dagger \phi_2|^2
    \nonumber \\[2mm]
    &+ 2 r_8 \left[ |\phi_1^\dagger \phi_3|^2 + |\phi_2^\dagger \phi_3|^2 \right]
    + r_{10} \left[ (\phi_1^\dagger \phi_2)^2 + h.c. \right]
    \nonumber \\[2mm]
    &+ 2 r_{11} \left[ (\phi_1^\dagger \phi_3)(\phi_2^\dagger \phi_3) + h.c. \right] \, ,
\end{align}
which can be easily achieved by setting from $\mathbb{Z}_4$
the constraints $r_2 \rightarrow r_1, r_6 \rightarrow r_5, r_9 \rightarrow r_8$.

Thus, we get the following scattering matrices.

\subsubsection*{The matrix $M_2^{++}$}

From $M_2^{++}$ we get
\begin{equation}
    \frac{1}{2} \, M_2^{++} = \mathrm{diag} \left\{
    \begin{pmatrix}
     r_4+r_7 & \sqrt{2} r_{11} \\
     \sqrt{2} r_{11} & r_3
    \end{pmatrix}
    ,
    \begin{pmatrix}
     r_1 & r_{10} \\
     r_{10} & r_1
    \end{pmatrix}
    ,
    (r_5 + r_8)
    , (r_5 + r_8) \right\} \, ,
\end{equation}
and thus we get the eigenvalues of $M_2^{++}$:
\begin{align}
    &\Lambda^{++}_{1,2} = \pm \sqrt{8 r_{11}^2 +\left(-r_3+r_4+r_7\right){}^2}+r_3+r_4+r_7 \, , \nonumber \\[2mm]
    &\Lambda^{++}_{3,4} = 2 \left(r_1 \pm r_{10} \right) 
    \, , \nonumber \\[2mm]
    &\Lambda^{++}_{5,6} = 2(r_5 + r_8) \, .
\end{align}

\subsubsection*{The matrix $M_2^+$}

From $M_2^{+}$ we get
\begin{equation}
    \frac{1}{2} \, M_2^{+} = \mathrm{diag} \left\{
    \begin{pmatrix}
     r_4 & r_7 & r_{11} \\
     r_7 & r_4 & r_{11} \\
     r_{11} & r_{11} & r_3
    \end{pmatrix}
    ,
    \begin{pmatrix}
     r_1 & r_{10} \\
     r_{10} & r_1
    \end{pmatrix}
    ,
    \begin{pmatrix}
     r_5 & r_8 \\
     r_8 & r_5
    \end{pmatrix}
    ,
    \begin{pmatrix}
     r_5 & r_8 \\
     r_8 & r_5
    \end{pmatrix} \right\} \, ,
\end{equation}
with eigenvalues of $M_2^{+}$:
\begin{align}
    &\Lambda^{+,2}_{1,2} = \Lambda^{++}_{1,2} \, , \nonumber \\[2mm]
    &\Lambda^{+,2}_{3} = 2 (r_4 - r_7) \, , \nonumber \\[2mm]
    &\Lambda^{+,2}_{4,5} = \Lambda^{++}_{3,4} \, , \nonumber \\[2mm]
    &\Lambda^{+,2}_{6-9} = 2(r_5 \pm r_8) \, .
\end{align}

\subsubsection*{The matrix $M_0^+$}

From $M_0^{+}$ we get
\begin{equation}
    \frac{1}{2} \, M_0^{+} = \mathrm{diag} \left\{
    \begin{pmatrix}
     r_1 & r_7 & r_8 \\
     r_7 & r_1 & r_8 \\
     r_8 & r_8 & r_3
    \end{pmatrix}
    ,
    \begin{pmatrix}
     r_4 & r_{10} \\
     r_{10} & r_4
    \end{pmatrix}
    ,
    \begin{pmatrix}
     r_5 & r_{11} \\
     r_{11} & r_5
    \end{pmatrix}
    ,
    \begin{pmatrix}
     r_5 & r_{11} \\
     r_{11} & r_5
    \end{pmatrix} \right\} \, ,
\end{equation}
with eigenvalues of $M_0^{+}$:
\begin{align}
    &\Lambda^{+,0}_{1,2} = \pm \sqrt{8 r_8^2 + \left(r_1-r_3+r_7\right){}^2}
    +r_1+r_3+r_7 \, , \nonumber \\[2mm]
    &\Lambda^{+,0}_{3} = 2 (r_1 - r_7 ) \, , \nonumber \\[2mm]
    &\Lambda^{+,0}_{4,5} = 2 \left(r_4 \pm r_{10} \right) \, , \nonumber \\[2mm]
    &\Lambda^{+,0}_{6-9} = 2 \left(r_5 \pm r_{11} \right) \, .
\end{align}

\subsubsection*{The matrix $M_2^0$}

From $M_2^{0}$ we get
\begin{align}
    \frac{1}{2} \, M_2^{0} = \frac{1}{2} \, M_2^{++} \ ,
\end{align}
with eigenvalues of $M_2^{0}$:
\begin{align}
    &\Lambda^{0,2}_{1-6} = \Lambda^{++}_{1-6} \, .
\end{align}

\subsubsection*{The matrix $M_0^0$}

From $M_0^{0}$ we get
\begin{align}
    \frac{1}{2} \, M_0^{0} \sim \, &
    \mathrm{diag} \left\{
    \frac{1}{2} M_0^{+}
    ,
    \begin{pmatrix}
     3 r_1 & 2 r_4+r_7 & 2 r_5+r_8 \\
     2 r_4+r_7 & 3 r_1 & 2 r_5+r_8 \\
     2 r_5+r_8 & 2 r_5+r_8 & 3 r_3
    \end{pmatrix}
    ,
    \begin{pmatrix}
     r_4+2 r_7 & 3 r_{10} \\
     3 r_{10} & r_4+2 r_7
    \end{pmatrix} , \right.
    \nonumber \\[3mm]
    & 
\hspace{15mm}
\left.
    \begin{pmatrix}
     r_5+2 r_8 & 3 r_{11} \\
     3 r_{11} & r_5+2 r_8
    \end{pmatrix}
    ,
    \begin{pmatrix}
     r_5+2 r_8 & 3 r_{11} \\
     3 r_{11} & r_5+2 r_8
    \end{pmatrix} \right\} \, ,
\end{align}
with eigenvalues of $M_0^{0}$:
\begin{align}
    &\Lambda^{0,0}_{1-9} = \Lambda^{+,0}_{1-9} \, , \nonumber \\[2mm]
    &\Lambda^{0,0}_{10,11} = \pm \sqrt{\left(3 r_1-3 r_3+2 r_4+r_7\right){}^2
    +8 \left(2 r_5+r_8\right){}^2}+3 r_1+3 r_3+2 r_4+r_7, \, , \nonumber \\[2mm]
    &\Lambda^{0,0}_{12} = 6 r_1-2 \left(2 r_4+r_7\right) \, , \nonumber \\[2mm]
    &\Lambda^{0,0}_{13,14} =  2 \left(\pm 3 r_{10}
    +r_4+2 r_7\right) \, , \nonumber \\[2mm]
    &\Lambda^{0,0}_{15-18} = 2 \left(\pm 3 r_{11}
    +r_5+2 r_8\right) \, .
\end{align}

\subsection{The $S_3$ symmetry}

By imposing $G =S_3$
we get the quartic potential
\begin{align}
    V_{S_3} =&
    r_1 \left[ (\phi_1^\dagger \phi_1)^2 + (\phi_2^\dagger \phi_2)^2 \right]
    + r_3 |\phi_3|^4 + 2 r_4 (\phi_1^\dagger \phi_1)(\phi_2^\dagger \phi_2)
    \nonumber \\[2mm]
    &+ 2 r_5 (\phi_1^\dagger \phi_1
    + \phi_2^\dagger \phi_2)(\phi_3^\dagger \phi_3) + 2 r_7 |\phi_1^\dagger \phi_2|^2
    + 2 r_8 \left[ |\phi_1^\dagger \phi_3|^2 + |\phi_2^\dagger \phi_3|^2 \right]
    \nonumber \\[2mm]
    &+ \Big[
	2 c_{11} (\phi_1^\dagger \phi_3)(\phi_2^\dagger \phi_3)
    + 2c_{12} \left[ (\phi_1^\dagger \phi_2) (\phi_3^\dagger \phi_2)
    + (\phi_2^\dagger \phi_1) (\phi_3^\dagger \phi_1) \right]
    + h.c. \Big] \, ,
\end{align}
which can be easily achieved by setting from $\mathbb{Z}_3$
the constraints $r_2 \rightarrow r_1, r_6 \rightarrow r_5, r_9 \rightarrow r_8,
c_{4} \rightarrow c_{12}^*$.

Thus, we get the following scattering matrices which
reproduce in the limit of real coefficients the conditions
(37a)--(37l) of ref.~\cite{Das:2014fea}.

\subsubsection*{The matrix $M_2^{++}$}

From $M_2^{++}$ we get
\begin{equation}
    \frac{1}{2} \, M_2^{++} = \mathrm{diag} \left\{
    \begin{pmatrix}
     r_1 & \sqrt{2} c_{12} \\
     \sqrt{2} c_{12}^* & r_5 + r_8
    \end{pmatrix}
    ,
    \begin{pmatrix}
     r_5 + r_8 & \sqrt{2} c_{12} \\
     \sqrt{2} c_{12}^* & r_1
    \end{pmatrix} 
    ,
    \begin{pmatrix}
     r_4 + r_7 & \sqrt{2} c_{11} \\
     \sqrt{2} c_{11}^* & r_3
    \end{pmatrix} \right\} \, ,
\end{equation}
and thus we get the eigenvalues of $M_2^{++}$:
\begin{align}
    &\Lambda^{++}_{1,2} = \pm \sqrt{8 |c_{12}|^2 +\left(-r_1+r_5+r_8\right){}^2}
    +r_1+r_5+r_8 \, , \nonumber \\[2mm]
    &\Lambda^{++}_{3,4} = \Lambda^{++}_{1,2} \, , \nonumber \\[2mm]
    &\Lambda^{++}_{5,6} = \pm \sqrt{8 |c_{11}|^2 +\left(-r_3+r_4+r_7\right){}^2}+r_3+r_4+r_7 \, .
\end{align}

\subsubsection*{The matrix $M_2^+$}

From $M_2^{+}$ we get
\begin{equation}
    \frac{1}{2} \, M_2^{+} = \mathrm{diag} \left\{
    \begin{pmatrix}
     r_1 & c_{12}^* & c_{12}^* \\
     c_{12} & r_5 & r_8 \\
     c_{12} & r_8 & r_5
    \end{pmatrix}
    ,
    \begin{pmatrix}
     r_5 & c_{12} & r_8 \\
     c_{12}^* & r_1 & c_{12}^* \\
     r_8 & c_{12} & r_5
    \end{pmatrix}
    ,
    \begin{pmatrix}
     r_4 & r_7 & c_{11} \\
     r_7 & r_4 & c_{11} \\
     c_{11}^* & c_{11}^* & r_3
    \end{pmatrix} \right\} \, ,
\end{equation}
with eigenvalues of $M_2^{+}$:
\begin{align}
    &\Lambda^{+,2}_{1} = 2(r_5 - r_8) \, , \nonumber \\[2mm]
    &\Lambda^{+,2}_{2,3} = \Lambda^{++}_{1,2} \, , \nonumber \\[2mm]
    &\Lambda^{+,2}_{4-6} = \Lambda^{+,2}_{1-3} \, , \nonumber \\[2mm]
    &\Lambda^{+,2}_{7} = 2(r_4 - r_7) \, \nonumber \\[2mm]
    &\Lambda^{+,2}_{8,9} = \Lambda^{++}_{5,6} \, .
\end{align}

\subsubsection*{The matrix $M_0^+$}

From $M_0^{+}$ we get
\begin{equation}
    \frac{1}{2} \, M_0^{+} = \mathrm{diag} \left\{
    \begin{pmatrix}
     r_1 & r_7 & r_8 \\
     r_7 & r_1 & r_8 \\
     r_8 & r_8 & r_3
    \end{pmatrix}
    ,
    \begin{pmatrix}
     r_4 & c_{12} & c_{12}^* \\
     c_{12}^* & r_5 & c_{11} \\
     c_{12} & c_{11}^* & r_5
    \end{pmatrix}
    ,
    \begin{pmatrix}
     r_5 & c_{12}^* & c_{11} \\
     c_{12} & r_4 & c_{12}^* \\
     c_{11}^* & c_{12} & r_5
    \end{pmatrix} \right\} \, ,
\end{equation}
with eigenvalues of $M_0^{+}$:
\begin{align}
    &\Lambda^{+,0}_{1} = 2 \left(r_1-r_7\right) \, , \nonumber \\[2mm]
    &\Lambda^{+,0}_{2,3} = \pm \sqrt{\left(r_1-r_3+r_7\right){}^2
    +8 r_8^2}+r_1+r_3+r_7 \, , \nonumber \\[2mm]
    &\Lambda^{+,0}_{4-6} =  \text{Roots of:} \nonumber \\
    &\quad \quad \quad \quad x^3 - 2(r_4 + 2 r_5)x^2
    + 4(- 2|c_{12}|^2 - |c_{11}|^2+2 r_4 r_5 + r_5^2 )x \nonumber \\
    &\quad \quad \quad \quad + 8 \left( 2 r_5 |c_{12}|^2 + r_4 |c_{11}|^2 
    - 2 \Re (c_{11} c_{12}^2)
    - r_4 r_5^2 \right) = 0 \, , \nonumber \\[2mm]
    &\Lambda^{+,0}_{7-9} = \Lambda^{+,0}_{4-6} \, .
\end{align}

\subsubsection*{The matrix $M_2^0$}

From $M_2^{0}$ we get
\begin{align}
    \frac{1}{2} \, M_2^{0} = \frac{1}{2} \, M_2^{++} \ ,
\end{align}
with eigenvalues of $M_2^{0}$:
\begin{align}
    &\Lambda^{0,2}_{1-6} = \Lambda^{++}_{1-6} \, .
\end{align}

\subsubsection*{The matrix $M_0^0$}

From $M_0^{0}$ we get
\begin{align}
    \frac{1}{2} \, M_0^{0} \sim \, &
    \mathrm{diag} \left\{
    \frac{1}{2} M_0^{+}
    ,
    \begin{pmatrix}
     3 r_1 & 2 r_4+r_7 & 2 r_5+r_8 \\
     2 r_4+r_7 & 3 r_1 & 2 r_5+r_8 \\
     2 r_5+r_8 & 2 r_5+r_8 & 3 r_3
    \end{pmatrix}
    ,
    \begin{pmatrix}
     r_4+2 r_7 & 3 c_{12} & 3 c_{12}^* \\
     3 c_{12}^* & r_5+2 r_8 & 3 c_{11} \\
     3 c_{12} & 3 c_{11}^* & r_5+2 r_8
    \end{pmatrix} , \right.
    \nonumber \\[3mm]
    & 
\hspace{15mm}
\left.
    \begin{pmatrix}
     r_5+2 r_8 & 3 c_{12}^* & 3 c_{11} \\
     3 c_{12} & r_4+2 r_7 & 3 c_{12}^* \\
     3 c_{11}^* & 3 c_{12} & r_5+2 r_8
    \end{pmatrix} \right\} \, ,
\end{align}
with eigenvalues of $M_0^{0}$:
\begin{align}
    &\Lambda^{0,0}_{1-9} = \Lambda^{+,0}_{1-9} \, , \nonumber \\[2mm]
    &\Lambda^{0,0}_{10} = 6 r_1-2 \left(2 r_4+r_7\right) \, , \nonumber \\[2mm]
    &\Lambda^{0,0}_{11,12} = \pm \sqrt{\left(3 r_1-3 r_3+2 r_4+r_7\right){}^2+8 
    \left(2 r_5+r_8\right){}^2}+3 r_1+3 r_3+2 r_4+r_7  \, , \nonumber \\[2mm]
    &\Lambda^{0,0}_{13-15} = \text{Roots of:} \nonumber \\
    &\quad \quad \quad \quad x^3 + 2 \left(-r_4-2r_5-2 r_7-4 r_8\right) x^2 
    + 4 \left(-18 |c_{12}|^2 -9 |c_{11}|^2
    \right. \nonumber \\
    &\quad \quad \quad \quad \left. + r_5^2+2 r_4 r_5+4 r_7 r_5
    +4 r_8 r_5+4 r_8^2+4 r_4 r_8+8 r_7 r_8 \right) x \nonumber \\
    &\quad \quad \quad \quad +
    8 \left[ \left(r_4+2 r_7\right) \left(9 |c_{11}|^2 -\left(r_5+2 r_8\right){}^2\right) \right. \nonumber \\
    &\quad \quad \quad \quad \left. -54 \Re(c_{11} c_{12}^2) + \left(r_5+2 r_8\right)|c_{12}|^2
    \right] = 0 \, , \nonumber \\[2mm]
    &\Lambda^{0,0}_{16-18} = \Lambda^{0,0}_{13-15} \, .
\end{align}

\subsection{The $S_3 \times \mathbb{Z}_2^{(\mathrm{CP})}$ symmetry}

By imposing $G =S_3 \times \mathbb{Z}_2^{(\mathrm{CP})}$
we get the quartic potential
\begin{align}
    V_{S_3 \times \mathbb{Z}_2^{(\mathrm{CP})}} =&
    r_1 \left[ (\phi_1^\dagger \phi_1)^2 + (\phi_2^\dagger \phi_2)^2 \right]
    + r_3 |\phi_3|^4 + 2 r_4 (\phi_1^\dagger \phi_1)(\phi_2^\dagger \phi_2)
    \nonumber \\[2mm]
    &+ 2 r_5 (\phi_1^\dagger \phi_1
    + \phi_2^\dagger \phi_2)(\phi_3^\dagger \phi_3) + 2 r_7 |\phi_1^\dagger \phi_2|^2
    + 2 r_8 \left[ |\phi_1^\dagger \phi_3|^2 + |\phi_2^\dagger \phi_3|^2 \right]
    \nonumber \\[2mm]
    &+ 2 r_{10} \left[ (\phi_1^\dagger \phi_3)(\phi_2^\dagger \phi_3) + h.c. \right]
    + 2r_{11} \left[ (\phi_1^\dagger \phi_2) (\phi_3^\dagger \phi_2)
    + (\phi_2^\dagger \phi_1) (\phi_3^\dagger \phi_1) + h.c. \right]
     \, ,
\end{align}
which can be easily achieved by setting from $S_3$
the constraints $\{c_{11} , c_{12}^* \} \in \mathbb{R}$.

Thus, we get the following scattering matrices which
reproduce the conditions
(37a)--(37l) of ref.~\cite{Das:2014fea}.

\subsubsection*{The matrix $M_2^{++}$}

From $M_2^{++}$ we get
\begin{equation}
    \frac{1}{2} \, M_2^{++} = \mathrm{diag} \left\{
    \begin{pmatrix}
     r_1 & \sqrt{2} r_{11} \\
     \sqrt{2} r_{11} & r_5 + r_8
    \end{pmatrix}
    ,
    \begin{pmatrix}
     r_5 + r_8 & \sqrt{2} r_{11} \\
     \sqrt{2} r_{11} & r_1
    \end{pmatrix} 
    ,
    \begin{pmatrix}
     r_4 + r_7 & \sqrt{2} r_{10} \\
     \sqrt{2} r_{10} & r_3
    \end{pmatrix} \right\} \, ,
\end{equation}
and thus we get the eigenvalues of $M_2^{++}$:
\begin{align}
    &\Lambda^{++}_{1,2} = \pm \sqrt{8 r_{11}^2 +\left(-r_1+r_5+r_8\right){}^2}
    +r_1+r_5+r_8 \, , \nonumber \\[2mm]
    &\Lambda^{++}_{3,4} = \Lambda^{++}_{1,2} \, , \nonumber \\[2mm]
    &\Lambda^{++}_{5,6} = \pm \sqrt{8 r_{10}^2 +\left(-r_3+r_4+r_7\right){}^2}+r_3+r_4+r_7 \, .
\end{align}

\subsubsection*{The matrix $M_2^+$}

From $M_2^{+}$ we get
\begin{equation}
    \frac{1}{2} \, M_2^{+} = \mathrm{diag} \left\{
    \begin{pmatrix}
     r_1 & r_{11} & r_{11} \\
     r_{11} & r_5 & r_8 \\
     r_{11} & r_8 & r_5
    \end{pmatrix}
    ,
    \begin{pmatrix}
     r_5 & r_{11} & r_8 \\
     r_{11} & r_1 & r_{11} \\
     r_8 & r_{11} & r_5
    \end{pmatrix}
    ,
    \begin{pmatrix}
     r_4 & r_7 & r_{10} \\
     r_7 & r_4 & r_{10} \\
     r_{10} & r_{10} & r_3
    \end{pmatrix} \right\} \, ,
\end{equation}
with eigenvalues of $M_2^{+}$:
\begin{align}
    &\Lambda^{+,2}_{1} = 2(r_5 - r_8) \, , \nonumber \\[2mm]
    &\Lambda^{+,2}_{2,3} = \Lambda^{++}_{1,2} \, , \nonumber \\[2mm]
    &\Lambda^{+,2}_{4-6} = \Lambda^{+,2}_{1-3} \, , \nonumber \\[2mm]
    &\Lambda^{+,2}_{7} = 2(r_4 - r_7) \, \nonumber \\[2mm]
    &\Lambda^{+,2}_{8,9} = \Lambda^{++}_{5,6} \, .
\end{align}

\subsubsection*{The matrix $M_0^+$}

From $M_0^{+}$ we get
\begin{equation}
    \frac{1}{2} \, M_0^{+} = \mathrm{diag} \left\{
    \begin{pmatrix}
     r_1 & r_7 & r_8 \\
     r_7 & r_1 & r_8 \\
     r_8 & r_8 & r_3
    \end{pmatrix}
    ,
    \begin{pmatrix}
     r_4 & r_{11} & r_{11} \\
     r_{11} & r_5 & r_{10} \\
     r_{11} & r_{10} & r_5
    \end{pmatrix}
    ,
    \begin{pmatrix}
     r_5 & r_{11} & r_{10} \\
     r_{11} & r_4 & r_{11} \\
     r_{10} & r_{11} & r_5
    \end{pmatrix} \right\} \, ,
\end{equation}
with eigenvalues of $M_0^{+}$:
\begin{align}
    &\Lambda^{+,0}_{1} = 2 \left(r_1-r_7\right) \, , \nonumber \\[2mm]
    &\Lambda^{+,0}_{2,3} = \pm \sqrt{\left(r_1-r_3+r_7\right){}^2
    +8 r_8^2}+r_1+r_3+r_7 \, , \nonumber \\[2mm]
    &\Lambda^{+,0}_{4-6} =  \pm \sqrt{\left(-r_4+r_5+r_{10}\right){}^2+8 r_{11}^2}
    +r_4+r_5+r_{10} \, , \nonumber \\[2mm]
    &\Lambda^{+,0}_{7-9} = \Lambda^{+,0}_{4-6} \, .
\end{align}

\subsubsection*{The matrix $M_2^0$}

From $M_2^{0}$ we get
\begin{align}
    \frac{1}{2} \, M_2^{0} = \frac{1}{2} \, M_2^{++} \ ,
\end{align}
with eigenvalues of $M_2^{0}$:
\begin{align}
    &\Lambda^{0,2}_{1-6} = \Lambda^{++}_{1-6} \, .
\end{align}

\subsubsection*{The matrix $M_0^0$}

From $M_0^{0}$ we get
\begin{align}
    \frac{1}{2} \, M_0^{0} \sim \, &
    \mathrm{diag} \left\{
    \frac{1}{2} M_0^{+}
    ,
    \begin{pmatrix}
     3 r_1 & 2 r_4+r_7 & 2 r_5+r_8 \\
     2 r_4+r_7 & 3 r_1 & 2 r_5+r_8 \\
     2 r_5+r_8 & 2 r_5+r_8 & 3 r_3
    \end{pmatrix}
    ,
    \begin{pmatrix}
     r_4+2 r_7 & 3 r_{11} & 3 r_{11} \\
     3 r_{11} & r_5+2 r_8 & 3 r_{10} \\
     3 r_{11} & 3 r_{10} & r_5+2 r_8
    \end{pmatrix} , \right.
    \nonumber \\[3mm]
    & 
\hspace{15mm}
\left.
    \begin{pmatrix}
     r_5+2 r_8 & 3 r_{11} & 3 r_{10} \\
     3 r_{11} & r_4+2 r_7 & 3 r_{11} \\
     3 r_{10} & 3 r_{11} & r_5+2 r_8
    \end{pmatrix} \right\} \, ,
\end{align}
with eigenvalues of $M_0^{0}$:
\begin{align}
    &\Lambda^{0,0}_{1-9} = \Lambda^{+,0}_{1-9} \, , \nonumber \\[2mm]
    &\Lambda^{0,0}_{10} = 6 r_1-2 \left(2 r_4+r_7\right) \, , \nonumber \\[2mm]
    &\Lambda^{0,0}_{11,12} = \pm \sqrt{\left(3 r_1-3 r_3+2 r_4+r_7\right){}^2+8 
    \left(2 r_5+r_8\right){}^2}+3 r_1+3 r_3+2 r_4+r_7  \, , \nonumber \\[2mm]
    &\Lambda^{0,0}_{13,14} = \pm \sqrt{\left(-r_4+r_5-2 r_7+2 r_8+3 r_{10}\right){}^2
    +72 r_{11}^2}+r_4+r_5+2 r_7+2 r_8+3 r_{10} \, , \nonumber \\[2mm]
    &\Lambda^{0,0}_{15} = 2 \left(r_5+2 r_8-3 r_{10}\right) \, , \nonumber \\[2mm]
    &\Lambda^{0,0}_{16-18} = \Lambda^{0,0}_{13-15} \, .
\end{align}

\subsection{The $CP4$ symmetry}

By imposing $G =CP4$
we get the quartic potential
\begin{align}
    V_{CP4} =& r_1 (\phi_1^\dagger \phi_1)^2
    +r_2 \left[ (\phi_2^\dagger \phi_2)^2 + (\phi_3^\dagger \phi_3)^2 \right]
    + 2 r_4 (\phi_1^\dagger \phi_1)(\phi_2^\dagger \phi_2 + \phi_3^\dagger \phi_3)
    \nonumber \\[2mm]
    &+ 2 r_6 (\phi_2^\dagger \phi_2)(\phi_3^\dagger \phi_3)
    + 2 r_7 \left[ |\phi_1^\dagger \phi_2|^2 + |\phi_1^\dagger \phi_3|^2 \right]
    + 2 r_9 |\phi_2^\dagger \phi_3|^2 
    \nonumber \\[2mm]
    &+ 2 r_{10} \left[ (\phi_2^\dagger \phi_1)(\phi_3^\dagger \phi_1) + h.c. \right]
    + r_{11} \left[ (\phi_1^\dagger \phi_2)^2 - (\phi_1^\dagger \phi_3)^2 + h.c. \right]
    \nonumber \\[2mm]
    &+ \Big[
	c_{17} (\phi_2^\dagger \phi_3)^2
    + 2c_{16} (\phi_2^\dagger \phi_3) (\phi_2^\dagger \phi_2 - \phi_3^\dagger \phi_3)
    + h.c. \Big] \, .
\end{align}
Thus, we get the following scattering matrices which
reproduce the conditions
(4.24)--(4.32) of ref.~\cite{Bento:2017eti}.

\subsubsection*{The matrix $M_2^{++}$}

From $M_2^{++}$ we get
\begin{equation}
    \frac{1}{2} \, M_2^{++} = \mathrm{diag} \left\{
    \begin{pmatrix}
     r_1 & r_{11} & \sqrt{2} r_{10} & -r_{11} \\
     r_{11} & r_2 & \sqrt{2} c_{16} & c_{17} \\
     \sqrt{2} r_{10} & \sqrt{2} c_{16}^* & r_6+r_9 & -\sqrt{2} c_{16} \\
     -r_{11} & c_{17}^* & -\sqrt{2} c_{16}^* & r_2
    \end{pmatrix}
    ,
    (r_4 + r_7)
    ,
    (r_4 + r_7) \right\} \, ,
\end{equation}
and thus we get the eigenvalues of $M_2^{++}$:
\begin{align}
    &\Lambda^{++}_{1-4} = \text{Roots of:} \nonumber \\
    &\quad \quad \quad \quad x^4 + 2 \left(-r_1-2 r_2-r_6-r_9\right) x^3 
    + 4 \left[-4 \left| c_{16}\right| {}^2-\left| c_{17}\right| {}^2+r_1 \left(2 r_2+r_6+r_9\right)
    \right. \nonumber \\
    &\quad \quad \quad \quad \left. +r_2 \left(r_2+2 \left(r_6+r_9\right)\right)-2 \left(r_{10}^2+r_{11}^2\right) \right] x^2
    + 8 \left[ 4 \left(r_1+r_2\right) \left| c_{16}\right| {}^2
    \right. \nonumber \\
    &\quad \quad \quad \quad +\left(r_1+r_6+r_9\right) \left| c_{17}\right| {}^2
    +4 \Re \left( c_{16}^2 c_{17}^* \right) 
    +2 r_{11}^2 \left(\Re\left(c_{17}\right)+r_2+r_6+r_9\right)
    \nonumber \\
    &\quad \quad \quad \quad \left. -8 r_{10} r_{11} \Re\left(c_{16}\right)-r_2 \left(-4 r_{10}^2+r_2 \left(r_6+r_9\right)+r_1 \left(r_2+2 \left(r_6+r_9\right)\right)\right) \right] x
    \nonumber \\
    &\quad \quad \quad \quad + 16 \left[
    4 r_{11}^2 \left| c_{16}\right| {}^2-4 r_1 r_2 \left| c_{16}\right| {}^2+2 r_{10}^2 \left| c_{17}\right| {}^2-r_1 r_6 \left| c_{17}\right| {}^2-r_1 r_9 \left| c_{17}\right| {}^2
    \right. \nonumber \\
    &\quad \quad \quad \quad -2 r_{11}^2 \left(c_{16}^*\right){}^2-2 c_{16}^2 r_1 c_{17}^*-2 c_{17} r_1 \left(c_{16}^*\right){}^2+4 c_{16} r_{10} r_{11} c_{17}^*
    \nonumber \\
    &\quad \quad \quad \quad +4 c_{17} r_{10} r_{11} c_{16}^*-2 r_6 r_{11}^2 \Re\left(c_{17}\right)-2 r_9 r_{11}^2 \Re\left(c_{17}\right)+8 r_2 r_{10} r_{11} \Re\left(c_{16}\right)
    \nonumber \\
    &\quad \quad \quad \quad \left. -2 c_{16}^2 r_{11}^2-2 r_2^2 r_{10}^2-2 r_2 r_6 r_{11}^2-2 r_2 r_9 r_{11}^2+r_1 r_2^2 r_6+r_1 r_2^2 r_9 \right] = 0 \, , \nonumber \\[2mm]
    &\Lambda^{++}_{5,6} = 2(r_4 + r_7) \, .
\end{align}

\subsubsection*{The matrix $M_2^+$}

From $M_2^{+}$ we get
\begin{equation}
    \frac{1}{2} \, M_2^{+} \sim \mathrm{diag} \left\{
    \frac{1}{2} \begin{pmatrix}
     r_1 & r_{11} & 2 r_{10} & -r_{11} \\
     r_{11} & r_2 & 2 c_{16} & c_{17} \\
     2 r_{10} & 2 c_{16}^* & 2(r_6+r_9) & -2 c_{16} \\
     -r_{11} & c_{17}^* & -2c_{16}^* & r_2
    \end{pmatrix}
    , r_6-r_9
	,
    \begin{pmatrix}
     r_4 & r_7 \\
     r_7 & r_4
    \end{pmatrix}
    ,
    \begin{pmatrix}
     r_4 & r_7 \\
     r_7 & r_4
    \end{pmatrix} \right\} \, .
\label{181}
\end{equation}
Here, the procedure we have mentioned thus far,
which includes the algorithm in appendix~\ref{app:block},
yields a $5\times5$ matrix.
Supplemented by a suitable rotation, it can be written in the $4\times4$ and $1\times1$
blocks given in eq.~\eqref{181}.
The eigenvalues of $M_2^{+}$ are:
\begin{align}
    &\Lambda^{+,2}_{1-4} = \Lambda^{++}_{1-4} \, , \nonumber \\[2mm]
    &\Lambda^{+,2}_{5} = 2(r_6 - r_9) \, , \nonumber \\[2mm]
    &\Lambda^{+,2}_{6,7} = 2(r_4 \pm r_7) \, , \nonumber \\[2mm]
    &\Lambda^{+,2}_{8,9} = \Lambda^{++}_{6,7} \, .
\end{align}

\subsubsection*{The matrix $M_0^+$}

From $M_0^{+}$ we get
\begin{equation}
    \frac{1}{2} \, M_0^{+} = 
\mathrm{diag} \left\{
    \begin{pmatrix}
     r_1 & r_7 & r_7 & 0 & 0 \\
     r_7 & r_2 & r_9 & c_{16}^* & c_{16} \\
     r_7 & r_9 & r_2 & -c_{16}^* & -c_{16} \\
     0 & c_{16} & -c_{16} & r_6 & c_{17} \\
     0 & c_{16}^* & -c_{16}^* & c_{17}^* & r_6
    \end{pmatrix},
 \begin{pmatrix}
     r_4 & r_{11} & r_{10} & 0 \\
     r_{11} & r_4 & 0 & r_{10} \\
     r_{10} & 0 & r_4 & -r_{11} \\
     0 & r_{10} & -r_{11} & r_4 
    \end{pmatrix}
\right\}
    \, ,
\end{equation}
As in eq.~\eqref{181}, each matrix can be further block diagonalized in an easy
way, leading to one $3\times3$ block, one $2\times2$ block, and two identical $2\times2$ blocks.
The eigenvalues of $M_0^{+}$ are:
\begin{align}
    &\Lambda^{+,0}_{1-3} = \text{Roots of:} \nonumber \\
    &\quad \quad \quad \quad x^3 - 2(-r_2-2 r_6+r_9)x^2
    + 4(-4 \left| c_{16}\right| {}^2-\left| c_{17}\right| {}^2+r_6^2+2 r_2 r_6-2 r_6 r_9)x 
    \nonumber \\
    &\quad \quad \quad \quad + 8 \left( 4 r_6 \left| c_{16}\right| {}^2+\left(r_2-r_9\right)
    \left(\left| c_{17}\right| {}^2-r_6^2\right)-4 \Re \left( c_{16}^2 c_{17}^* \right) \right) 
    = 0 \, , \nonumber \\[2mm]
    &\Lambda^{+,0}_{4,5} = \pm \sqrt{8 r_7^2+\left(-r_1+r_2+r_9\right){}^2}
    +r_1+r_2+r_9 \, , \nonumber \\[2mm]
    &\Lambda^{+,0}_{6-7} = \Lambda^{+,0}_{8-9} =2 \left(r_4 \pm \sqrt{r_{10}^2+r_{11}^2}\right) \, .
\end{align}

\subsubsection*{The matrix $M_2^0$}

From $M_2^{0}$ we get
\begin{align}
    \frac{1}{2} \, M_2^{0} = \frac{1}{2} \, M_2^{++} \ ,
\end{align}
with eigenvalues of $M_2^{0}$:
\begin{align}
    &\Lambda^{0,2}_{1-6} = \Lambda^{++}_{1-6} \, .
\end{align}

\subsubsection*{The matrix $M_0^0$}

From $M_0^{0}$ we get the eigenvalues of $M_0^{0}$:
\begin{align}
    &\Lambda^{0,0}_{1-9} = \Lambda^{+,0}_{1-9} \, , \nonumber \\[2mm]
    &\Lambda^{0,0}_{10-12} = \text{Roots of:} \nonumber \\
    &\quad \quad \quad \quad x^3 + 2 \left(-3 r_2-3 r_9\right) x^2 
    + 4 \left(-36 \left| c_{16}\right| {}^2-9 \left| c_{17}\right| {}^2 -3 r_6^2
    \right. \nonumber \\
    &\quad \quad \quad \quad \left. +6 r_2 r_6+12 r_2 r_9-6 r_6 r_9 \right) x 
    + 8 \left[ 36 \left(r_6+2 r_9\right) \left| c_{16}\right| {}^2 \right.
    \nonumber \\
    &\quad \quad \quad \quad
    -\left(3 r_2-2 r_6-r_9\right) \left(\left(r_6+2 r_9\right){}^2-9 \left| c_{17}\right| {}^2\right) \nonumber \\
    &\quad \quad \quad \quad \left.  
    -108 \Re \left( c_{16}^2 c_{17}^* \right) \right] = 0 \, , \nonumber \\[2mm]
    &\Lambda^{0,0}_{13,14} = \pm \sqrt{8 \left(2 r_4+r_7\right){}^2
    +\left(-3 r_1+3 r_2+2 r_6+r_9\right){}^2}+3 r_1+3 r_2+2 r_6+r_9 \, , \nonumber \\[2mm]
    &\Lambda^{0,0}_{15-18} =2 \left(\pm 3 \sqrt{r_{10}^2+r_{11}^2}+r_4+2 r_7\right)
    \, ,
\end{align}
where we have suppressed the form of the matrix due to its size.

\subsection{The $\mathrm{SU}(3)$ symmetry}

By imposing $G =\mathrm{SU}(3)$
we get the quartic potential
\begin{align}
    V_{\mathrm{SU}(3)} =&
    r_1 \left[ (\phi_1^\dagger \phi_1) + (\phi_2^\dagger \phi_2) +
    (\phi_3^\dagger \phi_3) \right]^2
    \nonumber \\[2mm]
    &+ 2 r_7 \left[ |\phi_1^\dagger \phi_2|^2 + |\phi_1^\dagger \phi_3|^2 
    + |\phi_2^\dagger \phi_3|^2 \right. \nonumber \\[2mm]
    &\left.- (\phi_1^\dagger \phi_1)(\phi_2^\dagger \phi_2)
    -(\phi_1^\dagger \phi_1)(\phi_3^\dagger \phi_3) 
    - (\phi_2^\dagger \phi_2)(\phi_3^\dagger \phi_3)\right] \, ,
\end{align}
with the following scattering matrices.

\subsubsection*{The matrix $M_2^{++}$}

From $M_2^{++}$ we get
\begin{equation}
    \frac{1}{2} \, M_2^{++} =
    \mathrm{diag} (r_1, r_1, r_1, r_1, r_1, r_1) \, ,
\end{equation}
and thus we get the eigenvalues of $M_2^{++}$:
\begin{align}
    &\Lambda^{++}_{1-6} = 2 r_1 \, .
\end{align}

\subsubsection*{The matrix $M_2^+$}

From $M_2^{+}$ we get
\begin{equation}
    \frac{1}{2} \, M_2^{+} = \mathrm{diag} \left\{
    \begin{pmatrix}
     r_1 - r_7 & r_7 \\
     r_7 & r_1 - r_7
    \end{pmatrix}
    ,
    \begin{pmatrix}
     r_1 - r_7 & r_7 \\
     r_7 & r_1 - r_7
    \end{pmatrix}
    ,
    \begin{pmatrix}
     r_1 - r_7 & r_7 \\
     r_7 & r_1 - r_7
    \end{pmatrix}
    ,
    r_1
    ,
    r_1
    ,
    r_1 \right\} \, ,
\end{equation}
with eigenvalues of $M_2^{+}$:
\begin{align}
    &\Lambda^{+,2}_{1} = 2 r_1 \, , \nonumber \\[2mm]
    &\Lambda^{+,2}_{2} = 2 \left(r_1-2 r_7\right) \, , \nonumber \\[2mm]
    &\Lambda^{+,2}_{3,4} = \Lambda^{+,2}_{1,2} \, , \nonumber \\[2mm]
    &\Lambda^{+,2}_{5,6} = \Lambda^{+,2}_{1,2} \, \nonumber \\[2mm]
    &\Lambda^{+,2}_{7-9} = \Lambda^{+,2}_{1} \, .
\end{align}

\subsubsection*{The matrix $M_0^+$}

From $M_0^{+}$ we get
\begin{equation}
    \frac{1}{2} \, M_0^{+} = \mathrm{diag} \left\{
    \begin{pmatrix}
     r_1 & r_7 & r_7 \\
     r_7 & r_1 & r_7 \\
     r_7 & r_7 & r_1
    \end{pmatrix}
    ,
    r_1 - r_7,r_1 - r_7,r_1 - r_7,r_1 - r_7,r_1 - r_7,r_1 - r_7 \right\} \, ,
\end{equation}
with eigenvalues of $M_0^{+}$:
\begin{align}
    &\Lambda^{+,0}_{1} = 2 \left(r_1+2 r_7\right) \, , \nonumber \\[2mm]
    &\Lambda^{+,0}_{2,3} = 2 (r_1 - r_7) \, , \nonumber \\[2mm]
    &\Lambda^{+,0}_{4-9} = 2 (r_1 - r_7) \, .
\end{align}

\subsubsection*{The matrix $M_2^0$}

From $M_2^{0}$ we get
\begin{align}
    \frac{1}{2} \, M_2^{0} = \frac{1}{2} \, M_2^{++} \ ,
\end{align}
with eigenvalues of $M_2^{0}$:
\begin{align}
    &\Lambda^{0,2}_{1-6} = \Lambda^{++}_{1-6} \, .
\end{align}

\subsubsection*{The matrix $M_0^0$}

From $M_0^{0}$ we get
\begin{align}
    \frac{1}{2} \, M_0^{0} \sim \, &
    \mathrm{diag} \left\{
    \frac{1}{2} M_0^{+}
    ,
    \begin{pmatrix}
     3 r_1 & 2 r_1+r_7 & 2 r_1-r_7 \\
     2 r_1-r_7 & 3 r_1 & 2 r_1-r_7 \\
     2 r_1-r_7 & 2 r_1-r_7 & 3 r_1
    \end{pmatrix}
    ,
    (r_1 + r_7) , \right.
    \nonumber \\[3mm]
    &
\hspace{15mm}
    (r_1 + r_7)
    ,
    (r_1 + r_7)
    ,
    (r_1 + r_7)
    ,
    (r_1 + r_7)
    ,
    (r_1 + r_7) \Big\} \, ,
\end{align}
with eigenvalues of $M_0^{0}$:
\begin{align}
    &\Lambda^{0,0}_{1-9} = \Lambda^{+,0}_{1-9} \, , \nonumber \\[2mm]
    &\Lambda^{0,0}_{10} = 14 r_1-4 r_7 \, , \nonumber \\[2mm]
    &\Lambda^{0,0}_{11,12} = 2 \left(r_1+r_7\right) \, , \nonumber \\[2mm]
    &\Lambda^{0,0}_{13-18} = 2 \left(r_1+r_7\right) \, .
\end{align}

\subsection{The $A_4$ symmetry}

Imposing $G = A_4$ we get the quartic potential
\begin{align}
    V_{A_4} =& \frac{r_1 + 2 r_4}{3} \left[ (\phi_1^\dagger \phi_1) 
    + (\phi_2^\dagger \phi_2) + (\phi_3^\dagger \phi_3) \right]^2
    + \frac{2 (r_1 - r_4)}{3} \left[ (\phi_1^\dagger \phi_1)^2 
    + (\phi_2^\dagger \phi_2)^2  \right.
    \nonumber \\[2mm]
    &\left. + (\phi_3^\dagger \phi_3)^2
    - (\phi_1^\dagger \phi_1) (\phi_2^\dagger \phi_2)
    - (\phi_2^\dagger \phi_2) (\phi_3^\dagger \phi_3)
    - (\phi_3^\dagger \phi_3) (\phi_1^\dagger \phi_1)\right]
    \nonumber \\[2mm]
    &+ 2 r_7 \left( |\phi_1^\dagger \phi_2|^2 + |\phi_2^\dagger \phi_3|^2 
    + |\phi_3^\dagger \phi_1|^2 \right)
    \nonumber \\[2mm]
    & + \Big[
	c_3 \left[ (\phi_1^\dagger \phi_2)^2 + (\phi_2^\dagger \phi_3)^2 
    + (\phi_3^\dagger \phi_1)^2 \right]
    + h.c. \Big] \, ,
\label{V4_A4_one}
\end{align}
which can be easily achieved by setting from $\mathbb{Z}_2 \times
\mathbb{Z}_2$ with
the constraints $r_2=r_3=r_1$, $r_5=r_6=r_4$,
$r_8=r_9=r_7$ and $c_5^*=c_{17}=c_3$.
It can not be easily achieved from $\mathbb{Z}_3$ due to
our choice of basis. Nevertheless,
we work it out in appendix~\ref{app:arrows}.

Thus, we get the following scattering matrices.

\subsubsection*{The matrix $M_2^{++}$}

From $M_2^{++}$ we get
\begin{equation}
    \frac{1}{2} \, M_2^{++} = \mathrm{diag} \left\{
    \begin{pmatrix}
     r_1 & c_3 & c_3^* \\
     c_3^* & r_1 & c_{3} \\
     c_3 & c_3^* & r_1
    \end{pmatrix}
    ,
    (r_4 + r_7)
    ,
    (r_4 + r_7)
    ,
    (r_4+r_7) \right\} \, ,
\end{equation}
and thus we get the eigenvalues of $M_2^{++}$:
\begin{align}
    &\Lambda^{++}_{1} = 2 \left(2 \mathrm{Re}(c_3) + r_1\right) \, , \nonumber \\[2mm]
    &\Lambda^{++}_{2,3} = 2 \left( \pm \sqrt{3} \left| \Im\left(c_3\right)\right|
    -\Re\left(c_3\right)+r_1\right) \, , \nonumber \\[2mm]
    &\Lambda^{++}_{4-6} =  2 (r_4+r_7) \, .
\end{align}

\subsubsection*{The matrix $M_2^+$}

From $M_2^{+}$ we get
\begin{equation}
    \frac{1}{2} \, M_2^{+} = \mathrm{diag} \left\{
    \begin{pmatrix}
     r_1 & c_3 & c_3^* \\
     c_3^* & r_1 & c_{3} \\
     c_3 & c_3^* & r_1
    \end{pmatrix}
    ,
    \begin{pmatrix}
     r_4 & r_7 \\
     r_7 & r_4
    \end{pmatrix}
    ,
    \begin{pmatrix}
     r_4 & r_7 \\
     r_7 & r_4
    \end{pmatrix}
    ,
    \begin{pmatrix}
     r_4 & r_7 \\
     r_7 & r_4
    \end{pmatrix} \right\} \, ,
\end{equation}
with eigenvalues of $M_2^{+}$:
\begin{align}
    &\Lambda^{+,2}_{1-3} = \Lambda^{++}_{1-3} \, , \nonumber \\[2mm]
    &\Lambda^{+,2}_{4-9} = 2 (r_4 \pm r_7) \, .
\end{align}

\subsubsection*{The matrix $M_0^+$}

From $M_0^{+}$ we get
\begin{equation}
    \frac{1}{2} \, M_0^{+} = \mathrm{diag} \left\{
    \begin{pmatrix}
     r_1 & r_7 & r_7 \\
     r_7 & r_1 & r_7 \\
     r_7 & r_7 & r_1
    \end{pmatrix}
    ,
    \begin{pmatrix}
     r_4 & c_3 \\
     c_3^* & r_4
    \end{pmatrix}
    ,
    \begin{pmatrix}
     r_4 & c_3^* \\
     c_3 & r_4
    \end{pmatrix}
    ,
    \begin{pmatrix}
     r_4 & c_3 \\
     c_3^* & r_4
    \end{pmatrix} \right\} \, ,
\end{equation}
with eigenvalues of $M_0^{+}$:
\begin{align}
    &\Lambda^{+,0}_{1} = 2 \left(r_1+2 r_7\right) \, , \nonumber \\[2mm]
    &\Lambda^{+,0}_{2,3} = 2 \left(r_1 - r_7\right) \, , \nonumber \\[2mm]
    &\Lambda^{+,0}_{4-9} = 2 \left(r_4 \pm \sqrt{c_3} \sqrt{c_3^*}\right)
    \, .
\end{align}

\subsubsection*{The matrix $M_2^0$}

From $M_2^{0}$ we get
\begin{align}
    \frac{1}{2} \, M_2^{0} = \frac{1}{2} \, M_2^{++} \ ,
\end{align}
with eigenvalues of $M_2^{0}$:
\begin{align}
    &\Lambda^{0,2}_{1-6} = \Lambda^{++}_{1-6} \, .
\end{align}

\subsubsection*{The matrix $M_0^0$}

From $M_0^{0}$ we get
\begin{align}
    \frac{1}{2} \, M_0^{0} \sim \, &
    \mathrm{diag} \left\{
    \frac{1}{2} M_0^{+}
    ,
    \begin{pmatrix}
     3 r_1 & 2 r_4+r_7 & 2 r_4+r_7 \\
     2 r_4+r_7 & 3 r_1 & 2 r_4+r_7 \\
     2 r_4+r_7 & 2 r_4+r_7 & 3 r_1
    \end{pmatrix}
    ,
    \begin{pmatrix}
     r_4+2 r_7 & 3 c_3 \\
     3 c_3^* & r_4+2 r_7
    \end{pmatrix} , \right.
    \nonumber \\[3mm]
    &
\hspace{15mm}
\left.
    \begin{pmatrix}
     r_4+2 r_7 & 3 c_3^* \\
     3 c_3 & r_4+2 r_7
    \end{pmatrix}
    ,
    \begin{pmatrix}
     r_4+2 r_7 & 3 c_3 \\
     3 c_3^* & r_4+2 r_7
    \end{pmatrix} \right\} \, .
\end{align}
with eigenvalues of $M_0^{0}$:
\begin{align}
    &\Lambda^{0,0}_{1-9} = \Lambda^{+,0}_{1-9} \, , \nonumber \\[2mm]
    &\Lambda^{0,0}_{10} = 6 r_1+8 r_4+4 r_7 \, , \nonumber \\[2mm]
    &\Lambda^{0,0}_{11,12} = 6 r_1 - 2 \left(2 r_4+r_7\right) \, , \nonumber \\[2mm]
    &\Lambda^{0,0}_{13-18} = \pm 6 \sqrt{c_3} \sqrt{c_3^*}+2 r_4+4 r_7
    \, .
\end{align}

\subsection{The $S_4$ symmetry}

Imposing $G = S_4$ we get the quartic potential
\begin{align}
    V_{S_4} =& \frac{r_1 + 2 r_4}{3} \left[ (\phi_1^\dagger \phi_1) 
    + (\phi_2^\dagger \phi_2) + (\phi_3^\dagger \phi_3) \right]^2
    + \frac{2 (r_1 - r_4)}{3} \left[ (\phi_1^\dagger \phi_1)^2 
    + (\phi_2^\dagger \phi_2)^2  \right.
    \nonumber \\[2mm]
    &\left. + (\phi_3^\dagger \phi_3)^2
    - (\phi_1^\dagger \phi_1) (\phi_2^\dagger \phi_2)
    - (\phi_2^\dagger \phi_2) (\phi_3^\dagger \phi_3)
    - (\phi_3^\dagger \phi_3) (\phi_1^\dagger \phi_1)\right]
    \nonumber \\[2mm]
    &+ 2 r_7 \left( |\phi_1^\dagger \phi_2|^2 + |\phi_2^\dagger \phi_3|^2 
    + |\phi_3^\dagger \phi_1|^2 \right)
    \nonumber \\[2mm]
    &+ r_{10} \left((\phi_1^\dagger \phi_2)^2 + (\phi_2^\dagger \phi_3)^2 
    + (\phi_3^\dagger \phi_1)^2 + (\phi_2^\dagger \phi_1)^2 + (\phi_3^\dagger \phi_2)^2 
    + (\phi_1^\dagger \phi_3)^2 \right) \, .
\end{align}
which can be easily achieved by setting from $A_4$ with
the constraint $c_3 \in \mathbb{R}$.
It can not be easily achieved from $D_4$
or $S_3$ due to our choice of basis.

Thus, we get the following scattering matrices.

\subsubsection*{The matrix $M_2^{++}$}

From $M_2^{++}$ we get
\begin{equation}
    \frac{1}{2} \, M_2^{++} = \mathrm{diag} \left\{
    \begin{pmatrix}
     r_1 & r_{10} & r_{10} \\
     r_{10} & r_1 & r_{10} \\
     r_{10} & r_{10} & r_1
    \end{pmatrix}
    ,
    (r_4 + r_7)
    ,
    (r_4 + r_7)
    ,
    (r_4+r_7) \right\} \, ,
\end{equation}
and thus we get the eigenvalues of $M_2^{++}$:
\begin{align}
    &\Lambda^{++}_{1} =2 \left(r_1+2 r_{10}\right) \, , \nonumber \\[2mm]
    &\Lambda^{++}_{2,3} = 2 \left(r_1-r_{10}\right) \, , \nonumber \\[2mm]
    &\Lambda^{++}_{4-6} =  2 (r_4+r_7) \, .
\end{align}

\subsubsection*{The matrix $M_2^+$}

From $M_2^{+}$ we get
\begin{equation}
    \frac{1}{2} \, M_2^{+} = \mathrm{diag} \left\{
    \begin{pmatrix}
     r_1 & r_{10} & r_{10} \\
     r_{10} & r_1 & r_{10} \\
     r_{10} & r_{10} & r_1
    \end{pmatrix}
    ,
    \begin{pmatrix}
     r_4 & r_7 \\
     r_7 & r_4
    \end{pmatrix}
    ,
    \begin{pmatrix}
     r_4 & r_7 \\
     r_7 & r_4
    \end{pmatrix}
    ,
    \begin{pmatrix}
     r_4 & r_7 \\
     r_7 & r_4
    \end{pmatrix} \right\} \, ,
\end{equation}
with eigenvalues of $M_2^{+}$:
\begin{align}
    &\Lambda^{+,2}_{1-3} = \Lambda^{++}_{1-3} \, , \nonumber \\[2mm]
    &\Lambda^{+,2}_{4-9} = 2 (r_4 \pm r_7) \, .
\end{align}

\subsubsection*{The matrix $M_0^+$}

From $M_0^{+}$ we get
\begin{equation}
    \frac{1}{2} \, M_0^{+} = \mathrm{diag} \left\{
    \begin{pmatrix}
     r_1 & r_7 & r_7 \\
     r_7 & r_1 & r_7 \\
     r_7 & r_7 & r_1
    \end{pmatrix}
    ,
    \begin{pmatrix}
     r_4 & r_{10} \\
     r_{10} & r_4
    \end{pmatrix}
    ,
    \begin{pmatrix}
     r_4 & r_{10} \\
     r_{10} & r_4
    \end{pmatrix}
    ,
    \begin{pmatrix}
     r_4 & r_{10} \\
     r_{10} & r_4
    \end{pmatrix} \right\} \, ,
\end{equation}
with eigenvalues of $M_0^{+}$:
\begin{align}
    &\Lambda^{+,0}_{1} = 2 \left(r_1+2 r_7\right) \, , \nonumber \\[2mm]
    &\Lambda^{+,0}_{2,3} = 2 \left(r_1 - r_7\right) \, , \nonumber \\[2mm]
    &\Lambda^{+,0}_{4-9} = 2 \left(r_4 \pm r_{10}\right)
    \, .
\end{align}

\subsubsection*{The matrix $M_2^0$}

From $M_2^{0}$ we get
\begin{align}
    \frac{1}{2} \, M_2^{0} = \frac{1}{2} \, M_2^{++} \ ,
\end{align}
with eigenvalues of $M_2^{0}$:
\begin{align}
    &\Lambda^{0,2}_{1-6} = \Lambda^{++}_{1-6} \, .
\end{align}

\subsubsection*{The matrix $M_0^0$}

From $M_0^{0}$ we get
\begin{align}
    \frac{1}{2} \, M_0^{0} \sim \, &
    \mathrm{diag} \left\{
    \frac{1}{2} M_0^{+}
    ,
    \begin{pmatrix}
     3 r_1 & 2 r_4+r_7 & 2 r_4+r_7 \\
     2 r_4+r_7 & 3 r_1 & 2 r_4+r_7 \\
     2 r_4+r_7 & 2 r_4+r_7 & 3 r_1
    \end{pmatrix}
    ,
    \begin{pmatrix}
     r_4+2 r_7 & 3 r_{10} \\
     3 r_{10} & r_4+2 r_7
    \end{pmatrix} , \right.
    \nonumber \\[3mm]
    &
\hspace{15mm}
\left.
    \begin{pmatrix}
     r_4+2 r_7 & 3 r_{10} \\
     3 r_{10} & r_4+2 r_7
    \end{pmatrix}
    ,
    \begin{pmatrix}
     r_4+2 r_7 & 3 r_{10} \\
     3 r_{10} & r_4+2 r_7
    \end{pmatrix} \right\} \, .
\end{align}
with eigenvalues of $M_0^{0}$:
\begin{align}
    &\Lambda^{0,0}_{1-9} = \Lambda^{+,0}_{1-9} \, , \nonumber \\[2mm]
    &\Lambda^{0,0}_{10} = 6 r_1+8 r_4+4 r_7 \, , \nonumber \\[2mm]
    &\Lambda^{0,0}_{11,12} = 6 r_1 - 2 \left(2 r_4+r_7\right) \, , \nonumber \\[2mm]
    &\Lambda^{0,0}_{13-18} = 2 \left(r_4+2 r_7 \pm 3 r_{10}\right)
    \, .
\end{align}

\subsection{The $\mathrm{SO}(3)$ symmetry}

Imposing $G = \mathrm{SO}(3)$ we get the quartic potential
\begin{align}
    V_{\mathrm{SO}(3)} =& \frac{r_1 + 2 r_4}{3} \left[ (\phi_1^\dagger \phi_1) 
    + (\phi_2^\dagger \phi_2) + (\phi_3^\dagger \phi_3) \right]^2
    + \frac{2 (r_1 - r_4)}{3} \left[ (\phi_1^\dagger \phi_1)^2 
    + (\phi_2^\dagger \phi_2)^2  \right.
    \nonumber \\[2mm]
    &\left. + (\phi_3^\dagger \phi_3)^2
    - (\phi_1^\dagger \phi_1) (\phi_2^\dagger \phi_2)
    - (\phi_2^\dagger \phi_2) (\phi_3^\dagger \phi_3)
    - (\phi_3^\dagger \phi_3) (\phi_1^\dagger \phi_1)\right]
    \nonumber \\[2mm]
    &+ 2 r_7 \left( |\phi_1^\dagger \phi_2|^2 + |\phi_2^\dagger \phi_3|^2 
    + |\phi_3^\dagger \phi_1|^2 \right)
    \nonumber \\[2mm]
    &+ (r_{1} - r_4 - r_7) \left((\phi_1^\dagger \phi_2)^2 + (\phi_2^\dagger \phi_3)^2 
    + (\phi_3^\dagger \phi_1)^2 + (\phi_2^\dagger \phi_1)^2 + (\phi_3^\dagger \phi_2)^2 
    + (\phi_1^\dagger \phi_3)^2 \right) \, .
\end{align}
which can be easily achieved by setting from $S_4$ with
the constraint $r_{10} \rightarrow r_1 - r_4 - r_7$.

Thus, we get the following scattering matrices.

\subsubsection*{The matrix $M_2^{++}$}

From $M_2^{++}$ we get
\begin{align}
    \frac{1}{2} \, M_2^{++} = \mathrm{diag} \left\{
    \begin{pmatrix}
     r_1 & r_1 - r_4 - r_7 & r_1 - r_4 - r_7 \\
     r_1 - r_4 - r_7 & r_1 & r_1 - r_4 - r_7 \\
     r_1 - r_4 - r_7 & r_1 - r_4 - r_7 & r_1
    \end{pmatrix}
    ,
    (r_4 + r_7)
    ,
    (r_4 + r_7)
    ,
    (r_4+r_7) \right\} \, ,
\end{align}
and thus we get the eigenvalues of $M_2^{++}$:
\begin{align}
    &\Lambda^{++}_{1} = 2 \left(3 r_1-2 r_4-2 r_7\right) \, , \nonumber \\[2mm]
    &\Lambda^{++}_{2,3} = 2 (r_4+r_7) \, , \nonumber \\[2mm]
    &\Lambda^{++}_{4-6} =  2 (r_4+r_7) \, .
\end{align}

\subsubsection*{The matrix $M_2^+$}

From $M_2^{+}$ we get
\begin{align}
    \frac{1}{2} \, M_2^{+} = \mathrm{diag} \left\{
    \begin{pmatrix}
     r_1 & r_1 - r_4 - r_7 & r_1 - r_4 - r_7 \\
     r_1 - r_4 - r_7 & r_1 & r_1 - r_4 - r_7 \\
     r_1 - r_4 - r_7 & r_1 - r_4 - r_7 & r_1
    \end{pmatrix}
    ,
    \begin{pmatrix}
     r_4 & r_7 \\
     r_7 & r_4
    \end{pmatrix}
    ,
    \begin{pmatrix}
     r_4 & r_7 \\
     r_7 & r_4
    \end{pmatrix}
    ,
    \begin{pmatrix}
     r_4 & r_7 \\
     r_7 & r_4
    \end{pmatrix} \right\} \, ,
\end{align}
with eigenvalues of $M_2^{+}$:
\begin{align}
    &\Lambda^{+,2}_{1-3} = \Lambda^{++}_{1-3} \, , \nonumber \\[2mm]
    &\Lambda^{+,2}_{4-9} = 2 (r_4 \pm r_7) \, .
\end{align}

\subsubsection*{The matrix $M_0^+$}

From $M_0^{+}$ we get
\begin{align}
    \frac{1}{2} \, M_0^{+} =&
    \mathrm{diag} \left\{
    \begin{pmatrix}
     r_1 & r_7 & r_7 \\
     r_7 & r_1 & r_7 \\
     r_7 & r_7 & r_1
    \end{pmatrix}
    ,
    \begin{pmatrix}
     r_4 & r_1 - r_4 - r_7 \\
     r_1 - r_4 - r_7 & r_4
    \end{pmatrix} , \right.     \nonumber \\[3mm]
    &
\hspace{12mm}
\left.
    \begin{pmatrix}
     r_4 & r_1 - r_4 - r_7 \\
     r_1 - r_4 - r_7 & r_4
    \end{pmatrix}
    ,
    \begin{pmatrix}
     r_4 & r_1 - r_4 - r_7 \\
     r_1 - r_4 - r_7 & r_4
    \end{pmatrix} \right\} \, ,
\end{align}
with eigenvalues of $M_0^{+}$:
\begin{align}
    &\Lambda^{+,0}_{1} = 2 \left(r_1+2 r_7\right) \, , \nonumber \\[2mm]
    &\Lambda^{+,0}_{2,3} = 2 \left(r_1 - r_7\right) \, , \nonumber \\[2mm]
    &\Lambda^{+,0}_{4-9} = 2 \left( r_4 \pm (r_1-r_4-r_7) \right)
    \, .
\end{align}

\subsubsection*{The matrix $M_2^0$}

From $M_2^{0}$ we get
\begin{align}
    \frac{1}{2} \, M_2^{0} = \frac{1}{2} \, M_2^{++} \ ,
\end{align}
with eigenvalues of $M_2^{0}$:
\begin{align}
    &\Lambda^{0,2}_{1-6} = \Lambda^{++}_{1-6} \, .
\end{align}

\subsubsection*{The matrix $M_0^0$}

From $M_0^{0}$ we get
\begin{align}
    \frac{1}{2} \, M_0^{0} \sim \, &
    \mathrm{diag} \left\{
    \frac{1}{2} M_0^{+}
    ,
    \begin{pmatrix}
     3 r_1 & 2 r_4+r_7 & 2 r_4+r_7 \\
     2 r_4+r_7 & 3 r_1 & 2 r_4+r_7 \\
     2 r_4+r_7 & 2 r_4+r_7 & 3 r_1
    \end{pmatrix} , \ 
    \begin{pmatrix}
     r_4+2 r_7 & 3 \left(r_1-r_4-r_7\right) \\
     3 \left(r_1-r_4-r_7\right) & r_4+2 r_7
    \end{pmatrix} , \right. 
    \nonumber \\[3mm]
    &
\hspace{8mm}
\left.
    \begin{pmatrix}
     r_4+2 r_7 & 3 \left(r_1-r_4-r_7\right) \\
     3 \left(r_1-r_4-r_7\right) & r_4+2 r_7
    \end{pmatrix} ,\ 
    \begin{pmatrix}
     r_4+2 r_7 & 3 \left(r_1-r_4-r_7\right) \\
     3 \left(r_1-r_4-r_7\right) & r_4+2 r_7
    \end{pmatrix} \right\} \, .
\end{align}
with eigenvalues of $M_0^{0}$:
\begin{align}
    &\Lambda^{0,0}_{1-9} = \Lambda^{+,0}_{1-9} \, , \nonumber \\[2mm]
    &\Lambda^{0,0}_{10} = 6 r_1+8 r_4+4 r_7 \, , \nonumber \\[2mm]
    &\Lambda^{0,0}_{11,12} = 6 r_1 - 2 \left(2 r_4+r_7\right) \, , \nonumber \\[2mm]
    &\Lambda^{0,0}_{13-18} = 2(r_4 + 2 r_7) \pm 6(r_1 - r_4 - r_7)
    \, .
\end{align}

\subsection{The $\Delta(54)$ symmetry}

Imposing $G = \Delta(54)$ we get the quartic potential
\begin{align}
    V_{\Delta(54)} =& \frac{r_1 + 2 r_4}{3} \left[ (\phi_1^\dagger \phi_1) 
    + (\phi_2^\dagger \phi_2) + (\phi_3^\dagger \phi_3) \right]^2
    + \frac{2 (r_1 - r_4)}{3} \left[ (\phi_1^\dagger \phi_1)^2 
    + (\phi_2^\dagger \phi_2)^2  \right.
    \nonumber \\[2mm]
    &\left. + (\phi_3^\dagger \phi_3)^2
    - (\phi_1^\dagger \phi_1) (\phi_2^\dagger \phi_2)
    - (\phi_2^\dagger \phi_2) (\phi_3^\dagger \phi_3)
    - (\phi_3^\dagger \phi_3) (\phi_1^\dagger \phi_1)\right]
    \nonumber \\[2mm]
    &+ 2 r_7 \left( |\phi_1^\dagger \phi_2|^2 + |\phi_2^\dagger \phi_3|^2 
    + |\phi_3^\dagger \phi_1|^2 \right)
    \nonumber \\[2mm]
    &+ \Big[
	2c_{11} \left((\phi_1^\dagger \phi_3)(\phi_2^\dagger \phi_3) 
    + (\phi_2^\dagger \phi_1)(\phi_3^\dagger \phi_1)
    + (\phi_3^\dagger \phi_2)(\phi_1^\dagger \phi_2) \right) + h.c. \Big] \, .
\end{align}
which can not be easily achieved from $S_3$ due to
our choice of basis. The details are contained in appendix~\ref{app:block}

We get the following scattering matrices.

\subsubsection*{The matrix $M_2^{++}$}

From $M_2^{++}$ we get
\begin{align}
    \frac{1}{2} \, M_2^{++} =
    \mathrm{diag} \left\{
    \begin{pmatrix}
     r_1 & \sqrt{2} c_{11}^* \\
     \sqrt{2} c_{11} & r_4+r_7
    \end{pmatrix}
    ,
    \begin{pmatrix}
     r_4+r_7 & \sqrt{2} c_{11} \\
     \sqrt{2} c_{11}^* & r_1
    \end{pmatrix}
    ,
    \begin{pmatrix}
     r_4+r_7 & \sqrt{2} c_{11} \\
     \sqrt{2} c_{11}^* & r_1
    \end{pmatrix} \right\} \, ,
\end{align}
and thus we get the eigenvalues of $M_2^{++}$:
\begin{align}
    &\Lambda^{++}_{1-6} = \pm \sqrt{8 \left| c_{11}\right| {}^2+\left(-r_1+r_4+r_7\right){}^2}
    +r_1+r_4+r_7 \, .
\end{align}

\subsubsection*{The matrix $M_2^+$}

From $M_2^{+}$ we get
\begin{align}
    \frac{1}{2} \, M_2^{+} = \mathrm{diag} \left\{
    \begin{pmatrix}
     r_1 & c_{11}^* & c_{11}^* \\
     c_{11} & r_4 & r_7 \\
     c_{11} & r_7 & r_4
    \end{pmatrix}
    ,
    \begin{pmatrix}
     r_4 & r_7 & c_{11} \\
     r_7 & r_4 & c_{11} \\
     c_{11}^* & c_{11}^* & r_1
    \end{pmatrix}
    ,
    \begin{pmatrix}
     r_4 & c_{11} & r_7 \\
     c_{11}^* & r_1 & c_{11}^* \\
     r_7 & c_{11} & r_4
    \end{pmatrix} \right\} \, ,
\end{align}
with eigenvalues of $M_2^{+}$:
\begin{align}
    &\Lambda^{+,2}_{1} = 2 \left(r_4-r_7\right) \, , \nonumber \\[2mm]
    &\Lambda^{+,2}_{2,3} = \pm \sqrt{8 \left| c_{11}\right| {}^2+\left(-r_1+r_4+r_7\right){}^2}
    +r_1+r_4+r_7 \, , \nonumber \\[2mm]
    &\Lambda^{+,2}_{4-6} = \Lambda^{+,2}_{1-3} \, , \nonumber \\[2mm]
    &\Lambda^{+,2}_{7-9} = \Lambda^{+,2}_{1-3} \, .
\end{align}

\subsubsection*{The matrix $M_0^+$}

From $M_0^{+}$ we get
\begin{align}
    \frac{1}{2} \, M_0^{+} = \mathrm{diag} \left\{
    \begin{pmatrix}
     r_1 & r_7 & r_7 \\
     r_7 & r_1 & r_7 \\
     r_7 & r_7 & r_1
    \end{pmatrix}
    ,
    \begin{pmatrix}
     r_4 & c_{11} & c_{11}^* \\
     c_{11}^* & r_4 & c_{11} \\
     c_{11} & c_{11}^* & r_4
    \end{pmatrix}
    ,
    \begin{pmatrix}
     r_4 & c_{11}^* & c_{11} \\
     c_{11} & r_4 & c_{11}^* \\
     c_{11}^* & c_{11} & r_4
    \end{pmatrix} \right\} \, ,
\end{align}
with eigenvalues of $M_0^{+}$:
\begin{align}
    &\Lambda^{+,0}_{1} = 2 \left(r_1+2 r_7\right) \, , \nonumber \\[2mm]
    &\Lambda^{+,0}_{2,3} = 2 \left(r_1 - r_7\right) \, , \nonumber \\[2mm]
    &\Lambda^{+,0}_{4} = 2 \left(2 \Re\left(c_{11}\right)+r_4\right) \, , \nonumber \\[2mm]
    &\Lambda^{+,0}_{5,6} = 2 \left(\pm \sqrt{3} \left| \Im\left(c_{11}\right)\right|
    -\Re\left(c_{11}\right)+r_4\right) \, , \nonumber \\[2mm]
    &\Lambda^{+,0}_{7-9} = \Lambda^{+,0}_{4-6} \, .
\end{align}

\subsubsection*{The matrix $M_2^0$}

From $M_2^{0}$ we get
\begin{align}
    \frac{1}{2} \, M_2^{0} = \frac{1}{2} \, M_2^{++} \ ,
\end{align}
with eigenvalues of $M_2^{0}$:
\begin{align}
    &\Lambda^{0,2}_{1-6} = \Lambda^{++}_{1-6} \, .
\end{align}

\subsubsection*{The matrix $M_0^0$}

From $M_0^{0}$ we get
\begin{align}
    \frac{1}{2} \, M_0^{0} \sim \, &
    \mathrm{diag} \left\{
    \frac{1}{2} M_0^{+}
    ,
    \begin{pmatrix}
     3 r_1 & 2 r_4+r_7 & 2 r_4+r_7 \\
     2 r_4+r_7 & 3 r_1 & 2 r_4+r_7 \\
     2 r_4+r_7 & 2 r_4+r_7 & 3 r_1
    \end{pmatrix} , \right. \nonumber \\[3mm]
    &
\hspace{10mm}
\left.   
 \begin{pmatrix}
     r_4+2 r_7 & 3 c_{11} & 3 c_{11}^* \\
     3 c_{11}^* & r_4+2 r_7 & 3 c_{11} \\
     3 c_{11} & 3 c_{11}^* & r_4+2 r_7
    \end{pmatrix}, \ 
    \begin{pmatrix}
     r_4+2 r_7 & 3 c_{11}^* & 3 c_{11} \\
     3 c_{11} & r_4+2 r_7 & 3 c_{11}^* \\
     3 c_{11}^* & 3 c_{11} & r_4+2 r_7
    \end{pmatrix} \right\} \, .
\end{align}
with eigenvalues of $M_0^{0}$:
\begin{align}
    &\Lambda^{0,0}_{1-9} = \Lambda^{+,0}_{1-9} \, , \nonumber \\[2mm]
    &\Lambda^{0,0}_{10} = 6 r_1+8 r_4+4 r_7 \, , \nonumber \\[2mm]
    &\Lambda^{0,0}_{11,12} = 6 r_1 - 2 \left(2 r_4+r_7\right) \, , \nonumber \\[2mm]
    &\Lambda^{0,0}_{13} = 2 \left(6 \Re\left(c_{11}\right)+r_4+2 r_7\right) \, , \nonumber \\[2mm]
    &\Lambda^{0,0}_{14,15} = 2 \left(\pm 3 \sqrt{3} \left| \Im\left(c_{11}\right)\right| 
    -3 \Re\left(c_{11}\right)+r_4+2 r_7\right) \, , \nonumber \\[2mm]
    &\Lambda^{0,0}_{16-18} = \Lambda^{0,0}_{13-15} \, .
\end{align}

\subsection{The $\Delta(54) \rtimes \mathbb{Z}_2^{(\mathrm{CP})}$ symmetry}

Imposing $G = \Delta(54) \rtimes \mathbb{Z}_2^{(\mathrm{CP})}$ we get the quartic potential
\begin{align}
    V_{\Delta(54) \rtimes \mathbb{Z}_2^{(\mathrm{CP})}} =& \frac{r_1 + 2 r_4}{3} \left[ (\phi_1^\dagger \phi_1) 
    + (\phi_2^\dagger \phi_2) + (\phi_3^\dagger \phi_3) \right]^2
    + \frac{2 (r_1 - r_4)}{3} \left[ (\phi_1^\dagger \phi_1)^2 
    + (\phi_2^\dagger \phi_2)^2  \right.
    \nonumber \\[2mm]
    &\left. + (\phi_3^\dagger \phi_3)^2
    - (\phi_1^\dagger \phi_1) (\phi_2^\dagger \phi_2)
    - (\phi_2^\dagger \phi_2) (\phi_3^\dagger \phi_3)
    - (\phi_3^\dagger \phi_3) (\phi_1^\dagger \phi_1)\right]
    \nonumber \\[2mm]
    &+ 2 r_7 \left( |\phi_1^\dagger \phi_2|^2 + |\phi_2^\dagger \phi_3|^2 
    + |\phi_3^\dagger \phi_1|^2 \right)
    \nonumber \\[2mm]
    &+ 2r_{10} \left[(\phi_1^\dagger \phi_3)(\phi_2^\dagger \phi_3) 
    + (\phi_2^\dagger \phi_1)(\phi_3^\dagger \phi_1)
    + (\phi_3^\dagger \phi_2)(\phi_1^\dagger \phi_2) + h.c.  \right] \, .
\end{align}
which can be easily achieved by setting from $\Delta(54)$
the constraints $c_{11} \in \mathbb{R}$.

Thus we get the following scattering matrices.

\subsubsection*{The matrix $M_2^{++}$}

From $M_2^{++}$ we get
\begin{align}
    \frac{1}{2} \, M_2^{++} =
    \mathrm{diag} \left\{
    \begin{pmatrix}
     r_1 & \sqrt{2} r_{10} \\
     \sqrt{2} r_{10} & r_4+r_7
    \end{pmatrix}
    ,
    \begin{pmatrix}
     r_4+r_7 & \sqrt{2} r_{10} \\
     \sqrt{2} r_{10} & r_1
    \end{pmatrix}
    ,
    \begin{pmatrix}
     r_4+r_7 & \sqrt{2} r_{10} \\
     \sqrt{2} r_{10} & r_1
    \end{pmatrix} \right\} \, ,
\end{align}
and thus we get the eigenvalues of $M_2^{++}$:
\begin{align}
    &\Lambda^{++}_{1-6} = \pm \sqrt{8 r_{10}^2+\left(-r_1+r_4+r_7\right){}^2}
    +r_1+r_4+r_7 \, .
\end{align}

\subsubsection*{The matrix $M_2^+$}

From $M_2^{+}$ we get
\begin{align}
    \frac{1}{2} \, M_2^{+} = \mathrm{diag} \left\{
    \begin{pmatrix}
     r_1 & r_{10}  & r_{10}  \\
     r_{10}  & r_4 & r_7 \\
     r_{10}  & r_7 & r_4
    \end{pmatrix}
    ,
    \begin{pmatrix}
     r_4 & r_7 & r_{10} \\
     r_7 & r_4 & r_{10} \\
     r_{10} & r_{10} & r_1
    \end{pmatrix}
    ,
    \begin{pmatrix}
     r_4 & r_{10} & r_7 \\
     r_{10} & r_1 & r_{10} \\
     r_7 & r_{10} & r_4
    \end{pmatrix} \right\} \, ,
\end{align}
with eigenvalues of $M_2^{+}$:
\begin{align}
    &\Lambda^{+,2}_{1} = 2 \left(r_4-r_7\right) \, , \nonumber \\[2mm]
    &\Lambda^{+,2}_{2,3} = \pm \sqrt{8 r_{10}^2+\left(-r_1+r_4+r_7\right){}^2}
    +r_1+r_4+r_7 \, , \nonumber \\[2mm]
    &\Lambda^{+,2}_{4-6} = \Lambda^{+,2}_{1-3} \, , \nonumber \\[2mm]
    &\Lambda^{+,2}_{7-9} = \Lambda^{+,2}_{1-3} \, .
\end{align}

\subsubsection*{The matrix $M_0^+$}

From $M_0^{+}$ we get
\begin{align}
    \frac{1}{2} \, M_0^{+} = \mathrm{diag} \left\{
    \begin{pmatrix}
     r_1 & r_7 & r_7 \\
     r_7 & r_1 & r_7 \\
     r_7 & r_7 & r_1
    \end{pmatrix}
    ,
    \begin{pmatrix}
     r_4 & r_{10} & r_{10} \\
     r_{10} & r_4 & r_{10} \\
     r_{10} & r_{10} & r_4
    \end{pmatrix}
    ,
    \begin{pmatrix}
     r_4 & r_{10} & r_{10} \\
     r_{10} & r_4 & r_{10} \\
     r_{10} & r_{10} & r_4
    \end{pmatrix} \right\} \, ,
\end{align}
with eigenvalues of $M_0^{+}$:
\begin{align}
    &\Lambda^{+,0}_{1} = 2 \left(r_1+2 r_7\right) \, , \nonumber \\[2mm]
    &\Lambda^{+,0}_{2,3} = 2 \left(r_1 - r_7\right) \, , \nonumber \\[2mm]
    &\Lambda^{+,0}_{4} = 2 \left(r_4 + 2 r_{10}\right) \, , \nonumber \\[2mm]
    &\Lambda^{+,0}_{5,6} = 2 \left( r_4-r_{10} \right) \, , \nonumber \\[2mm]
    &\Lambda^{+,0}_{7-9} = \Lambda^{+,0}_{4-6} \, .
\end{align}

\subsubsection*{The matrix $M_2^0$}

From $M_2^{0}$ we get
\begin{align}
    \frac{1}{2} \, M_2^{0} = \frac{1}{2} \, M_2^{++} \ ,
\end{align}
with eigenvalues of $M_2^{0}$:
\begin{align}
    &\Lambda^{0,2}_{1-6} = \Lambda^{++}_{1-6} \, .
\end{align}

\subsubsection*{The matrix $M_0^0$}

From $M_0^{0}$ we get
\begin{align}
    \frac{1}{2} \, M_0^{0} \sim \, &
    \mathrm{diag} \left\{
    \frac{1}{2} M_0^{+}
    ,
    \begin{pmatrix}
     3 r_1 & 2 r_4+r_7 & 2 r_4+r_7 \\
     2 r_4+r_7 & 3 r_1 & 2 r_4+r_7 \\
     2 r_4+r_7 & 2 r_4+r_7 & 3 r_1
    \end{pmatrix} , \right. \nonumber \\[3mm]
    &
\hspace{10mm}
\left.   
    \begin{pmatrix}
     r_4+2 r_7 & 3 r_{10} & 3 r_{10} \\
     3 r_{10} & r_4+2 r_7 & 3 r_{10} \\
     3 r_{10} & 3 r_{10} & r_4+2 r_7
    \end{pmatrix} , \ 
    \begin{pmatrix}
     r_4+2 r_7 & 3 r_{10} & 3 r_{10} \\
     3 r_{10} & r_4+2 r_7 & 3 r_{10} \\
     3 r_{10} & 3 r_{10} & r_4+2 r_7
    \end{pmatrix} \right\} \, .
\end{align}
with eigenvalues of $M_0^{0}$:
\begin{align}
    &\Lambda^{0,0}_{1-9} = \Lambda^{+,0}_{1-9} \, , \nonumber \\[2mm]
    &\Lambda^{0,0}_{10} = 6 r_1+8 r_4+4 r_7 \, , \nonumber \\[2mm]
    &\Lambda^{0,0}_{11,12} = 6 r_1 - 2 \left(2 r_4+r_7\right) \, , \nonumber \\[2mm]
    &\Lambda^{0,0}_{13} = 2 \left(6 r_{10} +r_4+2 r_7\right) \, , \nonumber \\[2mm]
    &\Lambda^{0,0}_{14,15} = 2 \left(r_4+2 r_7-3 r_{10}\right) \, , \nonumber \\[2mm]
    &\Lambda^{0,0}_{16-18} = \Lambda^{0,0}_{13-15} \, .
\end{align}

\subsection{The $\Sigma(36)$ symmetry}

Imposing $G = \Sigma(36)$ we get the quartic potential
\begin{align}
    V_{\Sigma(36)} =& \frac{r_1 + 2 r_4}{3} \left[ (\phi_1^\dagger \phi_1) 
    + (\phi_2^\dagger \phi_2) + (\phi_3^\dagger \phi_3) \right]^2
    + \frac{2 (r_1 - r_4)}{3} \left[ (\phi_1^\dagger \phi_1)^2 
    + (\phi_2^\dagger \phi_2)^2  \right.
    \nonumber \\[2mm]
    &\left. + (\phi_3^\dagger \phi_3)^2
    - (\phi_1^\dagger \phi_1) (\phi_2^\dagger \phi_2)
    - (\phi_2^\dagger \phi_2) (\phi_3^\dagger \phi_3)
    - (\phi_3^\dagger \phi_3) (\phi_1^\dagger \phi_1)\right]
    \nonumber \\[2mm]
    &+ 2 r_7 \left( |\phi_1^\dagger \phi_2|^2 + |\phi_2^\dagger \phi_3|^2 
    + |\phi_3^\dagger \phi_1|^2 \right)
    \nonumber \\[2mm]
    &+ (r_1-r_4-r_7) \left((\phi_1^\dagger \phi_3)(\phi_2^\dagger \phi_3) 
    + (\phi_2^\dagger \phi_1)(\phi_3^\dagger \phi_1)
    + (\phi_3^\dagger \phi_2)(\phi_1^\dagger \phi_2) + h.c. \right) \, .
\end{align}
which can be easily achieved by setting from $\Delta(54)$ with
the constraint $c_{11} \rightarrow r_1 - r_4 - r_7$.

Thus, we get the following scattering matrices.

\subsubsection*{The matrix $M_2^{++}$}

From $M_2^{++}$ we get
\begin{align}
    \frac{1}{2} \, M_2^{++} = \mathrm{diag} \left\{
    \begin{pmatrix}
     r_1 & \frac{r_1-r_4-r_7}{\sqrt{2}} \\
     \frac{r_1-r_4-r_7}{\sqrt{2}} & r_4+r_7
    \end{pmatrix}
    ,
    \begin{pmatrix}
     r_4+r_7 & \frac{r_1-r_4-r_7}{\sqrt{2}} \\
     \frac{r_1-r_4-r_7}{\sqrt{2}} & r_1
    \end{pmatrix} 
    ,
    \begin{pmatrix}
     r_4+r_7 & \frac{r_1-r_4-r_7}{\sqrt{2}} \\
     \frac{r_1-r_4-r_7}{\sqrt{2}} & r_1
    \end{pmatrix} \right\} \, ,
\end{align}
and thus we get the eigenvalues of $M_2^{++}$:
\begin{align}
    &\Lambda^{++}_{1-6} = \pm \sqrt{3} \left| -r_1+r_4+r_7\right| +(r_1+r_4+r_7) \, .
\end{align}

\subsubsection*{The matrix $M_2^+$}

From $M_2^{+}$ we get
\begin{align}
    \frac{1}{2} \, M_2^{+} =& \mathrm{diag} \left\{
    \begin{pmatrix}
     r_1 & \frac{1}{2} \left(r_1-r_4-r_7\right) & \frac{1}{2} \left(r_1-r_4-r_7\right) \\
     \frac{1}{2} \left(r_1-r_4-r_7\right) & r_4 & r_7 \\
     \frac{1}{2} \left(r_1-r_4-r_7\right) & r_7 & r_4
    \end{pmatrix} , \right. \nonumber \\[2mm]
    &
\hspace{10mm}
    \begin{pmatrix}
     r_4 & r_7 & \frac{1}{2} \left(r_1-r_4-r_7\right) \\
     r_7 & r_4 & \frac{1}{2} \left(r_1-r_4-r_7\right) \\
     \frac{1}{2} \left(r_1-r_4-r_7\right) & \frac{1}{2} \left(r_1-r_4-r_7\right) & r_1
    \end{pmatrix} , \nonumber \\[2mm]
    &
\hspace{10mm}
\left.
    \begin{pmatrix}
     r_4 & \frac{1}{2} \left(r_1-r_4-r_7\right) & r_7 \\
     \frac{1}{2} \left(r_1-r_4-r_7\right) & r_1 & \frac{1}{2} \left(r_1-r_4-r_7\right) \\
     r_7 & \frac{1}{2} \left(r_1-r_4-r_7\right) & r_4
    \end{pmatrix} \right\} \, ,
\end{align}
with eigenvalues of $M_2^{+}$:
\begin{align}
    &\Lambda^{+,2}_{1} = 2 \left(r_4-r_7\right) \, , \nonumber \\[2mm]
    &\Lambda^{+,2}_{2,3} = \Lambda^{++}_{1,2} \, , \nonumber \\[2mm]
    &\Lambda^{+,2}_{4-6} = \Lambda^{+,2}_{1-3} \, , \nonumber \\[2mm]
    &\Lambda^{+,2}_{7-9} = \Lambda^{+,2}_{1-3} \, .
\end{align}

\subsubsection*{The matrix $M_0^+$}

From $M_0^{+}$ we get
\begin{align}
    \frac{1}{2} \, M_0^{+} =& \mathrm{diag} \left\{
    \begin{pmatrix}
     r_1 & r_7 & r_7 \\
     r_7 & r_1 & r_7 \\
     r_7 & r_7 & r_1
    \end{pmatrix}
    ,
    \begin{pmatrix}
     r_4 & \frac{1}{2} \left(r_1-r_4-r_7\right) & \frac{1}{2} \left(r_1-r_4-r_7\right) \\
     \frac{1}{2} \left(r_1-r_4-r_7\right) & r_4 & \frac{1}{2} \left(r_1-r_4-r_7\right) \\
     \frac{1}{2} \left(r_1-r_4-r_7\right) & \frac{1}{2} \left(r_1-r_4-r_7\right) & r_4
    \end{pmatrix} , \right. \nonumber \\[2mm]
    &
\hspace{10mm}
    \left.
    \begin{pmatrix}
     r_4 & \frac{1}{2} \left(r_1-r_4-r_7\right) & \frac{1}{2} \left(r_1-r_4-r_7\right) \\
     \frac{1}{2} \left(r_1-r_4-r_7\right) & r_4 & \frac{1}{2} \left(r_1-r_4-r_7\right) \\
     \frac{1}{2} \left(r_1-r_4-r_7\right) & \frac{1}{2} \left(r_1-r_4-r_7\right) & r_4
    \end{pmatrix} \right\} \, ,
\end{align}
with eigenvalues of $M_0^{+}$:
\begin{align}
    &\Lambda^{+,0}_{1} = 2 \left(r_1+2 r_7\right) \, , \nonumber \\[2mm]
    &\Lambda^{+,0}_{2,3} = 2 \left(r_1 - r_7\right) \, , \nonumber \\[2mm]
    &\Lambda^{+,0}_{4} = \Lambda^{+,0}_{2} \, , \nonumber \\[2mm]
    &\Lambda^{+,0}_{5,6} = -r_1+3 r_4+r_7 \, , \nonumber \\[2mm]
    &\Lambda^{+,0}_{7-9} = \Lambda^{+,0}_{4-6} \, .
\end{align}

\subsubsection*{The matrix $M_2^0$}

From $M_2^{0}$ we get
\begin{align}
    \frac{1}{2} \, M_2^{0} = \frac{1}{2} \, M_2^{++} \ ,
\end{align}
with eigenvalues of $M_2^{0}$:
\begin{align}
    &\Lambda^{0,2}_{1-6} = \Lambda^{++}_{1-6} \, .
\end{align}

\subsubsection*{The matrix $M_0^0$}

From $M_0^{0}$ we get
\begin{align}
    \frac{1}{2} \, M_0^{0} \sim \, &
    \mathrm{diag} \left\{
    \frac{1}{2} M_0^{+}
    ,
    \begin{pmatrix}
     3 r_1 & 2 r_4+r_7 & 2 r_4+r_7 \\
     2 r_4+r_7 & 3 r_1 & 2 r_4+r_7 \\
     2 r_4+r_7 & 2 r_4+r_7 & 3 r_1
    \end{pmatrix} , \right. \nonumber \\[10mm]
    &
\hspace{12mm}
    \frac{1}{2}
    \begin{pmatrix}
     2(r_4+2 r_7) & 3 (r_1-r_4-r_7) & 3 (r_1-r_4-r_7) \\
     3 (r_1-r_4-r_7) & 2(r_4+2 r_7) & 3 (r_1-r_4-r_7) \\
     3 (r_1-r_4-r_7) & 3 (r_1-r_4-r_7) & 2(r_4+2 r_7)
    \end{pmatrix} , \nonumber \\[10mm]
    &
\hspace{12mm}
    \left.
    \frac{1}{2}
    \begin{pmatrix}
     2(r_4+2 r_7) & 3 (r_1-r_4-r_7) & 3 (r_1-r_4-r_7) \\
     3 (r_1-r_4-r_7) & 2(r_4+2 r_7) & 3 (r_1-r_4-r_7) \\
     3 (r_1-r_4-r_7) & 3 (r_1-r_4-r_7) & 2(r_4+2 r_7)
    \end{pmatrix} \right\} \, .
\end{align}
with eigenvalues of $M_0^{0}$:
\begin{align}
    &\Lambda^{0,0}_{1-9} = \Lambda^{+,0}_{1-9} \, , \nonumber \\[2mm]
    &\Lambda^{0,0}_{10} = 6 r_1+8 r_4+4 r_7 \, , \nonumber \\[2mm]
    &\Lambda^{0,0}_{11,12} = 6 r_1 - 2 \left(2 r_4+r_7\right) \, , \nonumber \\[2mm]
    &\Lambda^{0,0}_{13} = \Lambda^{0,0}_{11} \, , \nonumber \\[2mm]
    &\Lambda^{0,0}_{14,15} = -3 r_1+5 r_4+7 r_7 \, , \nonumber \\[2mm]
    &\Lambda^{0,0}_{16-18} = \Lambda^{0,0}_{13-15} \, .
\end{align}

\section{\label{sec:concl}Conclusions}

Having found one elementary scalar particle,
the most important issue is the determination of how many such scalars
exist in nature.
The possibility that there could be three Higgs doublets has several interesting features.

A 3HDM,
in what we denote here by the symmetry-constrained
$\mathbb{Z}_2 \times \mathbb{Z}_2$ version,
was originally proposed by Weinberg \cite{Weinberg:1976hu},
in order to have a model
which simultaneously allows for CP violation and for the natural
flavour conservation (NFC) mechanism \cite{Glashow:1976nt,Paschos:1976ay}
designed to preclude flavour-changing neutral scalar exchanges.
It is also the simplest NHDM where one can have the fifth type of fermion NFC
couplings to scalars.
Indeed,
one can show that the usual NFC is only stable under the renormalization group
if one single Higgs doublet has Yukawa couplings to the
right-handed fermions of each electric charge \cite{Ferreira:2010xe}.
This yields only five cases,
dubbed in \cite{Yagyu:2016whx} types I, II, X, Y, and Z.
The first four are possible in the 2HDM.
The fifth, 
where each charged fermion sector (up-type quarks, down-type quarks, and charged leptons)
couples to a different scalar,
becomes possible in 3HDM (and for $N >3$).
3HDM are also interesting because the list of all symmetry-constrained
limits is known
\cite{Ivanov:2012fp,Ivanov:2014doa,deMedeirosVarzielas:2019rrp,Darvishi:2019dbh},
while no such list exists (currently) for larger $N$.

Such models \textit{must} obey the theoretical bounds from
bounded from below potential, verification that the chosen solution
of the stationarity equations is the global minimum,
and perturbative unitarity.
This article lists explicitly and exhaustively
the perturbative unitarity conditions for all symmetry-constrained 3HDM.

We have explored the method advocated in \cite{Bento:2017eti}
of classifying the scattering matrices
by the charge $Q$ and hypercharge $\mathcal{Y}$
of the initial/final states.
If there is an additional substructure induced by the charges of
the symmetry group, it is identified easily via the new algorithm
we propose in appendix~\ref{app:block},
without the prior need to study the implications of each specific symmetry in detail.
Appendix~\ref{app:arrows} will be useful for those wishing to relate the conditions
in a large group with those in one of its subgroups,
when the former and the latter are naturally written
in different basis for the group generators.

An important part of this article is also the use of principal minors
in order to obtain unitarity bounds \textit{without} the need
to perform matrix diagonalizations.
This is explained in detail in section~\ref{sec:procedure},
with examples provided in section~\ref{sec:applications}.

Together, these results will be necessary for anyone interested in
the rich and varied landscape of properties and signals of 3HDM.

An interesting avenue for further exploration concerns
the relation between the unitarity bounds on the
quartics couplings $z_{ij,kl}$
of \eqref{eq:VH}, on the one hand, and physical scalar masses, on the other.
If the vacuum expectation values (vev) of the scalar fields are non-vanishing,
then, in general, the physical masses involve
also the $\mu_{ij}$ couplings (to be precise, those $\mu_{ij}$ not fixed by
the quartic couplings and vevs via the stationarity equations).
Thus, in general, there is no direct relation.
For example,
the soflty broken $\mathbb{Z}_2$ 2HDM has a $\mu_{12}$ coupling
which controls the decoupling limit for all masses heavier
than the 125 GeV Higgs.
Thus, in that case,
one cannot in general find bounds on masses arising from unitarity
bounds.\footnote{However,
as pointed out in \cite{Kanemura:2015ska},
the situation is changed if the $hWW$ coupling of the 125 GeV Higgs
with the charged vector bosons has a fixed difference
from the SM predictions. Such a difference,
usually parametrized by $k_V-1$, is constrained by experiment.}
In contrast,
in the $\mu_{12}=0$, exact $\mathbb{Z}_2$-symmetric 2HDM,
which does not have a decoupling limit \cite{Gunion:2002zf},
unitarity bounds do turn into bounds on scalar masses.
The connection between symmetries, decoupling, and the impact on masses
due to unitarity bounds could be fruitful,
especially given the fact that a symmetry-constrained NHDM
will exhibit decoupling if and only if the vacuum also satisfies
the same symmetry \cite{Carrolo:2021euy}.
This issue lies beyond the scope of the present study.

\vspace{2ex}

\acknowledgments
\noindent

JPS is grateful to Howard Haber and Igor Ivanov for discussions.
The work of MPB is supported in part by the Portuguese
Funda\c{c}\~{a}o para a Ci\^{e}ncia e Tecnologia\/ (FCT)
under contract SFRH/BD/146718/2019.
This work is supported in part by FCT under Contracts
CERN/FIS-PAR/0008/2019,
PTDC/FIS-PAR/29436/2017,
UIDB/00777/2020,
and UIDP/00777/2020;
these projects are partially funded through POCTI (FEDER),
COMPETE,
QREN,
and the EU.





\appendix

\section{Notations of the 3HDM}\label{app:notations}

%

\subsection{As in Ferreira and Silva}

The notation in Ref.~\cite{Ferreira:2008zy} is
\begin{equation}
z_{ij,kl}
=
\begin{pmatrix}
\begin{pmatrix}
 r_1 & c_1 & c_2 \\
 c_1^* & r_4 & c_6 \\
 c_2^* & c_6^* & r_5 \\
\end{pmatrix}
 & 
\begin{pmatrix}
 c_1 & c_3 & c_4 \\
 r_7 & c_7 & c_8 \\
 c_9^* & c_{12} & c_{13} \\
\end{pmatrix}
 & 
\begin{pmatrix}
 c_2 & c_4 & c_5 \\
 c_9 & c_{10} & c_{11} \\
 r_8 & c_{14} & c_{15} \\
\end{pmatrix}
 \\
\begin{pmatrix}
 c_1^* & r_7 & c_9 \\
 c_3^* & c_7^* & c_{12}^* \\
 c_4^* & c_8^* & c_{13}^* \\
\end{pmatrix}
 & 
\begin{pmatrix}
 r_4 & c_7 & c_{10} \\
 c_7^* & r_2 & c_{16} \\
 c_{10}^* & c_{16}^* & r_6 \\
\end{pmatrix}
 & 
\begin{pmatrix}
 c_6 & c_8 & c_{11} \\
 c_{12}^* & c_{16} & c_{17} \\
 c_{14}^* & r_9 & c_{18} \\
\end{pmatrix}
 \\
\begin{pmatrix}
 c_2^* & c_9^* & r_8 \\
 c_4^* & c_{10}^* & c_{14}^* \\
 c_5^* & c_{11}^* & c_{15}^* \\
\end{pmatrix}
 & 
\begin{pmatrix}
 c_6^* & c_{12} & c_{14} \\
 c_8^* & c_{16}^* & r_9 \\
 c_{11}^* & c_{17}^* & c_{18}^* \\
\end{pmatrix}
 & 
\begin{pmatrix}
 r_5 & c_{13} & c_{15} \\
 c_{13}^* & r_6 & c_{18} \\
 c_{15}^* & c_{18}^* & r_3 \\
\end{pmatrix}
 \\
\end{pmatrix}\, .
\label{eq:FS_param}
\end{equation}
In general, the parameters $c$ ($r$) are complex (real).

\subsection{As in Varzielas and Ivanov}

The (partial) notation in Ref.~\cite{deMedeirosVarzielas:2019rrp} is:
\begin{equation}
z_{ij,kl}
= \frac{1}{2}
\begin{pmatrix}
\begin{pmatrix}
 2\lambda_1 & \times & \times \\
 \times & \lambda_{12} & \times \\
 \times & \times & \lambda_{13} \\
\end{pmatrix}
 & 
\begin{pmatrix}
 \times & 2\bar{\lambda}_{12} & \lambda_{6}^* \\
 {\lambda'}_{12} & \times & \times \\
 \times & \lambda_{7} & \times \\
\end{pmatrix}
 & 
\begin{pmatrix}
 \times & \lambda_{6}^* & 2\bar{\lambda}_{31}^* \\
 \times & \times & \lambda_{5} \\
 {\lambda'}_{13} & {\bar{\lambda}'}_8 & \times \\
\end{pmatrix}
 \\
\begin{pmatrix}
 \times & {{\lambda}'}_{12} & \times \\
 2\bar{\lambda}_{12}^* & \times & \lambda_{7}^* \\
 \lambda_{6} & \times & \times \\
\end{pmatrix}
 & 
\begin{pmatrix}
 \lambda_{12} & \times & \times \\
 \times & 2\lambda_2 & \times \\
 \times & \times & \lambda_{23} \\
\end{pmatrix}
 & 
\begin{pmatrix}
 \times & \times & \lambda_{5} \\
 \lambda_{7}^* & \times & 2\bar{\lambda}_{23} \\
 \bar{\lambda'}_{8}^{*} & {\lambda'}_{23} & \times \\
\end{pmatrix}
 \\
\begin{pmatrix}
 \times & \times & {\lambda}'_{13} \\
 \lambda_{6} & \times & \bar{\lambda'}_{8}^{*} \\
 2\bar{\lambda}_{31} & \lambda_{5}^* & \times \\
\end{pmatrix}
 & 
\begin{pmatrix}
 \times & \lambda_{7} & {\bar{\lambda}'}_8 \\
 \times & \times & {\lambda'}_{23} \\
 \lambda_{5}^* & 2\bar{\lambda}_{23}^* & \times \\
\end{pmatrix}
 & 
\begin{pmatrix}
 \lambda_{13} & \times & \times \\
 \times & \lambda_{23} & \times \\
 \times & \times & 2\lambda_3 \\
\end{pmatrix}
\end{pmatrix}\, ,
\end{equation}
where the entries denoted here by ``$\times$'' have not been named in
Ref.~\cite{deMedeirosVarzielas:2019rrp}.

\section{\label{app:M20=M2++}Proof that $M_2^0 = M_2^{++}$}

The proof is trivial and it follows from the definition of the matrices.
Let
\begin{equation}
    V_4 = \lambda_{ij,kl} (\Phi_i^\dagger \Phi_j) (\Phi_k^\dagger \Phi_l) \, ,
\end{equation}
with $\Phi_i^T = \left( w_i^+ \, \, n_i \right)$. Because
\begin{align}
    (M_2^{++})_{\alpha \beta} &= \frac{\partial^2 V_4}{\partial S_\alpha^{--}\partial S_\beta^{++}}
    \, , \nonumber \\[2mm]
    (M_2^{0})_{\alpha \beta} &= \frac{\partial^2 V_4}{\partial {S_\alpha^{0}}^*
    \partial S_\beta^{0}}
    \, ,
\end{align}
where $S_\alpha^{0} = \{ n_1 n_1, n_1 n_2, n_2 n_2 , \cdots , n_3 n_3 \}$ and 
$S_\alpha^{++} = \{ w_1^+ w_1^+, w_1^+ w_2^+, w_2^+ w_2^+ , \cdots , w_3^+ w_3^+ \}$,
an exchange
$n \leftrightarrow w^+$ suffices to go from one matrix to another. Thus, we only need
to show that $V_4$ is invariant under the exchange in the doublet space.
Indeed, we have for every pair $(\Phi_i^\dagger \Phi_j)$
\begin{align}
    \Phi_i^\dagger \Phi_j &=
    w_i^- w_j^+ + n_i^* n_j
    =
    \begin{pmatrix}
     w^-_i & n_i^*
    \end{pmatrix}
    \begin{pmatrix}
     0 & 1 \\
     1 & 0
    \end{pmatrix}
    \begin{pmatrix}
     0 & 1 \\
     1 & 0
    \end{pmatrix}
    \begin{pmatrix}
     w^+_j \\
     n_j
    \end{pmatrix} \nonumber \\[2mm]
    &=
    \begin{pmatrix}
     n_i^* & w^-_i
    \end{pmatrix}
    \begin{pmatrix}
     n_j \\
     w^+_j
    \end{pmatrix} = \tilde{\Phi}^\dagger_i \tilde{\Phi}_j \, ,
\end{align}
where $\tilde{\Phi}$ is the doublet after the exchange of $n \leftrightarrow w^+$.
Thus, for every NHDM we have $M_2^0 = M_2^{++}$.

\section{\label{app:block}A generalized algorithm for block diagonalization}

There is a procedure in which we may not even care
about the hypercharge and electric charge. 
In section~\ref{sec:unitbound} we first separate the matrices into
its hypercharge and electric charge charges.
Then, we use an algorithm to put the matrices into block diagonal form
using only permutations.

The method described in this appendix block diagonalizes an Hermitian matrix of arbitrary size.
Let $M$ be the matrix created with all possible combinations of quadratic
forms $(w_i^- n_j)$, as we have done so far. The procedure is as follows.
\begin{itemize}
    \item Build the matrix $M$ from all combinations.
    \item Build a matrix $P$ of the same size with zeros everywhere.
    \item Go to the first line of $M$ and for every $M_{1j} \neq 0$,
    put $P_{k j} = 1$ in consecutive lines (where $k$ runs from 1 to the number
	of nonzero entries in $M_{1j}$).
    \item Repeat this process until every line of $P$ has exactly one entry equal
    to $1$.
    \item Compute $\tilde{M} = P M P^T$. This matrix $\tilde{M}$ is now
    block diagonalized up to permutations.
\end{itemize}

Let us consider as an explicit example
the matrix in eq.~\eqref{eq:example_alg}:
\begin{equation}
    M =
    \begin{pmatrix}
     2 r_1 & 2 \sqrt{2} c_1 & 0 & 2 c_3 & 0 & 2 c_5 \\
     2 \sqrt{2} c_1^* & 2 \left(r_4+r_7\right) & 0 & 2 \sqrt{2} c_7 & 0 & 2 \sqrt{2} c_{11} \\
     0 & 0 & 2 \left(r_5+r_8\right) & 0 & 2 \left(c_{13}+c_{14}\right) & 0 \\
     2 c_3^* & 2 \sqrt{2} c_7^* & 0 & 2 r_2 & 0 & 2 c_{17} \\
     0 & 0 & 2 \left(c_{13}^*+c_{14}^*\right) & 0 & 2 \left(r_6+r_9\right) & 0 \\
     2 c_5^* & 2 \sqrt{2} c_{11}^* & 0 & 2 c_{17}^* & 0 & 2 r_3
    \end{pmatrix} \, ,
\end{equation}
which arises in the $M_2^{++}$ scattering matrix of the $\mathbb{Z}_2$-symmetric 3HDM.
Then,
\begin{equation}
P = 
    \begin{pmatrix}
     1 & 0 & 0 & 0 & 0 & 0 \\
     0 & 1 & 0 & 0 & 0 & 0 \\
     0 & 0 & 0 & 1 & 0 & 0 \\
     0 & 0 & 0 & 0 & 0 & 1 \\
     0 & 0 & 1 & 0 & 0 & 0 \\
     0 & 0 & 0 & 0 & 1 & 0
    \end{pmatrix}\, ,
\end{equation}
where:
\begin{itemize}
    \item The first line of $M$ has non-zero entries in columns
    $\{ 1,2,4,6 \}$. Then for every line in $P$ we put $1$ for the columns
    $\{ 1,2,4,6 \}$.
    \item The second line is equal to the first.
    \item The third line of $M$ has non-zero entries in columns
    $\{ 3,5 \}$. Then for every remaining line in $P$ we put $1$ for the columns
    $\{ 3,5 \}$.
    \item We are done as there are no other unique lines in $M$ or (equivalently) more
    lines in $P$.
    \item Now we compute $\tilde{M} = P M P^T$. This matrix $\tilde{M}$ is now
    block diagonalized up to permutations.
\end{itemize}
Thus,
\begin{equation}
    \tilde{M} =
    \begin{pmatrix}
     r_1 & \sqrt{2} c_1 & c_3 & c_5 & 0 & 0 \\
     \sqrt{2} c_1^* & r_4+r_7 & \sqrt{2} c_7 & \sqrt{2} c_{11} & 0 & 0 \\
     c_3^* & \sqrt{2} c_7^* & r_2 & c_{17} & 0 & 0 \\
     c_5^* & \sqrt{2} c_{11}^* & c_{17}^* & r_3 & 0 & 0 \\
     0 & 0 & 0 & 0 & r_5+r_8 & c_{13}+c_{14} \\
     0 & 0 & 0 & 0 & c_{13}^*+c_{14}^* & r_6+r_9
    \end{pmatrix} \, .
\end{equation}
This technique allows us to separate the diagonal blocks that arise
from electric charge, hypercharge and global symmetries in general.

\section{Relating basis}\label{app:arrows}

The potentials are shown in section~\ref{sec:unitbound}
choosing some particular representation for the respective symmetry.
Typically, for each symmetry, we made the choice which simplifies the presentation
of the quartic part of the respective symmetry-constrained potential.
For example,
eq.~\eqref{V_Z3_one} for the $\mathbb{Z}_3$-symmetric 3HDM
was written in the basis where the $\mathbb{Z}_3$ generator is
represented by
$\mathrm{diag}(e^{\frac{2 \pi i}{3}},e^{\frac{-2 \pi i}{3}},1)$.

But we see from fig.~\ref{fig:tree} that the $A_4$-symmetric 3HDM can be
obtained from the $\mathbb{Z}_3$-symmetric 3HDM.
When the $\mathbb{Z}_3$-symmetric 3HDM is written as in eq.~\eqref{V_Z3_one},
the $A_4$ limit arises from a complicated relation among the parameters in
eq.~\eqref{V_Z3_one}, and, moreover, it does \textit{not} have the
simple form in eq.~\eqref{V4_A4_one}.

In contrast, had we written the $\mathbb{Z}_3$-symmetric 3HDM
potential in the basis where the generator is written as in
eq.~\eqref{b_Z3_for_A4} below,
then,
imposing invariance under the appropriate additional diagonal generator,
$\mathrm{diag} (1, -1 , -1)$,
the $A_4$ potential would have the simple form in eq.~\eqref{V4_A4_one}.
This is what we show next. The remaining subsections are intended to
facilitate the interpretation of other limiting cases shown in fig.~\ref{fig:tree}.
The limiting cases present in fig.~\ref{fig:tree} and not covered in this
appendix, are trivially found using the basis choices made in section~\ref{sec:unitbound}.

\subsection{$A_4$ from $\mathbb{Z}_3$}

Going to $A_4$ from $\mathbb{Z}_3$ is easier to see with a good
choice for the basis of the latter symmetry.

Let us choose the generator of $\mathbb{Z}_3$ to be
\begin{equation}
    b = 
    \begin{pmatrix}
     0 & 1 & 0 \\
     0 & 0 & 1 \\
     1 & 0 & 0
    \end{pmatrix} \, ,
\label{b_Z3_for_A4}
\end{equation}
instead of the usual diagonal form $\mathrm{diag} (\omega ,  \omega^2, 1)$.
Then, the quartic potential is given by
\begin{align}
    V_{\mathbb{Z}_3} =&
    r_1 \left[ (\phi_1^\dagger \phi_1)^2 + (\phi_2^\dagger \phi_2)^2
    + (\phi_3^\dagger \phi_3)^2 \right] \nonumber \\[2mm] 
    & +  2 r_4 \left[ (\phi_1^\dagger \phi_1)(\phi_2^\dagger \phi_2)
    + (\phi_1^\dagger \phi_1)(\phi_3^\dagger \phi_3)
    + (\phi_2^\dagger \phi_2)(\phi_3^\dagger \phi_3) \right]
    \nonumber \\[2mm]
    & + 2 r_7 \left[ |\phi_1^\dagger \phi_2|^2
    + |\phi_2^\dagger \phi_3|^2 + |\phi_3^\dagger \phi_1|^2 \right]
    \nonumber \\[2mm]
    &+ \Big[
	2 c_1 \left[ (\phi_1^\dagger \phi_1)(\phi_1^\dagger \phi_2)
    + (\phi_2^\dagger \phi_2)(\phi_2^\dagger \phi_3)
    + (\phi_3^\dagger \phi_3)(\phi_3^\dagger \phi_1) \right]
    \nonumber \\[2mm]
    &+ 2 c_2 \left[ (\phi_1^\dagger \phi_1)(\phi_1^\dagger \phi_3)
    + (\phi_2^\dagger \phi_2)(\phi_2^\dagger \phi_1)
    + (\phi_3^\dagger \phi_3)(\phi_3^\dagger \phi_2) \right]
    \nonumber \\[2mm]
    &+ c_3 \left[ (\phi_1^\dagger \phi_2)^2
    + (\phi_2^\dagger \phi_3)^2
    + (\phi_3^\dagger \phi_1)^2 \right]
    \nonumber \\[2mm]
    &+ 2 c_4 \left[ (\phi_1^\dagger \phi_2)(\phi_1^\dagger \phi_3)
    + (\phi_2^\dagger \phi_3)(\phi_2^\dagger \phi_1)
    + (\phi_3^\dagger \phi_1)(\phi_3^\dagger \phi_2) \right]
    \nonumber \\[2mm]
    &+ 2 c_6 \left[ (\phi_1^\dagger \phi_1)(\phi_2^\dagger \phi_3)
    + (\phi_2^\dagger \phi_2)(\phi_3^\dagger \phi_1)
    + (\phi_3^\dagger \phi_3)(\phi_1^\dagger \phi_2) \right]
    \nonumber \\[2mm]
    &+ 2 c_8 \left[ (\phi_1^\dagger \phi_2)(\phi_2^\dagger \phi_3)
    + (\phi_2^\dagger \phi_3)(\phi_3^\dagger \phi_1)
    + (\phi_3^\dagger \phi_1)(\phi_1^\dagger \phi_2) \right]
	+ h.c. \Big] \, .
\end{align}
This is, of course, equivalent to the usual basis for the
symmetry.
By enforcing, in addition,
the generator $\mathrm{diag} (1, -1 , -1)$,
or equivalently,
further removing the complex coefficients $\{ c_1 , c_2 , c_4 , c_6, c_8 \}$, we get
\begin{align}
    V_{A_4} =&
    r_1 \left[ (\phi_1^\dagger \phi_1)^2 + (\phi_2^\dagger \phi_2)^2
    + (\phi_3^\dagger \phi_3)^2 \right] \nonumber \\[2mm] 
    & +  2 r_4 \left[ (\phi_1^\dagger \phi_1)(\phi_2^\dagger \phi_2)
    + (\phi_1^\dagger \phi_1)(\phi_3^\dagger \phi_3)
    + (\phi_2^\dagger \phi_2)(\phi_3^\dagger \phi_3) \right]
    \nonumber \\[2mm]
    & + 2 r_7 \left[ |\phi_1^\dagger \phi_2|^2
    + |\phi_1^\dagger \phi_3|^2 + |\phi_2^\dagger \phi_3|^2 \right]
    \nonumber \\[2mm]
    &+ \Big[ c_3 \left( (\phi_1^\dagger \phi_2)^2
    + (\phi_2^\dagger \phi_3)^2 + (\phi_3^\dagger \phi_1)^2 \right)
    + h.c. \Big] \, ,
\end{align}
which coincides with eq.~\eqref{V4_A4_one}.

\subsection{$S_4$ from $S_3$}

Going to $S_4$ from $S_3$ is easier to see with a good
choice for the basis of the latter symmetry.

Let us choose the generators of $S_3$ to be
\begin{equation}
    b = 
    \begin{pmatrix}
     0 & 1 & 0 \\
     0 & 0 & 1 \\
     1 & 0 & 0
    \end{pmatrix} \, ,
    \quad
    c =
    \begin{pmatrix}
     0 & 1 & 0 \\
     1 & 0 & 0 \\
     0 & 0 & 1
    \end{pmatrix} \, ,
\end{equation}
instead of the usual diagonal form $\mathrm{diag} (\omega ,  \omega^2, 1)$
and $c$.\footnote{Notice that the rotation of the diagonal generator
$\mathrm{diag} (\omega ,  \omega^2, 1)$ to $b$ leaves $c$ invariant.}
Then, the quartic potential is given by
\begin{align}
    V_{S_3} =&
    r_1 \left[ (\phi_1^\dagger \phi_1)^2 + (\phi_2^\dagger \phi_2)^2
    + (\phi_3^\dagger \phi_3)^2 \right] \nonumber \\[2mm] 
    & +  2 r_4 \left[ (\phi_1^\dagger \phi_1)(\phi_2^\dagger \phi_2)
    + (\phi_1^\dagger \phi_1)(\phi_3^\dagger \phi_3)
    + (\phi_2^\dagger \phi_2)(\phi_3^\dagger \phi_3) \right]
    \nonumber \\[2mm]
    & + 2 r_7 \left[ |\phi_1^\dagger \phi_2|^2
    + |\phi_2^\dagger \phi_3|^2 + |\phi_3^\dagger \phi_1|^2 \right]
    \nonumber \\[2mm]
    &+ \Big[ 2 c_1 \left( (\phi_1^\dagger \phi_1)(\phi_1^\dagger \phi_2)
    + (\phi_2^\dagger \phi_2)(\phi_2^\dagger \phi_3)
    + (\phi_3^\dagger \phi_3)(\phi_3^\dagger \phi_1) \right.
    \nonumber \\[2mm]
    &\left. + (\phi_1^\dagger \phi_1)(\phi_1^\dagger \phi_3)
    + (\phi_2^\dagger \phi_2)(\phi_2^\dagger \phi_1)
    + (\phi_3^\dagger \phi_3)(\phi_3^\dagger \phi_2) \right) + h.c. \Big]
    \nonumber \\[2mm]
    &+ r_{10} \left[ (\phi_1^\dagger \phi_2)^2
    + (\phi_2^\dagger \phi_3)^2
    + (\phi_3^\dagger \phi_1)^2 + h.c. \right]
    \nonumber \\[2mm]
    &+ \Big[ 2 c_4 \left( (\phi_1^\dagger \phi_2)(\phi_1^\dagger \phi_3)
    + (\phi_2^\dagger \phi_3)(\phi_2^\dagger \phi_1)
    + (\phi_3^\dagger \phi_1)(\phi_3^\dagger \phi_2) \right) + h.c. \Big]
    \nonumber \\[2mm]
    &+ 2 r_{11} \left[ (\phi_1^\dagger \phi_1)(\phi_2^\dagger \phi_3)
    + (\phi_2^\dagger \phi_2)(\phi_3^\dagger \phi_1)
    + (\phi_3^\dagger \phi_3)(\phi_1^\dagger \phi_2) + h.c. \right]
    \nonumber \\[2mm]
    &+ 2 r_{12} \left[ (\phi_1^\dagger \phi_2)(\phi_2^\dagger \phi_3)
    + (\phi_2^\dagger \phi_3)(\phi_3^\dagger \phi_1)
    + (\phi_3^\dagger \phi_1)(\phi_1^\dagger \phi_2) + h.c. \right] \, .
\end{align}
By enforcing the generators $\mathrm{diag} (-1, 1 , 1)$
and $\mathrm{diag} (1, 1 , -1)$, or equivalently,
removing the coefficients $\{ c_1 , c_4 , r_{11} , r_{12} \}$
we get the potential of $S_4$
\begin{align}
    V_{S_4} =&
    r_1 \left[ (\phi_1^\dagger \phi_1)^2 + (\phi_2^\dagger \phi_2)^2
    + (\phi_3^\dagger \phi_3)^2 \right] \nonumber \\[2mm] 
    & +  2 r_4 \left[ (\phi_1^\dagger \phi_1)(\phi_2^\dagger \phi_2)
    + (\phi_1^\dagger \phi_1)(\phi_3^\dagger \phi_3)
    + (\phi_2^\dagger \phi_2)(\phi_3^\dagger \phi_3) \right]
    \nonumber \\[2mm]
    & + 2 r_7 \left[ |\phi_1^\dagger \phi_2|^2
    + |\phi_2^\dagger \phi_3|^2 + |\phi_3^\dagger \phi_1|^2 \right]
    \nonumber \\[2mm]
    &+ r_{10} \left[ (\phi_1^\dagger \phi_2)^2
    + (\phi_2^\dagger \phi_3)^2 + (\phi_3^\dagger \phi_1)^2 + h.c. \right] 
    \, .
\end{align}
The unitarity of $S_3$ in the new basis, from which
we go to $S_4$ is not trivial to compute. Although we know
what the result should be, the new scattering matrices are rotated
with an orthogonal transformation. Thus, they can not be
trivially block diagonalized.

\subsection{$D_4$ from $\mathbb{Z}_2 \times \mathbb{Z}_2$}

Going to $D_4$ from $\mathbb{Z}_2 \times \mathbb{Z}_2$ is easier to see with a good
choice for the basis of the latter symmetry.

Let us choose the generators of $\mathbb{Z}_2 \times \mathbb{Z}_2$ to be
\begin{equation}
    c' = 
    \begin{pmatrix}
     0 & 1 & 0 \\
     1 & 0 & 0 \\
     0 & 0 & -1
    \end{pmatrix} \, ,
    \quad
    c = -
    \begin{pmatrix}
     1 & 0 & 0 \\
     0 & 0 & 1 \\
     0 & 1 & 0
    \end{pmatrix} \, ,
\end{equation}
instead of the usual diagonal forms $\mathrm{diag} (-1, -1 ,  1)$
and $\mathrm{diag} (1, -1 ,  -1)$.
Then, the quartic potential is given by
\begin{align}
    V_{\mathbb{Z}_2 \times \mathbb{Z}_2} =&
    r_1 \left[ (\phi_1^\dagger \phi_1)^2 + (\phi_2^\dagger \phi_2)^2 \right]
    + r_3 |\phi_3|^4 + 2 r_4 (\phi_1^\dagger \phi_1)(\phi_2^\dagger \phi_2)
    \nonumber \\[2mm]
    &+ 2 r_5 (\phi_1^\dagger \phi_1
    + \phi_2^\dagger \phi_2)(\phi_3^\dagger \phi_3) + 2 r_7 |\phi_1^\dagger \phi_2|^2
    \nonumber \\[2mm]
    &+ 2 r_8 \left[ |\phi_1^\dagger \phi_3|^2 + |\phi_2^\dagger \phi_3|^2 \right]
    \nonumber \\[2mm]
    &+ \text{complex terms}
    \, .
\end{align}
By enforcing the generator of  $\mathbb{Z}_4$
given by $\mathrm{diag} (i, -i , 1)$ we remove all complex coefficients
except $c_3$ and $c_{11}$, which can be rephased to be real. Thus,
we get the potential
\begin{align}
    V_{D_4} =&
    r_1 \left[ (\phi_1^\dagger \phi_1)^2 + (\phi_2^\dagger \phi_2)^2 \right]
    + r_3 |\phi_3|^4 + 2 r_4 (\phi_1^\dagger \phi_1)(\phi_2^\dagger \phi_2)
    \nonumber \\[2mm]
    &+ 2 r_5 (\phi_1^\dagger \phi_1
    + \phi_2^\dagger \phi_2)(\phi_3^\dagger \phi_3) + 2 r_7 |\phi_1^\dagger \phi_2|^2
    \nonumber \\[2mm]
    &+ 2 r_8 \left[ |\phi_1^\dagger \phi_3|^2 + |\phi_2^\dagger \phi_3|^2 \right]
    + r_{10} \left[ (\phi_1^\dagger \phi_2)^2 + h.c. \right]
    \nonumber \\[2mm]
    &+ 2 r_{11} \left[ (\phi_1^\dagger \phi_3)(\phi_2^\dagger \phi_3) + h.c. \right] \, ,
\end{align}

\subsection{$S_3$ from $\mathbb{Z}_2$}

Going to $S_3$ from $\mathbb{Z}_2$ is easier to see with a good
choice for the basis of the latter symmetry.

Let us choose the generator of $\mathbb{Z}_2$ to be
\begin{equation}
    a_2' =
    \begin{pmatrix}
     0 & 1 & 0 \\
     1 & 0 & 0 \\
     0 & 0 & 1
    \end{pmatrix} \, ,
\end{equation}
instead of the usual diagonal form $\mathrm{diag} (1, 1 ,  -1)$.
Then, the quartic potential is given by
\begin{align}
    V_{\mathbb{Z}_2} =&
    r_1 \left[ (\phi_1^\dagger \phi_1)^2 + (\phi_2^\dagger \phi_2)^2 \right]
    + r_3 |\phi_3|^4 + 2 r_4 (\phi_1^\dagger \phi_1)(\phi_2^\dagger \phi_2)
    \nonumber \\[2mm]
    &+ 2 r_5 (\phi_1^\dagger \phi_1
    + \phi_2^\dagger \phi_2)(\phi_3^\dagger \phi_3) + 2 r_7 |\phi_1^\dagger \phi_2|^2
    \nonumber \\[2mm]
    &+ 2 r_8 \left[ |\phi_1^\dagger \phi_3|^2 + |\phi_2^\dagger \phi_3|^2 \right]
    \nonumber \\[2mm]
    &+ \text{complex terms}
    \, .
\end{align}
By enforcing the generator of  $\mathbb{Z}_3$
given by $\mathrm{diag} (\omega , \omega^2, 1)$ we remove all 
remaining complex coefficients
except $c_{11}$ and $c_{12}$. Thus,
we get the potential
\begin{align}
    V_{S_3} =&
    r_1 \left[ (\phi_1^\dagger \phi_1)^2 + (\phi_2^\dagger \phi_2)^2 \right]
    + r_3 |\phi_3|^4 + 2 r_4 (\phi_1^\dagger \phi_1)(\phi_2^\dagger \phi_2)
    \nonumber \\[2mm]
    &+ 2 r_5 (\phi_1^\dagger \phi_1
    + \phi_2^\dagger \phi_2)(\phi_3^\dagger \phi_3) + 2 r_7 |\phi_1^\dagger \phi_2|^2
    + 2 r_8 \left[ |\phi_1^\dagger \phi_3|^2 + |\phi_2^\dagger \phi_3|^2 \right]
    \nonumber \\[2mm]
    &+ \Big[ 2 c_{11} (\phi_1^\dagger \phi_3)(\phi_2^\dagger \phi_3)
    + 2c_{12} \left( (\phi_1^\dagger \phi_2) (\phi_3^\dagger \phi_2)
    + (\phi_2^\dagger \phi_1) (\phi_3^\dagger \phi_1) \right)
    + h.c. \Big]\, ,
\end{align}

\subsection{$\Sigma(36)$ from $\mathbb{Z}_4$}

Going to $\Sigma(36)$ from $\mathbb{Z}_4$ is easier to see with a good
choice for the basis of the latter symmetry.

Let us choose the generator of $\mathbb{Z}_4$ to be
\begin{equation}
    d = \frac{i}{\sqrt{3}}
    \begin{pmatrix}
     \omega^2 & \omega & 1 \\
     \omega & \omega^2 & 1 \\
     1 & 1 & 1
    \end{pmatrix} \, ,
\end{equation}
instead of the usual diagonal form $\mathrm{diag} (i, -i ,  1)$.
Then, by also using 
\begin{equation}
    b = 
    \begin{pmatrix}
     0 & 1 & 0 \\
     0 & 0 & 1 \\
     1 & 0 & 0
    \end{pmatrix} \, ,
    \quad
    c =
    \begin{pmatrix}
     0 & 1 & 0 \\
     1 & 0 & 0 \\
     0 & 0 & 1
    \end{pmatrix} \, ,
\end{equation}
and $\mathrm{diag} (\omega ,  \omega^2, 1)$, we get the
symmetry $\Sigma(36)$.

\bibliographystyle{JHEP}
\bibliography{bibliography}

\end{document}